\documentclass[superscriptaddress, aps, preprintnumbers,
amsmath, amssymb, sort&compress, nofootinbib, 10pt, paper, floatfix]{revtex4-2}

\usepackage{amsfonts,amsmath,amssymb,ascmac,bm,tensor, microtype}
\usepackage{fnpct} 
\usepackage{comment}
\usepackage{ifpdf}
\usepackage{slashed}
\usepackage{color}
\usepackage[mathscr]{eucal}
\usepackage[utf8]{inputenc}
\usepackage{cancel}

\ifpdf
  \usepackage{graphicx}     
  \usepackage[bookmarksopen,colorlinks=true,linkcolor=bblue,citecolor=bblue,urlcolor=ppink]{hyperref}
\else     
\fi

\definecolor{red}{rgb}{1,0,0}
\definecolor{darkred}{rgb}{0.6,0,0}
\definecolor{darkgreen}{rgb}{0.992447,0.623778,0.034597}
\definecolor{ppink}{rgb}{1,0.4,0.4}
\definecolor{bblue}{rgb}{0.284602,0.317763,0.963947}
\definecolor{purple}{rgb}{0.5 ,0, 0.7}

\newcommand{\vev}[1]{ \left< {#1} \right> }
\newcommand{\dd}{\mathrm{d}}

\newcommand{\GW}{\text{GW}}

\newcommand{\MeV}{\text{MeV}}
\newcommand{\tot}{\text{tot}}
\newcommand{\ee}{\text{e}}
\newcommand{\hh}{\text{h}}
\newcommand{\Hz}{\text{Hz}}

\newcommand{\Pl}{\text{Pl}}

\newcommand{\en}{{\text{end}}}
\newcommand{\eq}{{\text{eq}}}

\newcommand{\Mpc}{{\text{Mpc}}}

\newcommand{\rr}{\text{r}}
\newcommand{\cc}{\text{c}}

\newcommand{\kk}{\text{kin}}

\makeatletter
\newcommand\footnoteref[1]{\protected@xdef\@thefnmark{\ref{#1}}\@footnotemark}
\makeatother

\allowdisplaybreaks[1]

\begin{document}

\title{
Induced Gravitational Waves with Kination Era for Recent Pulsar Timing Array Signals
}

\author{Keisuke Harigaya}
\affiliation{Department of Physics, University of Chicago, Chicago, IL 60637, USA}
\affiliation{Kavli Institute for Cosmological Physics and Enrico Fermi Institute, University of Chicago, Chicago, IL 60637, USA}
\affiliation{Kavli Institute for the Physics and Mathematics of the Universe (WPI),
The University of Tokyo Institutes for Advanced Study,
The University of Tokyo, Kashiwa, Chiba 277-8583, Japan}

\author{Keisuke Inomata}
\affiliation{Kavli Institute for Cosmological Physics and Enrico Fermi Institute, University of Chicago, Chicago, IL 60637, USA}

\author{Takahiro Terada}
\affiliation{Particle Theory and Cosmology Group, Center for Theoretical Physics of the Universe, 
Institute for Basic Science (IBS), Daejeon, 34126, Korea}

\preprint{CTPU-PTC-23-40}

\begin{abstract}
\noindent
The evidence of the stochastic gravitational-wave background around the nano-hertz frequency range was recently found by worldwide pulsar timing array (PTA) collaborations. One of the cosmological explanations is the gravitational waves induced by enhanced curvature perturbations, but the issue of primordial black hole (PBH) overproduction in this scenario was pointed out in the literature. Motivated by this issue and the $\Omega_\text{GW} \sim f^2$ scaling suggested by the data, we study the gravitational waves induced in a cosmological epoch dominated by a stiff fluid ($w=1$) and find that they can safely explain the PTA data well without PBH overproduction. 
\end{abstract}

\maketitle

\section{Introduction} \label{sec:intro}

Gravitational waves (GWs) are a unique probe of the early Universe as well as astrophysical phenomena.  Following the first indirect measurement of GWs from the observations of periods of a binary pulsar~\cite{Taylor:1979zz}, GWs from binary black hole/neutron star mergers were finally detected directly by the ground-based laser interferometers~\cite{Abbott:2016blz, LIGOScientific:2021djp}. This was the beginning of GW (and multi-messenger) astronomy. Recently, the long-sought evidence of the Hellings-Downs curve~\cite{Hellings:1983fr} was observed by the worldwide pulsar timing array (PTA) collaborations, in particular, by NANOGrav~\cite{NANOGrav:2023gor, NANOGrav:2023hde, NANOGrav:2023hvm} and by EPTA/InPTA~\cite{Antoniadis:2023rey, Antoniadis:2023utw, Antoniadis:2023zhi}; see also the results of PPTA~\cite{Reardon:2023gzh, Zic:2023gta, Reardon:2023zen} and CPTA~\cite{Xu:2023wog}.  This indicates the first evidence of the isotropic~\cite{NANOGrav:2023tcn}, stochastic~\cite{NANOGrav:2023pdq, Antoniadis:2023bjw} GW background (SGWB) in the Universe.  Whatever its origin is, it will give us new insight into our Universe. While the arguably standard interpretation is astrophysical, namely due to merger events of supermassive black hole binaries (SMBHBs), it requires somewhat nonstandard parameters to fit the data~\cite{NANOGrav:2023hfp, Antoniadis:2023zhi, Ellis:2023dgf} (see also Ref.~\cite{Depta:2023qst} for the interpretation as primordial SMBHBs). For instance, circular SMBHBs that lose their energy dominantly by the GW emission are excluded at 3.9$\sigma$~\cite{Figueroa:2023zhu}.  Though it is premature to draw any conclusions, and new-physics (or cosmological) interpretations always bring some non-minimal assumptions in the early Universe, some new-physics interpretations can fit the PTA data better~\cite{NANOGrav:2023hvm, Figueroa:2023zhu, Ellis:2023oxs} (see, however, also Ref.~\cite{Bian:2023dnv}).  Thus, we may be at the beginning of the era of observational GW cosmology.

It is intriguing that the Bayesian analyses by NANOGrav~\cite{NANOGrav:2023hvm} favor some particular new-physics interpretations.  One of the best scenarios is the so-called scalar-induced GWs (SIGWs)~\cite{NANOGrav:2023hvm, Figueroa:2023zhu}, namely, GWs secondarily induced from the curvature perturbations~\cite{10.1143/PTP.37.831, Matarrese:1993zf, Matarrese:1997ay, Ananda:2006af, Baumann:2007zm}. 
One reason why the SIGW interpretation is favored is that it is relatively easy to fit the observed slope of the spectrum $n_\text{T} \equiv \mathrm{d} \ln \Omega_\text{GW} / \mathrm{d}\ln f \sim 2$,  which we discuss further later. 
Another reason is that the large amplitude of the signal can be easily realized by simply assuming the large amplitude of the primordial curvature perturbations.
Of course, deriving such a power spectrum from a natural UV-complete model of inflation without any fine-tuning is much more difficult than just fitting the data phenomenologically. For the inflationary model building approach in the context of the induced GWs, see, e.g., Refs.~\cite{Inomata:2016rbd, Orlofsky:2016vbd, Di:2017ndc, Ando:2017veq,Drees:2019xpp, Fu:2022ypp}.  In this paper, we follow the phenomenological approach as a first step to narrow down possible underlying inflation models. For recent attempts to explain the PTA data by SIGWs, see Refs.~\cite{Vaskonen:2020lbd, DeLuca:2020agl, Kohri:2020qqd, Domenech:2020ers, Inomata:2020xad, Kawasaki:2021ycf, Dandoy:2023jot, Madge:2023cak, Franciolini:2023pbf,Cai:2023dls, Wang:2023ost, Liu:2023ymk,  Abe:2023yrw, Zhu:2023faa, Firouzjahi:2023lzg,  Bari:2023rcw, HosseiniMansoori:2023mqh, Balaji:2023ehk, Zhao:2023joc, Liu:2023pau, Yi:2023tdk, Frosina:2023nxu, Choudhury:2023wrm, Kawasaki:2023rfx,Yi:2023npi}.

Typically associated to substantial SIGWs are primordial black holes (PBHs)~\cite{Saito:2008jc, Saito:2009jt, Bugaev:2009zh}. This is because strong SIGWs are produced by substantial curvature perturbations, which also produce PBHs. This fact can be seen both as advantages and disadvantages. For example, PBHs can serve as a dark matter candidate, and in some scenarios, one can explain the dark matter abundance by PBHs and simultaneously predict observable SIGWs~\cite{Saito:2008jc, Garcia-Bellido:2017aan, Inomata:2017vxo,Bartolo:2018rku}. However, in some cases, the PBH abundance exceeds the observed dark matter abundance~\cite{Kawasaki:2015ppx,Pattison:2017mbe,Ezquiaga:2019ftu, Tada:2023pue} and then the model is excluded. This is because the intensity of the SIGWs depends on the square of the power spectrum of the curvature perturbations whereas the abundance of PBHs is exponentially sensitive to the inverse of the power spectrum of the curvature perturbations. On top of that, the PBH abundance is subject to orders-of-magnitude uncertainties depending on the calculation method (see, e.g., Refs.~\cite{Ando:2018qdb, Young:2019osy, Yoo:2020dkz, DeLuca:2023tun} and references therein). Nevertheless, multiple recent works~\cite{Dandoy:2023jot, Franciolini:2023pbf, Liu:2023ymk} pointed out the PBH overproduction issue when one fits the PTA data by SIGWs and discussed the necessity of non-Gaussianity of primordial curvature perturbations to avoid the overproduction.
To evade this issue, enhanced sound speed $c_\text{s}=1$ beyond the perfect fluid was studied in Ref.~\cite{Balaji:2023ehk}. With a similar motivation, Ref.~\cite{Gorji:2023sil} studied GWs induced from an extra spectator tensor field.

As mentioned above, the latest data favors the spectral index $n_\text{T}$ around 2. The NANOGrav data show $n_\text{T} = 1.8 \pm 0.6$ (90\% credible region)~\cite{NANOGrav:2023gor}. The EPTA/InPTA data have an internal tension between the full data set leading to $n_\text{T}  = 0.81 ^{+0.63}_{-0.73}$ and the new data set leading to $n_\text{T} = 2.29 ^{+0.73}_{-1.18}$ (90\% credible regions for both), but the spectral index being equal to 2 is consistent with both data sets at the 3$\sigma$ level~\cite{Antoniadis:2023rey}. Also, the analysis in Ref.~\cite{Figueroa:2023zhu} combining the NANOGrav 15-year data and the new data set of EPTA shows $n_\text{T} = 2.08^{+0.32}_{-0.30}$ (68\% confidence level). In our previous work with Kohri~\cite{Inomata:2023zup}, we delineated three ways to explain the  $\Omega_\text{GW} \sim f^2$ scaling:\footnote{Apart from SIGWs, the $\Omega_\GW \propto f^2$ scaling can also be realized with melting domain walls characterized by a time-dependent tension~\cite{Babichev:2023pbf}.} (1) SIGWs from the linear spectrum of the curvature perturbations $\mathcal{P}_\zeta (k) \sim k$, (2) the infrared (IR) tail~\cite{Cai:2019cdl, Yuan:2019wwo, Domenech:2020kqm} of SIGWs generated from a very sharply peaked spectrum $\mathcal{P}_\zeta (k) \sim \delta (\ln (k/k_*))$, and (3) the IR tail of SIGWs generated in a cosmological epoch dominated by a stiff fluid that has the equation-of-state parameter $w=1$\footnote{
$w=1/9$ is also a possible solution for the $f^2$ scaling of $\Omega_\GW$, but the PBH abundance is enhanced in contrast to the case of $w = 1$.
} rather than in the standard radiation-dominated (RD) era with $w=1/3$.
The second option was the main focus of Ref.~\cite{Inomata:2023zup}.
The last option is independent of the details of the underlying power spectrum of the curvature perturbations and is theoretically appealing. 
In this paper, we focus on the last option, i.e., the GWs induced by curvature perturbations in a cosmological era with $w = 1$. Such an era is realized, e.g., when the kinetic energy of a scalar field dominates the energy density of the Universe.  The kinetic-energy dominated era (kination)~\cite{Spokoiny:1993kt, Joyce:1996cp, Ferreira:1997hj} has been studied in a variety of contexts (see Refs.~\cite{Co:2021lkc, Gouttenoire:2021jhk} and references therein; Refs.~\cite{Haque:2021dha,Chowdhury:2023opo,Ben-Dayan:2023lwd} studied kination in the context of explaining the PTA data), and particle-physics UV completion by axion-like fields exist~\cite{Co:2019wyp}.  Another virtue of the induced GWs from a kination era beyond the fact that it leads to the $f^2$ IR tail to fit the PTA data is suppression of the PBH abundance.  As we explain in detail in the rest of this paper, there are two reasons for the suppression of the PBH abundance. The first reason is that we need a relatively small amplitude of the curvature perturbations to fit the PTA data since the energy fraction of GWs $\Omega_\text{GW}$ increases during a kination era. The second reason is that the threshold of the PBH formation is presumably greater in a kination era than in the RD era~\cite{Carr:1975qj, Musco:2012au, Harada:2013epa, Escriva:2020tak}.  Both of these facts exponentially suppress the PBH abundance.  

For the study of SIGWs beyond the RD era, we refer the reader to Refs.~\cite{Domenech:2019quo, Domenech:2020kqm,Witkowski:2021raz} in general. 
Furthermore, Refs.~\cite{Zhao:2023joc, Liu:2023pau} discussed SIGWs for the recent PTA data in general $w$, focusing on the narrow power spectrum of curvature perturbations with $\Delta \leq 0.1$ (see Eq.~(\ref{eq:p_zeta_pro}) for the definition of $\Delta$). 
In particular, Ref.~\cite{Zhao:2023joc} focused on the delta-function power spectrum ($\Delta \rightarrow 0$).
In the case of the narrow power spectrum with $\Delta \ll 1$, the slope of the SIGW power spectrum changes around $k/k_* \simeq \Delta$~\cite{Domenech:2020kqm}, where $k_*$ is the peak scale of the power spectrum (see Eq.~(\ref{eq:p_zeta_pro}) for the definition).
Refs.~\cite{Zhao:2023joc, Liu:2023pau} used the slope in $\Delta \lesssim k/k_* \lesssim 1$ to fit the PTA signals, where $\Omega_{\text{GW}} \propto f$ on those scales for $w=1$ if the scale of $k_* \Delta$ is inside the horizon at the end of the kination era~\cite{Domenech:2020kqm}.
Then, one of the papers concluded that the case of $w=1$ is outside the $90\,\%$ credible region~\cite{Liu:2023pau}.
In contrast, we consider $\Delta \geq 1$ throughout this work and focus on the IR slope of the GW spectrum that is different from those discussed in the previous works.

The structure of the paper is as follows. In Sec.~\ref{sec:GW}, we show that the induced GWs produced in a cosmological era with $w=1$ can fit the recent PTA data well.  In particular, the $f^2$ scaling is well explained by the IR tail of the induced GWs. In Sec.~\ref{sec:PBH}, we show that the PBH abundance is significantly suppressed in our scenario.  Our conclusions are given in Sec.~\ref{sec:conclusion}. 
We use the natural unit $c = \hbar = k_\text{B} = 1$.

\section{Induced Gravitational Waves in kination era} \label{sec:GW}

\subsection{Cosmological scenario and intuitive picture}
\label{subsec:intuitive_pic}

We consider a cosmological scenario in which there is a transient cosmological era with its equation-of-state parameter $w = 1$.  We assume this era ends before the big bang nucleosynthesis (BBN). 
We consider a transition of the era with $w=1$ into the standard RD era for concreteness,
although such an era can be in principle followed by, e.g., an early matter-dominated era. Depending on the underlying particle-physics model, the transition from the era with $w=1$ to the standard RD era may or may not involve entropy production. Since the dominant energy density redshifts faster than radiation in the era with $w=1$, a natural scenario is a smooth crossover without entropy production~\cite{Co:2021lkc, Gouttenoire:2021jhk}.  For definiteness, we consider such a scenario as a fiducial setup throughout this paper.  Alternatively, the transition by the decay of the field responsible for the era with $w=1$ into radiation, i.e., with entropy production, is conceivable. Given this possibility, we also mention possible modifications of equations in this case. 
The GW signals in our scenario are not sensitive to this difference.  Though we sometimes call the era with $w=1$ kination for short, we are agnostic about the details of the underlying model in this work.
Similarly, we do not specify the beginning of the kination era.  It may be preceded by, e.g., inflation~\cite{Spokoiny:1993kt} or an early matter-dominated era~\cite{Co:2019wyp,Co:2021lkc, Gouttenoire:2021jhk}. We only assume that the induced GWs on the relevant frequency ranges are produced in the kination era and the relevant scalar source modes enter the Hubble horizon during the kination era.

Now, let us discuss in an intuitive manner how the desired slope $n_\text{T}=2$ of the induced GWs can arise in our scenario. A more detailed formulation is given in the next subsection.
If the curvature power spectrum does not sharply increase on small scales, the spectrum of the GWs induced during the era with $w$ has the rough dependence $\Omega_\text{GW}(f) \sim (k/k_w)^{-2\beta} \mathcal{P}^2_\zeta(k)$ with the relation $2\pi f = k$, where $\beta \equiv (1-3w)/(1+3w)$ and $k_w$ is the inverse of the horizon scale at the end of the era with $w$, followed by the RD era.
Note that the $(k/k_w)^{-2\beta}$ factor comes from the redshift difference from the GWs and the total energy density in the era with $w$.
However, when $\mathcal{P}_\zeta (k)$ sharply increases on some scales, GWs are not necessarily induced dominantly by the scalar mode on the same $k$ but may be induced dominantly by smaller-scale scalar modes.
For example, if the curvature power spectrum sharply increases in $k < k_*$ with $n_s (\equiv \dd \ln \mathcal P_\zeta/\dd \ln k) > n_\beta (\equiv 3/2 + \beta - |\beta|)$ in the era with $w$ and has a (non-exceptionally) narrow peak\footnote{If $\mathcal P_\zeta$ has an exceptionally sharp peak at $k_*$ such as that with the Dirac delta function, the prediction changes~\cite{Espinosa:2018eve, Pi:2020otn, Domenech:2020kqm,Liu:2023pau}.} around $k \sim k_*$ or a flatter region $k \gtrsim k_*$ with $n_s < n_\beta$, the IR tail of the induced GWs in $k < k_*$ is determined only by the perturbations around $k\sim k_*$, where the IR tail is given by $\Omega_\GW \propto k^{3-2|\beta|}$ for the sharply increasing power spectrum~\cite{Domenech:2020kqm}.
This indicates that the IR tail in that case is determined by the cosmological background dynamics of the Universe, independently of the details of the shape of $\mathcal{P}_\zeta (k)$ in $k < k_*$. 
Note that $\beta=0$ and $n_\beta = 3/2$ in the RD era ($w=1/3$) and $\beta = -1/2$ and $n_\beta = 1/2$ in the kination era ($w=1$). 
For the sharply increasing power spectrum, this leads to the IR tails $\Omega_\GW \propto f^3$ in the RD era and $\Omega_\GW \propto f^2$ in the kination era.

Here, we briefly explain the physical interpretation of the frequency dependence of the IR tail of $\Omega_\GW$ in the RD and the kination era.
Causality tells us there is no correlation, except by chance, outside the Hubble horizon. 
Given that the superhorizon tensor perturbations with the absolute value of their wavenumber $k$ are composed of the superposition of $\sim \mathcal O((aH/k)^3)$ independent Hubble horizon patches,
the central limit theorem tells us that the power spectrum of the tensor modes is volume-suppressed leading to $\mathcal{P}_h(f) \propto f^3$ on superhorizon scales.  This cubic scaling is inherited by $\Omega_\text{GW}(f)$ as the universal IR tail $\Omega_\text{GW}(f) \propto f^3$ in the RD era even though the energy density of GWs is proportional to $f^2 \mathcal P_h(f)$. 
This is because the additional $f^2$ factor for the energy density of GWs is canceled by the redshift factor of the tensor modes, which are constant outside the horizon but get redshifted as $h \propto 1/a$ inside the horizon.
Note that the tensor modes with different scales enter the horizon at different times.
On the other hand, there are extra factors in the kination Universe. 
During the kination era, the decrease of the source terms with the subhorizon perturbations is slower than the decrease of the Hubble parameter unlike during the RD era. 
This leads to the growth of the superhorizon tensor perturbations as $\mathcal{P}_h \propto a^4$ while keeping the frequency dependence $\mathcal P_h \propto f^3$.
Once they enter the horizon, they decouple from the source terms and behave as freely propagating waves, which redshift as $h\propto 1/a$. 
The redshift difference between the energy densities of the kination fluid background and the subhorizon GWs leads to the behavior of $\Omega_\GW \propto a^2$ on subhorizon scales. 
From the above relations, we can see $\Omega_\GW \sim \mathcal P_h|_{\text{hc}} (a/a_\text{hc})^2 \propto f^2$ on the IR tail, where the subscript `hc' denotes the value at the horizon crossing and we have used $\Omega_\text{GW}|_\text{hc} \sim \mathcal P_h|_{\text{hc}}$, $\mathcal P_h|_{\text{hc}} \propto f^3 a_\text{hc}^4 \propto f$, and $a \propto \eta^{1/2}$.
See Refs.~\cite{Cai:2019cdl, Yuan:2019wwo, Domenech:2020kqm} for more details on the IR tail.

\subsection{Basic formulas for induced gravitational waves}

In this subsection, we summarize the equations for the GWs induced by scalar perturbations during the kination era, followed by the RD era. See also Refs.~\cite{Domenech:2019quo, Domenech:2020kqm, Harigaya:2023ecg} for the details.

Throughout this work, we take Newtonian gauge, where the metric perturbations are expressed as
\begin{align}
  \dd s^2 = a^2 \left[ -(1+2 \Phi)\dd \eta^2 + \left((1-2\Psi) \delta_{ij} + \frac{h_{ij}}{2} \right) \dd x^i \dd x^j \right].
\end{align}
$\Phi$ and $\Psi$ are the first-order scalar perturbations and $h_{ij}$ is the second-order tensor perturbation. 
We have neglected the vector perturbations because their contribution to GW production is negligible.
In this work, we consider a perfect fluid and set $\Phi= \Psi$ throughout the evolution for simplicity. 
We expand the tensor perturbations as 
\begin{align}
  h_{ij}(\bm x) =  \sum_{\lambda = +,\times}\int \frac{\dd^3 k}{(2\pi)^3} e^\lambda_{ij}(\hat{\bm{k}})\, h^\lambda_{\bm k} \, \ee^{i \bm{k} \cdot \bm{x}},
\end{align}
where $e^\lambda_{ij}$ is the polarization tensor that satisfies $k^i e^\lambda_{ij} = 0$, $\delta^{ij} e^\lambda_{ij}=0$, and $\delta^{ik} \delta^{jl}e^\lambda_{kl} e^{\lambda'}_{ij}=\delta^{\lambda \lambda'}$.
With their Fourier modes, we can express the power spectrum of the tensor perturbations as 
\begin{align}
  \vev{h^\lambda_{\bm k} h^{\lambda'}_{\bm k'}} = (2\pi)^3 \delta(\bm k + \bm k') \delta^{\lambda \lambda'} \frac{2\pi^2}{k^3} \mathcal P_h(\eta, k).
\end{align}
The equation of motion for the tensor perturbation during the kination era is given by~\cite{Ananda:2006af, Baumann:2007zm, Domenech:2021ztg} 
\begin{align}
  {h^\lambda_{\bm k}}'' + 2 \mathcal H {h^\lambda_{\bm k}}' + k^2 h^\lambda_{\bm k} = 4 \mathcal S^\lambda_{\bm k},
\end{align}
where $\mathcal{H}$ is the conformal Hubble parameter, a prime denotes differentiation with respect to the conformal time $\eta$, and the source term $\mathcal S$ is given by 
\begin{align}
\mathcal S^\lambda_{\bm k}(\eta) 
&= \int \frac{\dd^3 k'}{(2\pi)^3} \ee^{\lambda\, lm}(\hat k) k_l' k_m' \Biggr[ 2\Phi_{\bm k'}(\eta) \Phi_{ \bm k - \bm k'}(\eta) + \frac{2}{3}\left( \Phi_{\bm k'}(\eta) + \mathcal H^{-1}\Phi'_{\bm k'}(\eta)\right)\left( \Phi_{\bm k - \bm k'}(\eta) + \mathcal H^{-1}{\Phi'_{\bm k-\bm k'}}(\eta)\right) \Biggr].
\end{align}

After some calculation based on the Green function method, we finally obtain the following expression for the time-averaged power spectrum of the induced GWs during a kination era, which is used when we discuss their energy density:
\begin{align}
        \overline{\mathcal P_h(\eta, k)} = 4 \int^\infty_0 \dd v \int^{1+v}_{|1-v|} \dd u \left( \frac{4v^2 - (1+v^2-u^2)^2}{4 u v} \right)^2 \overline {I^2(u,v,x)} \mathcal P_\zeta(k u) \mathcal P_\zeta(k v),
        \label{eq:p_h_uv}
\end{align}
where $x = k \eta$, the overline denotes the time average over oscillations of the tensor modes, and 
\begin{align}
  I(u,v,x) = \int^x_0 \dd \bar x\, k G_k(\eta, \bar \eta) f(u,v,\bar x).
  \label{eq:I_def}
\end{align}
$G_k$ is the Green function satisfying $\hat {\mathcal N} G_k(\eta,\bar \eta) = \delta (\eta - \bar \eta)$, where 
\begin{align}
        \hat{\mathcal N} \equiv \frac{\partial^2}{\partial \eta^2} + 2 \mathcal H \frac{\partial}{\partial \eta} + k^2.
\end{align}
Specifically, the Green function for the GWs induced during a kination era is given by 
\begin{align}
  k G_k(\eta,\bar \eta) &= \frac{\pi}{2} \bar x (J_0(\bar x) Y_0(x) - J_0(x) Y_0(\bar x)),
\end{align}
where $\bar x = k\bar \eta$, $J_\nu(x)$ and $Y_\nu(x)$ with $\nu$ arbitrary are the Bessel functions of the first and the second kind, and we have used $\mathcal H = 1/(2\eta)$ during a kination era. 
The source function $f$ is given by 
\begin{align}
        f(u,v,\bar x) = 2 T(u\bar{x})T(v\bar{x}) + \frac{2}{3} \left( T(u\bar{x}) + 2 u \bar x \frac{\dd T(u\bar{x})}{\dd (u \bar x)} \right) \left( T(v\bar{x}) + 2 v \bar x \frac{\dd T(v\bar{x})}{\dd (v \bar x)}\right),
\label{eq:f_def}
\end{align}
where $\Phi_{\bm k}(\eta) = T(k \eta) \zeta_{\bm k}$ with $\zeta_{\bm k}$ being the curvature perturbation in the superhorizon limit. 
To obtain the concrete expression of the transfer function $T(k\eta)$, we need to solve the equation of motion for the gravitational potential:

\begin{align}
        \Phi''_{\bm k} + 6\mathcal H \Phi'_{\bm k} + k^2 \Phi_{\bm k} = 0.
        \label{eq:phi_eom_w}
\end{align}
The solution of this equation is
\begin{align}
        T(k\eta) = -\frac{3}{2}(k \eta)^{-1} J_1 (k \eta),
        \label{eq:transfer_function_KD}
\end{align}
where we have imposed the normalization of $T$ to satisfy the superhorizon-limit relation during a kination era, $\Phi_{\bm k}(\eta \ll 1/k) = -\frac{3}{4}\zeta_{\bm k}$, i.e., $T(k\eta \ll 1) = - \frac{3}{4}$.

For convenience, we rewrite Eq.~(\ref{eq:I_def}) as 
\begin{align}
         I(u,v,x) &= \int^x_0 \dd \bar x\, k G_k(\eta, \bar \eta) f(u,v,\bar x) \nonumber \\
         & = \frac{\pi}{2} \left[ Y_0(x) \mathcal I_{J}(u,v,x) -  J_0(x) \mathcal I_{Y}(u,v,x) \right],
         \label{eq:i_kernel}
\end{align}
where $\cal I_J$ and $\cal I_Y$ are defined as 
\begin{align}
       \label{eq:calI_J}
       \mathcal I_{J}(u,v,x) &\equiv  \int^x_{0} \dd \bar x \, \bar x J_0(\bar x) f(u,v,\bar x), \\ 
       \mathcal I_{Y}(u,v,x) &\equiv  \int^x_{0} \dd \bar x \, \bar x Y_0(\bar x) f(u,v,\bar x).
       \label{eq:calI_Y}
\end{align}
In the subhorizon limit ($x \gg 1$), $I$ can be approximated as 
\begin{align}
        I(u,v,x(\gg 1))
         & \simeq -\sqrt{\frac{\pi}{2}} x^{-1/2} \left[ \sin\left( \frac{\pi}{4} - x \right) \mathcal I_{J}(u,v,x) + \cos\left( \frac{\pi}{4} - x \right) \mathcal I_{Y}(u,v,x) \right].
\end{align}
Then, we can express the oscillation average of $I^2$ as 
\begin{align}
       \overline{I^2(u,v,x(\gg 1))}
         & \simeq \frac{\pi}{4} x^{-1} \left[ \mathcal I_{J}^2(u,v,x) + \mathcal I_{Y}^2(u,v,x) \right].
\end{align}
The analytic expression of this equation in the late-time limit $x\gg 1$ is given by~\cite{Domenech:2019quo}
\begin{align}
  \overline{I^2(u,v,x(\gg 1))} \simeq \frac{9}{16 \pi u^4 v^4 x} \left\{ \frac{(3(u^2 + v^2 -1)^2 - 4 u^2 v^2)^2}{4 u^2 v^2 - (u^2 + v^2-1)^2} + 9(u^2 + v^2-1)^2 \right\}.
\end{align}

The energy density of GWs per log bin ($\ln k$) is given by
\begin{align}
  \rho_\GW(\eta,k) = \frac{k^2}{8a^2}\overline{\mathcal P_h(\eta,k)},
\end{align}
and the energy density parameter of GWs is given by 
\begin{align}
        \Omega_\GW(\eta,k) &\equiv \frac{\rho_\GW(\eta,k)}{\rho_\tot(\eta)}= \frac{1}{24} \left( \frac{k}{\mathcal H}\right)^2 \overline{\mathcal P_h(\eta,k)} = \frac{x^2}{6} \overline{\mathcal P_h(\eta,k)}, 
\end{align}
where $\rho_\tot$ is the total energy density of the Universe and we have used $\mathcal H = 1/(2 \eta)$ again.
Unlike $\Omega_\GW$ during a RD era, $\Omega_\GW$ continues to grow proportionally to $a^2$ during a kination era even for the subhorizon GWs. 
This is because the redshift evolution of $\rho_\GW (\propto a^{-4})$ and $\rho_\tot (\propto a^{-6})$ is different during a kination era.

We here relate the energy density parameter during the kination era and the following RD era. 
For definiteness, we consider the case without the entropy production (the energy transfer from the kination fluid to the radiation). 
In that case, $\Omega_\GW$ during the following RD era is the same as the GW-radiation energy ratio $\rho_\GW/\rho_\rr$ during a kination era, which becomes constant for subhorizon GWs even during a kination era.
Specifically, we can express the energy density parameter during the following RD era as\footnote{If the kination era ends with the entropy production by the decay of the kination fluid at $\eta_\en$, the relation is modified as 
\begin{align}
        \Omega_{\GW}(\eta_\text{end,+},k)
        &\simeq \Omega_\rr(\eta_{\en,-}) \frac{\rho_\kk(\eta)}{\rho_\rr(\eta)} \Omega_\GW(\eta,k) \quad (\text{for } \eta \text{ that leads to } \Omega_\rr(\eta) \ll 1)\nonumber \\
        &= (1-\Omega_\rr(\eta_{\en,-}))\left(\frac{a(\eta_{\en,-})}{a(\eta)} \right)^2\frac{x^2}{6}\overline{\mathcal P_h(\eta,k)} \nonumber \\
         &= (1-\Omega_\rr(\eta_{\en,-}))^{1/2} \frac{k}{12  \mathcal H(\eta_{\en,-})} x\, \overline{\mathcal P_h(\eta,k)},\nonumber
\end{align}
where $\eta_{\text{end,}\pm}$ is the time right after/before the entropy production, respectively, and $\Omega_\rr(\eta_{\en,-})$ denotes the radiation energy density parameter just before the entropy production.
We have used $\rho_\kk(\eta)/\rho_\rr(\eta)= (a(\eta_{\en,-})/a(\eta))^2 (1-\Omega_\rr(\eta_{\en,-}))/\Omega_\rr(\eta_{\en,-})$ and $a(\eta)/a(\eta_{\en,-}) \simeq (2 \mathcal H(\eta_{\en,-}) (1-\Omega_\rr(\eta_{\en,-}))^{1/2}\eta)^{1/2}$ for $\eta$ that leads to  $\Omega_\rr(\eta)\ll 1$.
We can see that, if the entropy production terminates the kination era at $\eta_{\eq,\kk}$ ($\Omega_\rr(\eta_{\eq,\kk}) = 1/2$), $\Omega_\GW(\eta_\text{end,+},k)$ indeed becomes one half of Eq.~(\ref{eq:gw_r_ratio}).
In this entropy production case, $f_\text{UV}$ in Eq.~\eqref{f_UV} below is also modified similarly. 
} 
\begin{align}
         \Omega_{\GW}(\eta_\cc,k) = \frac{\rho_\GW(\eta,k)}{\rho_\rr(\eta)} &\simeq \frac{\rho_\kk(\eta)}{\rho_\rr(\eta)} \Omega_\GW(\eta,k) \quad (\eta \ll \eta_{\eq,\kk})\nonumber \\
         &= \left(\frac{a_{\eq,\kk}}{a(\eta)} \right)^2\frac{x^2}{6}\overline{\mathcal P_h(\eta,k)} \nonumber \\
         &= \frac{k}{6\sqrt{2} \mathcal H_{\eq,\kk}} x\, \overline{\mathcal P_h(\eta,k)},
        \label{eq:gw_r_ratio}
\end{align}
where $\rho_\kk$ is the energy density of the kination fluid, the subscript `$\eq,\kk$' denotes the value at the time when $\rho_\kk = \rho_\rr$, and 
$\eta_\cc$ is the time when the kination energy density becomes negligible compared to the radiation energy density and $\Omega_\GW$ becomes constant except for its evolution due to the change of the degrees of freedom ($\eta_\cc > \eta_{\eq,\kk}$).
We have also used the relation $a(\eta)/a_{\eq,\kk} \simeq (\sqrt{2} \mathcal H_{\eq,\kk} \eta)^{1/2}$ for $\eta \ll \eta_{\eq,\kk}$.
Note that $x\, \overline{\mathcal P_h(\eta,k)}$ becomes constant even during a kination era.

Taking into account the late-time MD era, which follows the RD era and dilutes $\Omega_\GW$, we find the current $\Omega_\GW$ as~\cite{Inomata:2020lmk}
\begin{align}
        \Omega_\GW(\eta_0, k) h^2 \simeq 0.84 \left( \frac{g^\rho_{\hh}}{10.75} \right) \left( \frac{g^s_{\hh}}{10.75} \right)^{-4/3} \Omega_{\rr,0} h^2  \Omega_\GW(\eta_\cc,k),
        \label{eq:omega_gw0}
\end{align}
where $g^\rho$ ($g^s$) is the effective relativistic degrees of freedom for the energy (entropy) density, the subscript `h' denotes the value at the horizon entry of the mode with $k$, $h = H_0/(100 \, \mathrm{km} \, \mathrm{s}^{-1} \, \mathrm{Mpc}^{-1})$ with $H_0$ being the current Hubble parameter, and $\Omega_{\rr,0}$ is the current energy density parameter for radiation ($\Omega_{\rr,0} h^2 \simeq 4.2 \times 10^{-5}$). We use the temperature dependence of $g^\rho$ and $g^s$ in Ref.~\cite{Saikawa:2018rcs}.
The argument of $\Omega_\GW$ in Eq.~(\ref{eq:omega_gw0}) is the scale of the tensor mode $k$, but it can be converted to the frequency $f$ through $f=k/(2\pi)$ because GWs propagate at the speed of light.

\subsection{Explaining the PTA data}

Before the application of our formulas to the PTA data, let us discuss constraints on our parameter space. 
The transition from the kination era to the RD era must occur at $T>\mathcal O(1)\,\MeV$ for the BBN not to be disturbed~\cite{Co:2021lkc,AristizabalSierra:2023bah}.
For simplicity, we regard $\eta_{\eq,\kk}$ as the end of the kination era, $\eta_\text{end}$, in the following.

The relation between the horizon scale and the temperature at $\eta_\en$ is given by 
\footnote{
When entropy production occurs at $\eta_\en$ instantaneously, $T_\en$, $g^s_\en$, and $g^\rho_\en$ in Eq.~(\ref{eq:horizon_tdec}) should be regarded as those right after the entropy production while $\Omega_\rr(\eta_\en)$ in it should be regarded as that just before the entropy production.
} 
\begin{align}
  a_\en H_\en &=  \left( \frac{g^s_\en}{g^s_0} \right)^{-1/3} \left(\frac{T_\en}{T_0}\right)^{-1} \left( \frac{\pi^2 g^\rho_\en T^4_\en}{90 M_\Pl^2} \right)^{1/2} \Omega^{-1/2}_\rr(\eta_\en) \nonumber \\
  &= 1.7\times 10^{5}\,\Mpc^{-1} \left( \frac{g^s_\en}{10.75} \right)^{-1/3} \left( \frac{g^\rho_\en}{10.75} \right)^{1/2} \left(\frac{T_\en}{10\, \MeV}\right) (2 \Omega_\rr(\eta_\en))^{-1/2},
  \label{eq:horizon_tdec}
\end{align}
where the subscript `end' denotes the value at $\eta_\en$ and the subscript `$0$' denotes the value at present as $g^s_0 = 3.91$~\cite{Kolb:206230} and $T_0 = 2.72\,\text{K} = 2.35\times 10^{-4}\,\text{eV}$~\cite{Planck:2018vyg}.
Using this and the relation $f_\en= a_\en H_\en/(2\pi)$, we obtain the corresponding frequency as 
\begin{align}
    f_\en = 2.6 \times 10^{-10} \, \mathrm{Hz} \, \left( \frac{g^s_\en}{10.75} \right)^{-1/3}\left( \frac{g^\rho_\en}{10.75} \right)^{1/2}  \left( \frac{T_\en}{10 \, \mathrm{MeV}} \right) (2\Omega_\text{r}(\eta_\en))^{-1/2}.
\end{align}
Note again that $T_\en> \mathcal O(1)\,\MeV$ must be satisfied not to affect the BBN.

There is also an upper bound on the characteristic frequency scale as we will see soon below.  To be concrete, we consider the log-normal power spectrum as an example,
\begin{align}
    \mathcal{P}_\zeta (k) = & \frac{A}{\sqrt{2\pi \Delta^2}} \exp \left( -\frac{(\ln (k/k_*))^2}{2 \Delta^2} \right),
    \label{eq:p_zeta_pro}
\end{align}
where $A$ is the amplitude, $\Delta$ is the width, and $k_*$ is the peak wave number.  This is a toy model to represent a smooth peak in the power spectrum.  Note that the dynamics during inflation is parameterized by a natural time variable during inflation, which is the e-folding number $N$.  It is related to the wave number $k$ via $k = a H \sim H e^{N}$, explaining the logarithmic dependence in the argument of the Gaussian. As a benchmark point, we take $\Delta = 1$ to reduce the number of free parameters unless otherwise noted.

$f_*= k_*/(2\pi)$ is upper-bounded from the requirement that the induced GWs are not overproduced as dark radiation, $\Omega_\text{GW}h^2 < 1.8 \times 10^{-6}$~\cite{Kohri:2018awv, Clarke:2020bil}, 
\begin{align}
    f_* \leq f_\text{UV} = & 2.8 \times 10^{-7} \, \mathrm{Hz} \, \left( \frac{f_\text{end}}{2.6 \times 10^{-10} \, \mathrm{Hz}} \right) \left( \frac{A}{1.3 \times 10^{-2}} \right)^{-2}. \label{f_UV}
\end{align}
The characteristic scale $f_*$ should also satisfy $(f_\text{end} < )f_\text{PTA} \lesssim f_*$ to explain the PTA results with the IR tail, where $f_\text{PTA} \approx 6 \times 10^{-8} \, \mathrm{Hz}$ is the highest frequency the PTAs are sensitive to. Thus, the viable range of $f_*$ is $f_\text{PTA} \lesssim f_* \leq f_\text{UV}$.

\begin{figure}
        \centering \includegraphics[width=0.5\columnwidth]{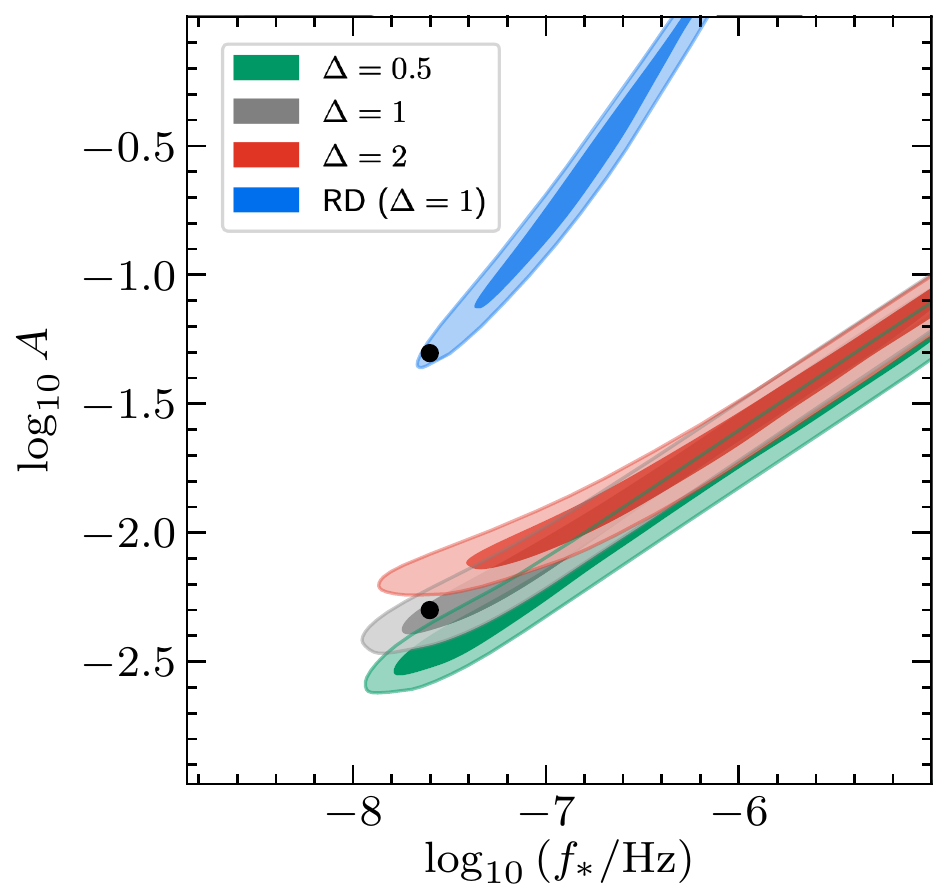}
        \caption{ The posteriors for the NANOGrav 15-year results~\cite{NANOGrav:2023gor,NANOGrav:2023hvm}.
        The power spectrum is given by Eq.~(\ref{eq:p_zeta_pro}).
        We take $\Delta = 0.5$ (green), $1$ (gray), and $2$ (red) with $T_\text{end} = 10\,\MeV$.
        For comparison, we also show the posterior for the GWs induced during the RD era with $\Delta = 1$ (blue).
         The dark and light color regions in the posteriors denote the $68\%$ and the $95\%$ credible regions. 
         The black dots correspond to the fiducial parameters taken to compare the PBH abundances in the kination and the RD era cases in Sec.~\ref{sec:PBH}.
         Specifically, the black dot in the gray region corresponds to $\log_{10} A = -2.3$ and $\log_{10}(f_*/\text{Hz}) = -7.6$, which is used in Fig.~\ref{fig:gw}, and that in the blue region $\log_{10} A = -1.3$ and $\log_{10}(f_*/\text{Hz}) = -7.6$.
         As priors, we take the log-uniform distribution in the range $[-11,-5]$ for $f_*/\text{Hz}$ and the log-uniform distribution in the range $[-5,0]$ for $A$.
         We used \texttt{PTArcade}~\cite{andrea_mitridate_2023, Mitridate:2023oar} with $\mathtt{Ceffyl}$~\cite{lamb2023need} to make this figure.  The dark radiation constraint is not imposed in this figure.
        }
        \label{fig:posteri}
\end{figure}

Since the NANOGrav among all the PTA observations has the strongest signals, we compare our results with the NANOGrav 15-year data~\cite{NANOGrav:2023gor,NANOGrav:2023hvm}. 
Figure~\ref{fig:posteri} shows the posteriors for the data~\cite{NANOGrav:2023gor,NANOGrav:2023hvm}. 
In this figure, we also show the posterior for the GWs induced during the RD era for comparison.
We can see that the kination era decreases the curvature amplitudes required to fit the NANOGrav results.
This is because $\Omega_\GW$ for the GWs induced during the kination era grows proportionally to $a^2$ from their horizon entry until the end of the kination era.

\begin{figure}
        \centering \includegraphics[width=0.7\columnwidth]{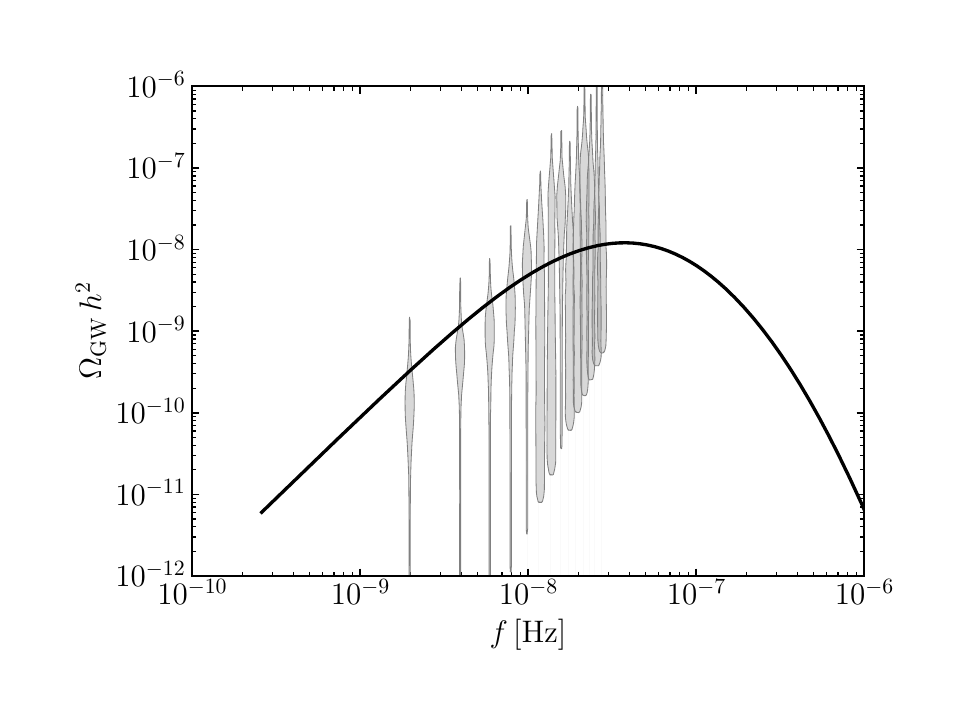}
        \caption{ 
        The GW spectrum with $\log_{10} A = -2.3$ and $\log_{10}(f_*/\text{Hz})= -7.6$ (the black dot in the gray region in Fig.~\ref{fig:posteri}).
        The gray violins are from the NANOGrav results~\cite{NANOGrav:2023hvm}.
        The lowest frequency of the black line corresponds to $f_\text{end}$ with $T_\text{end} = 10\,\MeV$.
        }
        \label{fig:gw}
\end{figure}

As long as the power spectrum of the curvature perturbations is not sharply peaked ($\Delta \gtrsim 1$), the resultant induced GWs produced in a kination era have the $f^2$ IR tail. The strength of the induced GWs has the parametric dependence $\Omega_\text{GW}(f) \sim A^2 f^2 /(f_\text{end} f_*) $. This is qualitatively the same for other generic (non-sharply-peaked) choices of the functional form of $\mathcal{P}_\zeta (k)$. Because of this dependence, there is degeneracy among parameters to fit the PTA data
\begin{align}
    A \approx 
    8 \times 10^{-3} \left( \frac{f_*}{1 \times 10^{-7} \, \mathrm{Hz}} \right)^{1/2}\left( \frac{f_\en}{2.6 \times 10^{-10} \, \mathrm{Hz}}\right)^{1/2} \ \ \text{for } f_* > \mathcal O(10^{-8})\,\Hz . \label{degeneracy_relation}
\end{align}
We can see this relation in Fig.~\ref{fig:posteri}. 
Setting $f_* = f_\text{UV}$ and combining the above relations, we obtain the maximal value of $A$ (or, equivalently, that of $f_*$) for a given $f_\text{end}$. 
For example, if we fix $f_\text{end} = 2.6 \times 10^{-10} \, \mathrm{Hz}$, the parameters should be $A \lesssim 1.3 \times 10^{-2}$ and, correspondingly, $f_* \lesssim 2.8 \times 10^{-7} \, \mathrm{Hz}$ to be consistent with the NANOGrav results.
Figure~\ref{fig:gw} shows the GW spectrum for the fiducial parameter values, $\log_{10} A = -2.3$ and $\log_{10}(f_*/\text{Hz}) = -7.6$.

\section{Suppression of Primordial black hole abundance} \label{sec:PBH}

In this section, we point out two reasons why the PBH abundance is suppressed in our scenario. 

The first reason is the reduction of the necessary amplitude of the curvature perturbation to fit the PTA data, as shown in Fig.~\ref{fig:posteri}. 
This is related to the enhancement of the induced GWs by the redshift factor $(a_\en/a_\text{h})^2$ relative to the kination fluid, where $a_\en$ and $a_\hh$ denote the scale factor at $\eta_\en$ and the horizon entry of the mode.  
This implies that the necessary amplitude of the curvature power spectrum to fit the PTA data is reduced by $(a_\en/a_\text{h})^{-1}$. Since the PBH abundance is exponentially sensitive to the inverse of the amplitude of the curvature power spectrum, the PBH abundance in our scenario is significantly reduced.

The second reason is a rise of the critical value $\delta_\text{c}$ in the density perturbations for the PBH formation. The threshold of overdensity for the PBH formation in a cosmological era with generic values of $w(= c_\text{s}^2)$ was studied in the literature~\cite{Carr:1975qj, Musco:2012au, Harada:2013epa, Escriva:2020tak}. The pioneering work by Carr used $\delta_\text{c} \sim w$, which shows the tendency that the critical value increases as $w$ increases. (However, a non-monotonic feature was found in Ref.~\cite{Harada:2013epa} though it was not in Refs.~\cite{Musco:2012au,  Escriva:2020tak}.)  As summarized in Fig.~11 of the recent work~\cite{Escriva:2020tak}, there is substantial uncertainty (as in the standard $w=1/3$ case), but overall, we see $0.4 \lesssim \delta_\text{c} \lesssim 0.75$ for the case of $w=1$. The PBH abundance is also sensitive to this parameter; it is exponentially sensitive to the square of $\delta_\text{c}$. Due to this, the PBH abundance in our scenario is further significantly suppressed compared to the PBH formation in the RD era.

To discuss these points quantitatively, we introduce some more formulas in the following.  The PBH energy density fraction at their formation, which can also be interpreted (up to the factor $\gamma$) as the PBH formation probability in each Hubble patch, can be calculated with Carr's formula (the Press-Schechter formalism)~\cite{Carr:1975qj} 
\begin{align}
    \beta(M) \equiv & \left. \frac{\rho_\text{PBH}(M)}{\rho_\tot} \right|_\text{form}= \gamma \int_{\delta_\text{c}}^\infty \frac{\mathrm{d} \delta}{\sqrt{2\pi} \sigma (M)} \exp \left( - \frac{\delta^2}{2 \sigma^2 (M)} \right) \nonumber \\
    =& \frac{\gamma}{2} \mathrm{Erfc} \left( \frac{\delta_\text{c}}{\sqrt{2}\sigma(M)} \right), 
\label{eq:beta}
\end{align}
where the subscript `form' denotes the value at the PBH formation, $\rho_\text{PBH}(M)$ is the energy density of PBHs per log $M$ bin, $\gamma \approx c_\text{s}^3$ with $c_\text{s}$ being the sound speed is the fraction of the horizon mass that goes into a PBH, $\sigma(M)$ is the coarse-grained density perturbations, and Erfc is the complementary error function.  In the era with $w(=c_\text{s})=1$, $\gamma \approx 1$. 
Although the dependence of $\beta(M)$ on $\mathcal P_\zeta(k)$ could be approximately understood as $\beta(M) \sim \gamma \frac{\sqrt{\mathcal P_\zeta(k(M))}}{\sqrt{2\pi} \delta_\cc} \exp\left( - \frac{\delta_\text{c}^2}{2 \mathcal{P}_\zeta (k(M))}\right)$ with a rough relation $\sigma^2 (M) \sim \mathcal{P}_\zeta (k(M))$, we calculate $\sigma^2(M)$ with a more specific formula in this paper:
\begin{align}
    \sigma^2(k) = & \frac{4}{9} \int \frac{\mathrm{d}q}{q} \left(\frac{q}{k}\right)^4 W^2 \left( \frac{q}{k}\right) \mathcal{T}^2 \left(q, \frac{1}{2k} \right) \mathcal{P}_\zeta (q), \label{sigma}
\end{align}
where $W(z)$ is a window function and $\mathcal{T}(q, 1/(2k)) \equiv T(q/(2k))$ (see Eq.~\eqref{eq:transfer_function_KD}) is the transfer function in the kination era evaluated at the Horizon reentry $k = \mathcal{H}$. 
For the window function, we take the real-space top-hat window function, whose form in Fourier space is given by
\begin{align}
    W(q r) = 3 \frac{\sin(q r) -(qr) \cos(qr)}{(qr)^3},
    \label{eq:window}
\end{align}
where $r$ is the smoothing scale and we substitute $r = 1/k$.
In Eq.~(\ref{sigma}), we have used the relation between the density perturbations and the curvature perturbations in comoving slices, $\delta = (8/3) (k\eta)^2 T(k\eta)\zeta$ (see also Ref.~\cite{Young:2014ana} for the expression in the superhorizon limit).
From the appearance of $\delta_\text{c}$ and $\mathcal{P}_\zeta$ in the exponential factor in Eq.~\eqref{eq:beta}, we confirm the above two reasons why the PBH abundance is suppressed significantly. 

The relation between the PBH mass $M$ and the corresponding wavenumber $k$ of the perturbations that produce the PBH is given by
\begin{align}
    k =& (a H)|_\text{form} \nonumber \\
      =& (1 - \Omega_\text{r} (\eta_{\text{end,}-}))^{1/6} \left( \frac{H_\text{form}}{H_\text{end}} \right)^{2/3} (a H)|_\text{end} \nonumber \\
      = & 7.1 \times 10^7 \, \mathrm{Mpc}^{-1} \times \left( \frac{g^s_{\en,+}}{10.75} \right)^{-1/3} \left( \frac{g^\rho_{\en,+}}{10.75} \right)^{1/6} \left( \frac{M_H}{10^{-1} M_\odot} \right)^{-2/3} \left( \frac{T_{\en,+}}{10 \, \mathrm{MeV}} \right)^{-1/3} \left(\frac{1-\Omega_\text{r}(\eta_{\text{end},-})}{\Omega_\text{r}(\eta_{\text{end},+})} \right)^{1/6},
\end{align}
where this equation is valid for both the cases with/without the entropy production.
For the case with the entropy production, we have assumed that the entropy production occurs instantaneously at $\eta_\en$ and we denote the time just before and after the entropy production by $\eta_{\en,-}$ and $\eta_{\en,+}$.
Note that the subscript `$\en,+$' denotes the value at $\eta_{\en,+}$.
On the other hand, for the case without the entropy production, which is the main focus of this paper, we set $\eta_{\en,+}=\eta_{\en,-}=\eta_\en = \eta_{\eq,\kk}$.
In this equation, we have assumed $k \gg k_\text{end}$ and used the fact that $a \propto H^{-1/3}$ in a kination era and that the PBH mass $M$ is related to the horizon mass $M_H = 4\pi M_\Pl^2 / H$ via $M = \gamma M_H$. 
From the above relation and the fact that the scale of $k=7.1\times 10^7\,\Mpc^{-1}$ corresponds to the frequency of $f = k/(2\pi)=1.1\times 10^{-7}\,\Hz$, we see that the typical mass of the PBH is $\mathcal{O}(10^{-1}) \, M_\odot$ for our fiducial parameter set in Fig.~\ref{fig:gw}.\footnote{
The possibility that the induced GWs associated with much smaller PBHs explain the PTA signals was considered in Ref.~\cite{Papanikolaou:2020qtd} and studied more in Ref.~\cite{Bhaumik:2023wmw} based on the Poltergeist mechanism for GW production~\cite{Inomata:2019ivs, Inomata:2020lmk} assuming instantaneous evaporation of PBHs. It will be interesting to see if the PTA signals could be explained even after taking into account suppression effects~\cite{Inomata:2019zqy, Inomata:2020lmk} during a  non-instantaneous transition from the matter (PBH) dominated era to the RD era in this scenario because these suppression effects always exist even in the case of the monochromatic PBH mass function~\cite{Inomata:2020lmk}.
}

The differential PBH abundance relative to the cold dark matter $\bar{f}_\text{PBH}(M) \equiv \rho_\text{PBH}(M) / \rho_\text{CDM}$ is
\begin{align}
    \bar{f}_\text{PBH}(M) &= \beta(M)  \left( \frac{a_\text{end}}{a_\text{form}} \right)^3 \frac{\Omega_{\text{m},0}}{\Omega_{\text{CDM},0}} \frac{g^\rho _{\en,+}}{g^\rho_\eq} \frac{g^s_\eq}{g^s_{\en,+}} \frac{T_\text{end,+}}{T_\text{eq}} \frac{1-\Omega_\text{r}(\eta_{\text{end},-})}{\Omega_\text{r}(\eta_{\text{end},+})} \nonumber \\
    &= \beta(M)  \frac{4\pi M_\Pl}{M_H} \left(\frac{\pi^2 g^\rho_{\en,+} T_{\en,+}^4}{90 M_\Pl^4} \right)^{-1/2} \frac{\Omega_{\text{m},0}}{\Omega_{\text{CDM},0}} \frac{g^\rho _{\en,+}}{g^\rho_\eq} \frac{g^s_\eq}{g^s_{\en,+}} \frac{T_\text{end,+}}{T_\text{eq}} \left(\frac{1-\Omega_\text{r}(\eta_{\text{end},-})}{\Omega_\text{r}(\eta_{\text{end},+})}\right)^{1/2},
\end{align}
where $\Omega_{X,0} \equiv \rho_{X,0} / \rho_{\text{tot},0}$ ($X=$ m (non-relativistic matter) or CDM (cold dark matter)) is the current fractional energy-density abundance of a cosmological component $X$ and the subscript `eq' means the value at the matter-radiation equality (at $z_\eq\simeq 3400$ with $z$ the redshift).
Compared to the standard expression in the case of PBH production in the RD era, we have a \emph{power} enhancement factor $(a_\text{end}/a_\text{form})^3 (T_\text{end}/T_\text{form}) \sim (a_\text{end}/a_\text{form})^2$.  However, $\beta(M)$ is \emph{exponentially} suppressed as we discussed above.  The total PBH abundance is obtained by integral of $\bar{f}_\text{PBH}$
\begin{align}
    f_\text{PBH} \equiv \int \frac{\mathrm{d}M}{M} \bar{f}_\text{PBH}(M).
\end{align}

\begin{figure}
        \centering \includegraphics[width=0.49\columnwidth]{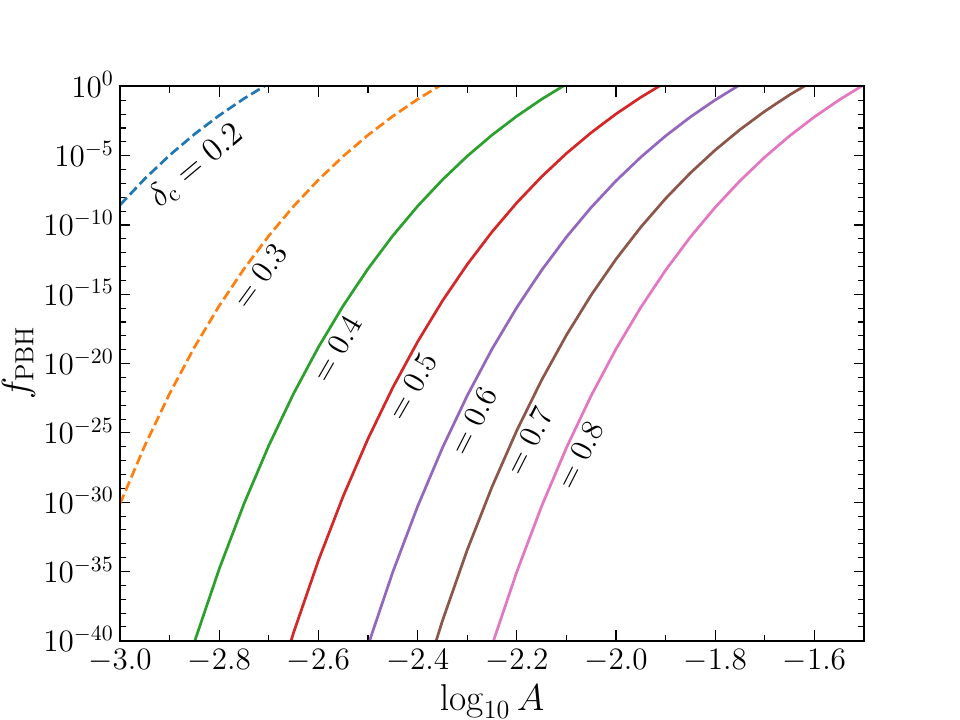}
        \includegraphics[width=0.49\columnwidth]{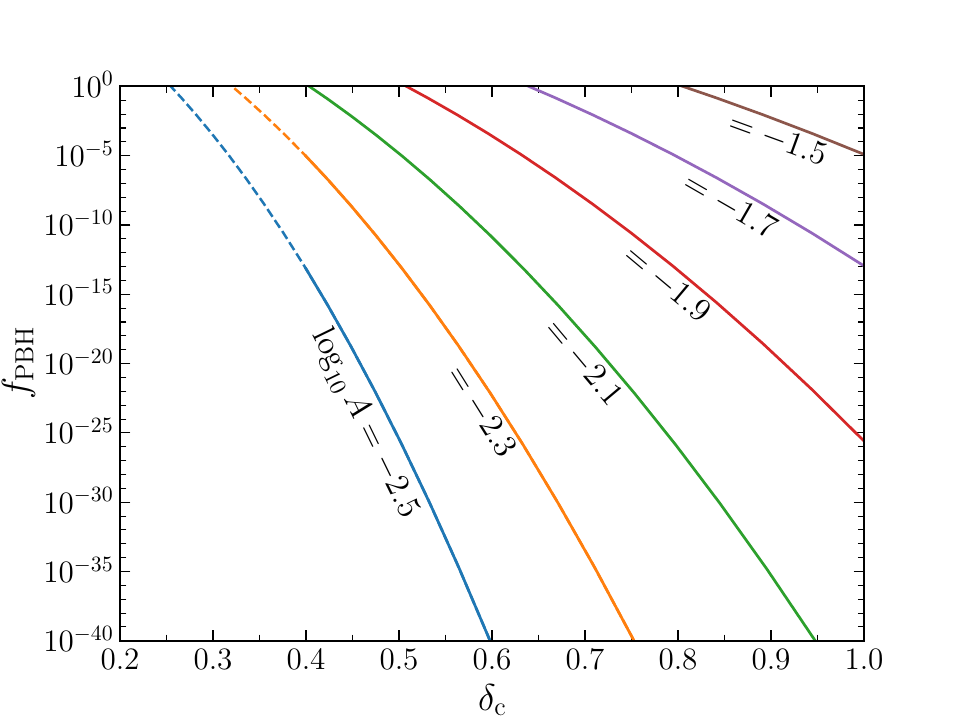}
        \caption{ 
        Dependence of the fraction of dark matter in the form of PBHs $f_\text{PBH}$ on the amplitude of the power spectrum of the curvature perturbations $A$ (left panel) and the critical density of PBH formation $\delta_\text{c}$ (right panel). 
        In the left panel, $\delta_\text{c} = 0.2$ (blue), 0.3 (orange), 0.4 (green), 0.5 (red), 0.6 (purple), 0.7 (brown), and 0.8 (pink) from top to bottom.  In the right panel, $\log_{10} A = -2.5$ (blue), $-2.3$ (orange), $-2.1$ (green), $-1.9$ (red), $-1.7$ (purple), and $-1.5$ (brown) from bottom to top.         
        In both panels, the dashed (parts of the) lines correspond to $\delta_\text{c} < 0.4$. Other parameters are fixed to $\Delta = 1$, $T_\text{end} = 10\, \mathrm{MeV}$, and $\log_{10}(f_*/\text{Hz}) = -7.6$.
        }
        \label{fig:f_PBH}
\end{figure}

Let us demonstrate that the PBH abundance is significantly suppressed in our scenario. From the above discussion, the two most important parameters are the critical density $\delta_\text{c}$ and the magnitude of $\mathcal{P}_\zeta$, i.e., the amplitude $A$. 
As a fiducial parameter set, we take $\log_{10}A = -2.3$ and $\log_{10} (f_*/\text{Hz}) = -7.6$, which is in the $68\%$ credible region in Fig.~\ref{fig:posteri} (the black dot in the gray shaded region).
In this case, we find
\begin{align}
    \log_{10} f_\text{PBH} = \begin{cases}
-5 & (\delta_\text{c} = 0.4) \\
-13 & (\delta_\text{c} = 0.5) \\
-22 & (\delta_\text{c} = 0.6) \\
-33 & (\delta_\text{c} = 0.7)
    \end{cases}\ ,
\end{align}
for $\Delta = 1$ and $T_\text{end} = 10 \, \mathrm{MeV}$, where we reported the values in logarithm since the uncertainty of the exponent propagates substantially.  
Thus, even for a conservative choice of $\delta_\text{c}$ in the Universe with $w=1$, the PBH abundance is significantly suppressed. 
For comparison, if we take $\log_{10} A = -1.3$ and $\log_{10}(f_*/\text{Hz}) = -7.6$ in the RD era case as a conservative choice (the black dot around the edge of the $95\%$ credible (light blue) region in Fig.~\ref{fig:posteri}), we obtain $\log_{10}f_\text{PBH} \simeq 5$ with the same formulas in Ref.~\cite{Inomata:2023zup} except for the window function, for which we use Eq.~(\ref{eq:window}) instead of the Gaussian window function.
\footnote{
This is because the setup in Ref.~\cite{Inomata:2023zup} does not necessarily lead to the PBH overproduction unlike that in Ref.~\cite{Franciolini:2023pbf}, which points out the possibility of the PBH overproduction with the use of the real space top-hat window function, Eq.~(\ref{eq:window}).
For example, if we adopt the Gaussian window function ($W(qr) = \ee^{-(qr)^2/2}$) as in Ref.~\cite{Inomata:2023zup}, we obtain $\log_{10} f_\text{PBH} = -55$ with $\delta_\cc = 0.4$ for the kination era and $\log_{10} f_\text{PBH} = -20$ with the same setup in Ref.~\cite{Inomata:2023zup} for the RD era, where the other fiducial parameters are the same as in the main text.
Thus, the PBH abundance is further suppressed, strengthening our conclusion.
}
This shows that the existence of the kination era helps avoid the PBH overproduction, which occurs in the RD era case.
The plots of $f_\text{PBH}$ for other values of $A$ and $\delta_\text{c}$ are shown in Fig.~\ref{fig:f_PBH}. Note that some parameters in Fig.~\ref{fig:f_PBH} are extreme: for example, $\log_{10} A = -1.5$ for $f_* \leq f_\text{UV}$ is too large at more than the $2 \sigma$ level in view of Fig.~\ref{fig:posteri}.  We nevertheless include such values too to show the scale of the uncertainty of $f_\text{PBH}$ (or the sensitivity of $f_\text{PBH}$ on $A$).
 
So far, we have focused on the dependence of $f_\text{PBH}$ on $A$ (or the correlated parameter $f_*$) and $\delta_\text{c}$, but let us comment on the dependence on the other parameters $\Delta$ and $T_\text{end}$.  Since we discuss the IR tail of the induced GWs, the precise value of $\Delta$ is unimportant for sufficiently large values of $f_*$ unless $\Delta \ll 1$, which can be seen in Fig.~\ref{fig:gw}. 
On the other hand, the dominant effect of changing $T_\text{end}$ is through changing the value of $A$ as shown in Eq.~\eqref{degeneracy_relation}. Since a lower (higher) $T_\text{end}$ means a longer (shorter) kination era, it leads to a smaller (larger) amount of PBHs. 

We also comment that, in our analysis for the PBH abundance, we do not take into account the effect of the nonlinear relation between the density perturbations and the curvature perturbations~\cite{Kawasaki:2019mbl,Young:2019yug,DeLuca:2019qsy}. 
This nonlinear relation leads to an intrinsic non-Gaussianity that decreases the PBH abundance with the curvature amplitude fixed, which strengthens our conclusion.

\section{Conclusions} \label{sec:conclusion}

In this paper, we have studied the GWs induced from curvature perturbations in a cosmological era with $w=1$ such as kination.  
In such a scenario, generic (not exceptionally narrow) power spectra of the curvature perturbations whose tilt is $n_s > 1/2$ on the PTA scales but is $n_s < 1/2$ on the adjacent smaller scales lead to the $\Omega_\text{GW} \sim f^2$ scaling at the IR tail of the induced GW spectrum on the PTA scales (see Sec.~\ref{subsec:intuitive_pic}).
The IR tail nicely fits the recent PTA data.  Also, the rapid redshift of the kination fluid leads to the relative enhancement of the induced GWs. Thanks to this effect, the required amplitude of the curvature perturbations is reduced.  Moreover, the critical value of the density perturbation for PBH formation is larger than that in the RD era. These two effects exponentially suppress the PBH abundance in our scenario. We have confirmed that the PBH abundance is suppressed by tens of orders of magnitude compared to the standard case, i.e., the PBH abundance associated with the GWs that are induced in the RD era and fit the NANOGrav data.

While the uncertainty of the spectrum of the induced GWs is relatively small as it is based on perturbative physics,\footnote{
For possible corrections, see Refs.~\cite{Nakama:2016gzw, Garcia-Bellido:2017aan,  Cai:2018dig, Unal:2018yaa, Yuan:2020iwf, Atal:2021jyo, Adshead:2021hnm,Garcia-Saenz:2022tzu, Li:2023qua, Yuan:2023ofl} for the effects of primordial non-Gaussianity of curvature perturbations, Refs.~\cite{Abe:2020sqb, Franciolini:2023wjm, Abe:2023yrw} for changes of the equation-of-state parameter and sound speed in the standard thermal history of the early Universe, Refs.~\cite{Yuan:2019udt, Zhou:2021vcw, Chang:2022nzu, Zhu:2023faa} for the third-order effects in the cosmological perturbation theory.
} the uncertainty of the PBH abundance is huge as the PBH formation is completely non-perturbative physics. We have taken almost the simplest method to compute the PBH abundance to clearly show the basic ideas of the suppression mechanisms. It will be interesting to improve the calculation method in our scenario to estimate the more accurate and precise PBH abundance, which becomes important when we discuss the observability of the PBHs through the microlensing~\cite{2017Natur.548..183M,Niikura:2017zjd,Niikura:2019kqi} and/or the GWs from their mergers~\cite{Wang:2019kaf,Inomata:2023zup}.
We leave it for future work.  

Another direction for future work is to study concrete and realistic particle physics models to realize kination.  For example, kination of an axion(-like) field is an interesting possibility since it is related to the dark matter abundance (\textit{kinetic misalignment})~\cite{Co:2019jts,Eroncel:2022vjg}, baryon asymmetry of the Universe (\textit{axiogenesis})~\cite{Co:2019wyp}, and possible enhancement of the induced GWs (\textit{axion Poltergeist})~\cite{Harigaya:2023ecg}.

In conclusion, the secondary GWs induced by curvature perturbations in a cosmological epoch with $w=1$ can fit the latest PTA data without the PBH overproduction issue.

\acknowledgments
We thank Kazunori Kohri for the discussion on the CMB and BBN constraints in the kination scenario. 
K.I.~was supported by the Kavli Institute
for Cosmological Physics at the University of Chicago
through an endowment from the Kavli Foundation and
its founder Fred Kavli.
K.H.~was partly supported by Grant-in-Aid for Scientific Research from the Ministry of Education, Culture, Sports, Science, and Technology (MEXT), Japan (20H01895) and by World Premier International Research Center Initiative (WPI), MEXT, Japan (Kavli IPMU).
The work of T.T.~was supported by IBS under the project code, IBS-R018-D1.

\small
\bibliographystyle{apsrev4-1}
\bibliography{pta_kination}

\begin{thebibliography}{136}%
\makeatletter
\providecommand \@ifxundefined [1]{%
 \@ifx{#1\undefined}
}%
\providecommand \@ifnum [1]{%
 \ifnum #1\expandafter \@firstoftwo
 \else \expandafter \@secondoftwo
 \fi
}%
\providecommand \@ifx [1]{%
 \ifx #1\expandafter \@firstoftwo
 \else \expandafter \@secondoftwo
 \fi
}%
\providecommand \natexlab [1]{#1}%
\providecommand \enquote  [1]{``#1''}%
\providecommand \bibnamefont  [1]{#1}%
\providecommand \bibfnamefont [1]{#1}%
\providecommand \citenamefont [1]{#1}%
\providecommand \href@noop [0]{\@secondoftwo}%
\providecommand \href [0]{\begingroup \@sanitize@url \@href}%
\providecommand \@href[1]{\@@startlink{#1}\@@href}%
\providecommand \@@href[1]{\endgroup#1\@@endlink}%
\providecommand \@sanitize@url [0]{\catcode `\\12\catcode `\$12\catcode
  `\&12\catcode `\#12\catcode `\^12\catcode `\_12\catcode `\%12\relax}%
\providecommand \@@startlink[1]{}%
\providecommand \@@endlink[0]{}%
\providecommand \url  [0]{\begingroup\@sanitize@url \@url }%
\providecommand \@url [1]{\endgroup\@href {#1}{\urlprefix }}%
\providecommand \urlprefix  [0]{URL }%
\providecommand \Eprint [0]{\href }%
\providecommand \doibase [0]{http://dx.doi.org/}%
\providecommand \selectlanguage [0]{\@gobble}%
\providecommand \bibinfo  [0]{\@secondoftwo}%
\providecommand \bibfield  [0]{\@secondoftwo}%
\providecommand \translation [1]{[#1]}%
\providecommand \BibitemOpen [0]{}%
\providecommand \bibitemStop [0]{}%
\providecommand \bibitemNoStop [0]{.\EOS\space}%
\providecommand \EOS [0]{\spacefactor3000\relax}%
\providecommand \BibitemShut  [1]{\csname bibitem#1\endcsname}%
\let\auto@bib@innerbib\@empty
\bibitem [{\citenamefont {Taylor}\ \emph {et~al.}(1979)\citenamefont {Taylor},
  \citenamefont {Fowler},\ and\ \citenamefont {McCulloch}}]{Taylor:1979zz}%
  \BibitemOpen
  \bibfield  {author} {\bibinfo {author} {\bibfnamefont {J.~H.}\ \bibnamefont
  {Taylor}}, \bibinfo {author} {\bibfnamefont {L.~A.}\ \bibnamefont {Fowler}},
  \ and\ \bibinfo {author} {\bibfnamefont {P.~M.}\ \bibnamefont {McCulloch}},\
  }\href {\doibase 10.1038/277437a0} {\bibfield  {journal} {\bibinfo  {journal}
  {Nature}\ }\textbf {\bibinfo {volume} {277}},\ \bibinfo {pages} {437}
  (\bibinfo {year} {1979})}\BibitemShut {NoStop}%
\bibitem [{\citenamefont {Abbott}\ \emph {et~al.}(2016)\citenamefont {Abbott}
  \emph {et~al.}}]{Abbott:2016blz}%
  \BibitemOpen
  \bibfield  {author} {\bibinfo {author} {\bibfnamefont {B.~P.}\ \bibnamefont
  {Abbott}} \emph {et~al.} (\bibinfo {collaboration} {Virgo, LIGO
  Scientific}),\ }\href {\doibase 10.1103/PhysRevLett.116.061102} {\bibfield
  {journal} {\bibinfo  {journal} {Phys. Rev. Lett.}\ }\textbf {\bibinfo
  {volume} {116}},\ \bibinfo {pages} {061102} (\bibinfo {year} {2016})},\
  \Eprint {http://arxiv.org/abs/1602.03837} {arXiv:1602.03837 [gr-qc]}
  \BibitemShut {NoStop}%
\bibitem [{\citenamefont {Abbott}\ \emph {et~al.}(2021)\citenamefont {Abbott}
  \emph {et~al.}}]{LIGOScientific:2021djp}%
  \BibitemOpen
  \bibfield  {author} {\bibinfo {author} {\bibfnamefont {R.}~\bibnamefont
  {Abbott}} \emph {et~al.} (\bibinfo {collaboration} {LIGO Scientific, VIRGO,
  KAGRA}),\ }\href@noop {} {\  (\bibinfo {year} {2021})},\ \Eprint
  {http://arxiv.org/abs/2111.03606} {arXiv:2111.03606 [gr-qc]} \BibitemShut
  {NoStop}%
\bibitem [{\citenamefont {Hellings}\ and\ \citenamefont
  {Downs}(1983)}]{Hellings:1983fr}%
  \BibitemOpen
  \bibfield  {author} {\bibinfo {author} {\bibfnamefont {R.~w.}\ \bibnamefont
  {Hellings}}\ and\ \bibinfo {author} {\bibfnamefont {G.~s.}\ \bibnamefont
  {Downs}},\ }\href {\doibase 10.1086/183954} {\bibfield  {journal} {\bibinfo
  {journal} {Astrophys. J.}\ }\textbf {\bibinfo {volume} {265}},\ \bibinfo
  {pages} {L39} (\bibinfo {year} {1983})}\BibitemShut {NoStop}%
\bibitem [{\citenamefont {Agazie}\ \emph
  {et~al.}(2023{\natexlab{a}})\citenamefont {Agazie} \emph
  {et~al.}}]{NANOGrav:2023gor}%
  \BibitemOpen
  \bibfield  {author} {\bibinfo {author} {\bibfnamefont {G.}~\bibnamefont
  {Agazie}} \emph {et~al.} (\bibinfo {collaboration} {NANOGrav}),\ }\href
  {\doibase 10.3847/2041-8213/acdac6} {\bibfield  {journal} {\bibinfo
  {journal} {Astrophys. J. Lett.}\ }\textbf {\bibinfo {volume} {951}},\
  \bibinfo {pages} {L8} (\bibinfo {year} {2023}{\natexlab{a}})},\ \Eprint
  {http://arxiv.org/abs/2306.16213} {arXiv:2306.16213 [astro-ph.HE]}
  \BibitemShut {NoStop}%
\bibitem [{\citenamefont {Agazie}\ \emph
  {et~al.}(2023{\natexlab{b}})\citenamefont {Agazie} \emph
  {et~al.}}]{NANOGrav:2023hde}%
  \BibitemOpen
  \bibfield  {author} {\bibinfo {author} {\bibfnamefont {G.}~\bibnamefont
  {Agazie}} \emph {et~al.} (\bibinfo {collaboration} {NANOGrav}),\ }\href
  {\doibase 10.3847/2041-8213/acda9a} {\bibfield  {journal} {\bibinfo
  {journal} {Astrophys. J. Lett.}\ }\textbf {\bibinfo {volume} {951}},\
  \bibinfo {pages} {L9} (\bibinfo {year} {2023}{\natexlab{b}})},\ \Eprint
  {http://arxiv.org/abs/2306.16217} {arXiv:2306.16217 [astro-ph.HE]}
  \BibitemShut {NoStop}%
\bibitem [{\citenamefont {Afzal}\ \emph {et~al.}(2023)\citenamefont {Afzal}
  \emph {et~al.}}]{NANOGrav:2023hvm}%
  \BibitemOpen
  \bibfield  {author} {\bibinfo {author} {\bibfnamefont {A.}~\bibnamefont
  {Afzal}} \emph {et~al.} (\bibinfo {collaboration} {NANOGrav}),\ }\href
  {\doibase 10.3847/2041-8213/acdc91} {\bibfield  {journal} {\bibinfo
  {journal} {Astrophys. J. Lett.}\ }\textbf {\bibinfo {volume} {951}},\
  \bibinfo {pages} {L11} (\bibinfo {year} {2023})},\ \Eprint
  {http://arxiv.org/abs/2306.16219} {arXiv:2306.16219 [astro-ph.HE]}
  \BibitemShut {NoStop}%
\bibitem [{\citenamefont {Antoniadis}\ \emph
  {et~al.}(2023{\natexlab{a}})\citenamefont {Antoniadis} \emph
  {et~al.}}]{Antoniadis:2023rey}%
  \BibitemOpen
  \bibfield  {author} {\bibinfo {author} {\bibfnamefont {J.}~\bibnamefont
  {Antoniadis}} \emph {et~al.},\ }\href@noop {} {\  (\bibinfo {year}
  {2023}{\natexlab{a}})},\ \Eprint {http://arxiv.org/abs/2306.16214}
  {arXiv:2306.16214 [astro-ph.HE]} \BibitemShut {NoStop}%
\bibitem [{\citenamefont {Antoniadis}\ \emph
  {et~al.}(2023{\natexlab{b}})\citenamefont {Antoniadis} \emph
  {et~al.}}]{Antoniadis:2023utw}%
  \BibitemOpen
  \bibfield  {author} {\bibinfo {author} {\bibfnamefont {J.}~\bibnamefont
  {Antoniadis}} \emph {et~al.},\ }\href {\doibase 10.1051/0004-6361/202346841}
  {\  (\bibinfo {year} {2023}{\natexlab{b}}),\ 10.1051/0004-6361/202346841},\
  \Eprint {http://arxiv.org/abs/2306.16224} {arXiv:2306.16224 [astro-ph.HE]}
  \BibitemShut {NoStop}%
\bibitem [{\citenamefont {Antoniadis}\ \emph
  {et~al.}(2023{\natexlab{c}})\citenamefont {Antoniadis} \emph
  {et~al.}}]{Antoniadis:2023zhi}%
  \BibitemOpen
  \bibfield  {author} {\bibinfo {author} {\bibfnamefont {J.}~\bibnamefont
  {Antoniadis}} \emph {et~al.},\ }\href@noop {} {\  (\bibinfo {year}
  {2023}{\natexlab{c}})},\ \Eprint {http://arxiv.org/abs/2306.16227}
  {arXiv:2306.16227 [astro-ph.CO]} \BibitemShut {NoStop}%
\bibitem [{\citenamefont {Reardon}\ \emph
  {et~al.}(2023{\natexlab{a}})\citenamefont {Reardon} \emph
  {et~al.}}]{Reardon:2023gzh}%
  \BibitemOpen
  \bibfield  {author} {\bibinfo {author} {\bibfnamefont {D.~J.}\ \bibnamefont
  {Reardon}} \emph {et~al.},\ }\href {\doibase 10.3847/2041-8213/acdd02}
  {\bibfield  {journal} {\bibinfo  {journal} {Astrophys. J. Lett.}\ }\textbf
  {\bibinfo {volume} {951}},\ \bibinfo {pages} {L6} (\bibinfo {year}
  {2023}{\natexlab{a}})},\ \Eprint {http://arxiv.org/abs/2306.16215}
  {arXiv:2306.16215 [astro-ph.HE]} \BibitemShut {NoStop}%
\bibitem [{\citenamefont {Zic}\ \emph {et~al.}(2023)\citenamefont {Zic} \emph
  {et~al.}}]{Zic:2023gta}%
  \BibitemOpen
  \bibfield  {author} {\bibinfo {author} {\bibfnamefont {A.}~\bibnamefont
  {Zic}} \emph {et~al.},\ }\href@noop {} {\  (\bibinfo {year} {2023})},\
  \Eprint {http://arxiv.org/abs/2306.16230} {arXiv:2306.16230 [astro-ph.HE]}
  \BibitemShut {NoStop}%
\bibitem [{\citenamefont {Reardon}\ \emph
  {et~al.}(2023{\natexlab{b}})\citenamefont {Reardon} \emph
  {et~al.}}]{Reardon:2023zen}%
  \BibitemOpen
  \bibfield  {author} {\bibinfo {author} {\bibfnamefont {D.~J.}\ \bibnamefont
  {Reardon}} \emph {et~al.},\ }\href {\doibase 10.3847/2041-8213/acdd03}
  {\bibfield  {journal} {\bibinfo  {journal} {Astrophys. J. Lett.}\ }\textbf
  {\bibinfo {volume} {951}},\ \bibinfo {pages} {L7} (\bibinfo {year}
  {2023}{\natexlab{b}})},\ \Eprint {http://arxiv.org/abs/2306.16229}
  {arXiv:2306.16229 [astro-ph.HE]} \BibitemShut {NoStop}%
\bibitem [{\citenamefont {Xu}\ \emph {et~al.}(2023)\citenamefont {Xu} \emph
  {et~al.}}]{Xu:2023wog}%
  \BibitemOpen
  \bibfield  {author} {\bibinfo {author} {\bibfnamefont {H.}~\bibnamefont {Xu}}
  \emph {et~al.},\ }\href {\doibase 10.1088/1674-4527/acdfa5} {\bibfield
  {journal} {\bibinfo  {journal} {Res. Astron. Astrophys.}\ }\textbf {\bibinfo
  {volume} {23}},\ \bibinfo {pages} {075024} (\bibinfo {year} {2023})},\
  \Eprint {http://arxiv.org/abs/2306.16216} {arXiv:2306.16216 [astro-ph.HE]}
  \BibitemShut {NoStop}%
\bibitem [{\citenamefont {Agazie}\ \emph
  {et~al.}(2023{\natexlab{c}})\citenamefont {Agazie} \emph
  {et~al.}}]{NANOGrav:2023tcn}%
  \BibitemOpen
  \bibfield  {author} {\bibinfo {author} {\bibfnamefont {G.}~\bibnamefont
  {Agazie}} \emph {et~al.} (\bibinfo {collaboration} {NANOGrav}),\ }\href@noop
  {} {\  (\bibinfo {year} {2023}{\natexlab{c}})},\ \Eprint
  {http://arxiv.org/abs/2306.16221} {arXiv:2306.16221 [astro-ph.HE]}
  \BibitemShut {NoStop}%
\bibitem [{\citenamefont {Agazie}\ \emph
  {et~al.}(2023{\natexlab{d}})\citenamefont {Agazie} \emph
  {et~al.}}]{NANOGrav:2023pdq}%
  \BibitemOpen
  \bibfield  {author} {\bibinfo {author} {\bibfnamefont {G.}~\bibnamefont
  {Agazie}} \emph {et~al.} (\bibinfo {collaboration} {NANOGrav}),\ }\href
  {\doibase 10.3847/2041-8213/ace18a} {\bibfield  {journal} {\bibinfo
  {journal} {Astrophys. J. Lett.}\ }\textbf {\bibinfo {volume} {951}},\
  \bibinfo {pages} {L50} (\bibinfo {year} {2023}{\natexlab{d}})},\ \Eprint
  {http://arxiv.org/abs/2306.16222} {arXiv:2306.16222 [astro-ph.HE]}
  \BibitemShut {NoStop}%
\bibitem [{\citenamefont {Antoniadis}\ \emph
  {et~al.}(2023{\natexlab{d}})\citenamefont {Antoniadis} \emph
  {et~al.}}]{Antoniadis:2023bjw}%
  \BibitemOpen
  \bibfield  {author} {\bibinfo {author} {\bibfnamefont {J.}~\bibnamefont
  {Antoniadis}} \emph {et~al.},\ }\href@noop {} {\  (\bibinfo {year}
  {2023}{\natexlab{d}})},\ \Eprint {http://arxiv.org/abs/2306.16226}
  {arXiv:2306.16226 [astro-ph.HE]} \BibitemShut {NoStop}%
\bibitem [{\citenamefont {Agazie}\ \emph
  {et~al.}(2023{\natexlab{e}})\citenamefont {Agazie} \emph
  {et~al.}}]{NANOGrav:2023hfp}%
  \BibitemOpen
  \bibfield  {author} {\bibinfo {author} {\bibfnamefont {G.}~\bibnamefont
  {Agazie}} \emph {et~al.} (\bibinfo {collaboration} {NANOGrav}),\ }\href@noop
  {} {\  (\bibinfo {year} {2023}{\natexlab{e}})},\ \Eprint
  {http://arxiv.org/abs/2306.16220} {arXiv:2306.16220 [astro-ph.HE]}
  \BibitemShut {NoStop}%
\bibitem [{\citenamefont {Ellis}\ \emph
  {et~al.}(2023{\natexlab{a}})\citenamefont {Ellis}, \citenamefont {Fairbairn},
  \citenamefont {H\"utsi}, \citenamefont {Raidal}, \citenamefont {Urrutia},
  \citenamefont {Vaskonen},\ and\ \citenamefont {Veerm\"ae}}]{Ellis:2023dgf}%
  \BibitemOpen
  \bibfield  {author} {\bibinfo {author} {\bibfnamefont {J.}~\bibnamefont
  {Ellis}}, \bibinfo {author} {\bibfnamefont {M.}~\bibnamefont {Fairbairn}},
  \bibinfo {author} {\bibfnamefont {G.}~\bibnamefont {H\"utsi}}, \bibinfo
  {author} {\bibfnamefont {J.}~\bibnamefont {Raidal}}, \bibinfo {author}
  {\bibfnamefont {J.}~\bibnamefont {Urrutia}}, \bibinfo {author} {\bibfnamefont
  {V.}~\bibnamefont {Vaskonen}}, \ and\ \bibinfo {author} {\bibfnamefont
  {H.}~\bibnamefont {Veerm\"ae}},\ }\href@noop {} {\  (\bibinfo {year}
  {2023}{\natexlab{a}})},\ \Eprint {http://arxiv.org/abs/2306.17021}
  {arXiv:2306.17021 [astro-ph.CO]} \BibitemShut {NoStop}%
\bibitem [{\citenamefont {Depta}\ \emph {et~al.}(2023)\citenamefont {Depta},
  \citenamefont {Schmidt-Hoberg},\ and\ \citenamefont
  {Tasillo}}]{Depta:2023qst}%
  \BibitemOpen
  \bibfield  {author} {\bibinfo {author} {\bibfnamefont {P.~F.}\ \bibnamefont
  {Depta}}, \bibinfo {author} {\bibfnamefont {K.}~\bibnamefont
  {Schmidt-Hoberg}}, \ and\ \bibinfo {author} {\bibfnamefont {C.}~\bibnamefont
  {Tasillo}},\ }\href@noop {} {\  (\bibinfo {year} {2023})},\ \Eprint
  {http://arxiv.org/abs/2306.17836} {arXiv:2306.17836 [astro-ph.CO]}
  \BibitemShut {NoStop}%
\bibitem [{\citenamefont {Figueroa}\ \emph {et~al.}(2023)\citenamefont
  {Figueroa}, \citenamefont {Pieroni}, \citenamefont {Ricciardone},\ and\
  \citenamefont {Simakachorn}}]{Figueroa:2023zhu}%
  \BibitemOpen
  \bibfield  {author} {\bibinfo {author} {\bibfnamefont {D.~G.}\ \bibnamefont
  {Figueroa}}, \bibinfo {author} {\bibfnamefont {M.}~\bibnamefont {Pieroni}},
  \bibinfo {author} {\bibfnamefont {A.}~\bibnamefont {Ricciardone}}, \ and\
  \bibinfo {author} {\bibfnamefont {P.}~\bibnamefont {Simakachorn}},\
  }\href@noop {} {\  (\bibinfo {year} {2023})},\ \Eprint
  {http://arxiv.org/abs/2307.02399} {arXiv:2307.02399 [astro-ph.CO]}
  \BibitemShut {NoStop}%
\bibitem [{\citenamefont {Ellis}\ \emph
  {et~al.}(2023{\natexlab{b}})\citenamefont {Ellis}, \citenamefont {Fairbairn},
  \citenamefont {Franciolini}, \citenamefont {H\"utsi}, \citenamefont {Iovino},
  \citenamefont {Lewicki}, \citenamefont {Raidal}, \citenamefont {Urrutia},
  \citenamefont {Vaskonen},\ and\ \citenamefont {Veerm\"ae}}]{Ellis:2023oxs}%
  \BibitemOpen
  \bibfield  {author} {\bibinfo {author} {\bibfnamefont {J.}~\bibnamefont
  {Ellis}}, \bibinfo {author} {\bibfnamefont {M.}~\bibnamefont {Fairbairn}},
  \bibinfo {author} {\bibfnamefont {G.}~\bibnamefont {Franciolini}}, \bibinfo
  {author} {\bibfnamefont {G.}~\bibnamefont {H\"utsi}}, \bibinfo {author}
  {\bibfnamefont {A.}~\bibnamefont {Iovino}}, \bibinfo {author} {\bibfnamefont
  {M.}~\bibnamefont {Lewicki}}, \bibinfo {author} {\bibfnamefont
  {M.}~\bibnamefont {Raidal}}, \bibinfo {author} {\bibfnamefont
  {J.}~\bibnamefont {Urrutia}}, \bibinfo {author} {\bibfnamefont
  {V.}~\bibnamefont {Vaskonen}}, \ and\ \bibinfo {author} {\bibfnamefont
  {H.}~\bibnamefont {Veerm\"ae}},\ }\href@noop {} {\  (\bibinfo {year}
  {2023}{\natexlab{b}})},\ \Eprint {http://arxiv.org/abs/2308.08546}
  {arXiv:2308.08546 [astro-ph.CO]} \BibitemShut {NoStop}%
\bibitem [{\citenamefont {Bian}\ \emph {et~al.}(2023)\citenamefont {Bian},
  \citenamefont {Ge}, \citenamefont {Shu}, \citenamefont {Wang}, \citenamefont
  {Yang},\ and\ \citenamefont {Zong}}]{Bian:2023dnv}%
  \BibitemOpen
  \bibfield  {author} {\bibinfo {author} {\bibfnamefont {L.}~\bibnamefont
  {Bian}}, \bibinfo {author} {\bibfnamefont {S.}~\bibnamefont {Ge}}, \bibinfo
  {author} {\bibfnamefont {J.}~\bibnamefont {Shu}}, \bibinfo {author}
  {\bibfnamefont {B.}~\bibnamefont {Wang}}, \bibinfo {author} {\bibfnamefont
  {X.-Y.}\ \bibnamefont {Yang}}, \ and\ \bibinfo {author} {\bibfnamefont
  {J.}~\bibnamefont {Zong}},\ }\href@noop {} {\  (\bibinfo {year} {2023})},\
  \Eprint {http://arxiv.org/abs/2307.02376} {arXiv:2307.02376 [astro-ph.HE]}
  \BibitemShut {NoStop}%
\bibitem [{\citenamefont {Tomita}(1967)}]{10.1143/PTP.37.831}%
  \BibitemOpen
  \bibfield  {author} {\bibinfo {author} {\bibfnamefont {K.}~\bibnamefont
  {Tomita}},\ }\href {\doibase 10.1143/PTP.37.831} {\bibfield  {journal}
  {\bibinfo  {journal} {Progress of Theoretical Physics}\ }\textbf {\bibinfo
  {volume} {37}},\ \bibinfo {pages} {831} (\bibinfo {year} {1967})},\ \Eprint
  {http://arxiv.org/abs/https://academic.oup.com/ptp/article-pdf/37/5/831/5234391/37-5-831.pdf}
  {https://academic.oup.com/ptp/article-pdf/37/5/831/5234391/37-5-831.pdf}
  \BibitemShut {NoStop}%
\bibitem [{\citenamefont {Matarrese}\ \emph {et~al.}(1994)\citenamefont
  {Matarrese}, \citenamefont {Pantano},\ and\ \citenamefont
  {Saez}}]{Matarrese:1993zf}%
  \BibitemOpen
  \bibfield  {author} {\bibinfo {author} {\bibfnamefont {S.}~\bibnamefont
  {Matarrese}}, \bibinfo {author} {\bibfnamefont {O.}~\bibnamefont {Pantano}},
  \ and\ \bibinfo {author} {\bibfnamefont {D.}~\bibnamefont {Saez}},\ }\href
  {\doibase 10.1103/PhysRevLett.72.320} {\bibfield  {journal} {\bibinfo
  {journal} {Phys. Rev. Lett.}\ }\textbf {\bibinfo {volume} {72}},\ \bibinfo
  {pages} {320} (\bibinfo {year} {1994})},\ \Eprint
  {http://arxiv.org/abs/astro-ph/9310036} {arXiv:astro-ph/9310036} \BibitemShut
  {NoStop}%
\bibitem [{\citenamefont {Matarrese}\ \emph {et~al.}(1998)\citenamefont
  {Matarrese}, \citenamefont {Mollerach},\ and\ \citenamefont
  {Bruni}}]{Matarrese:1997ay}%
  \BibitemOpen
  \bibfield  {author} {\bibinfo {author} {\bibfnamefont {S.}~\bibnamefont
  {Matarrese}}, \bibinfo {author} {\bibfnamefont {S.}~\bibnamefont
  {Mollerach}}, \ and\ \bibinfo {author} {\bibfnamefont {M.}~\bibnamefont
  {Bruni}},\ }\href {\doibase 10.1103/PhysRevD.58.043504} {\bibfield  {journal}
  {\bibinfo  {journal} {Phys. Rev. D}\ }\textbf {\bibinfo {volume} {58}},\
  \bibinfo {pages} {043504} (\bibinfo {year} {1998})},\ \Eprint
  {http://arxiv.org/abs/astro-ph/9707278} {arXiv:astro-ph/9707278} \BibitemShut
  {NoStop}%
\bibitem [{\citenamefont {Ananda}\ \emph {et~al.}(2007)\citenamefont {Ananda},
  \citenamefont {Clarkson},\ and\ \citenamefont {Wands}}]{Ananda:2006af}%
  \BibitemOpen
  \bibfield  {author} {\bibinfo {author} {\bibfnamefont {K.~N.}\ \bibnamefont
  {Ananda}}, \bibinfo {author} {\bibfnamefont {C.}~\bibnamefont {Clarkson}}, \
  and\ \bibinfo {author} {\bibfnamefont {D.}~\bibnamefont {Wands}},\ }\href
  {\doibase 10.1103/PhysRevD.75.123518} {\bibfield  {journal} {\bibinfo
  {journal} {Phys. Rev.}\ }\textbf {\bibinfo {volume} {D75}},\ \bibinfo {pages}
  {123518} (\bibinfo {year} {2007})},\ \Eprint
  {http://arxiv.org/abs/gr-qc/0612013} {arXiv:gr-qc/0612013 [gr-qc]}
  \BibitemShut {NoStop}%
\bibitem [{\citenamefont {Baumann}\ \emph {et~al.}(2007)\citenamefont
  {Baumann}, \citenamefont {Steinhardt}, \citenamefont {Takahashi},\ and\
  \citenamefont {Ichiki}}]{Baumann:2007zm}%
  \BibitemOpen
  \bibfield  {author} {\bibinfo {author} {\bibfnamefont {D.}~\bibnamefont
  {Baumann}}, \bibinfo {author} {\bibfnamefont {P.~J.}\ \bibnamefont
  {Steinhardt}}, \bibinfo {author} {\bibfnamefont {K.}~\bibnamefont
  {Takahashi}}, \ and\ \bibinfo {author} {\bibfnamefont {K.}~\bibnamefont
  {Ichiki}},\ }\href {\doibase 10.1103/PhysRevD.76.084019} {\bibfield
  {journal} {\bibinfo  {journal} {Phys. Rev.}\ }\textbf {\bibinfo {volume}
  {D76}},\ \bibinfo {pages} {084019} (\bibinfo {year} {2007})},\ \Eprint
  {http://arxiv.org/abs/hep-th/0703290} {arXiv:hep-th/0703290 [hep-th]}
  \BibitemShut {NoStop}%
\bibitem [{\citenamefont {Inomata}\ \emph {et~al.}(2017)\citenamefont
  {Inomata}, \citenamefont {Kawasaki}, \citenamefont {Mukaida}, \citenamefont
  {Tada},\ and\ \citenamefont {Yanagida}}]{Inomata:2016rbd}%
  \BibitemOpen
  \bibfield  {author} {\bibinfo {author} {\bibfnamefont {K.}~\bibnamefont
  {Inomata}}, \bibinfo {author} {\bibfnamefont {M.}~\bibnamefont {Kawasaki}},
  \bibinfo {author} {\bibfnamefont {K.}~\bibnamefont {Mukaida}}, \bibinfo
  {author} {\bibfnamefont {Y.}~\bibnamefont {Tada}}, \ and\ \bibinfo {author}
  {\bibfnamefont {T.~T.}\ \bibnamefont {Yanagida}},\ }\href {\doibase
  10.1103/PhysRevD.95.123510} {\bibfield  {journal} {\bibinfo  {journal} {Phys.
  Rev.}\ }\textbf {\bibinfo {volume} {D95}},\ \bibinfo {pages} {123510}
  (\bibinfo {year} {2017})},\ \Eprint {http://arxiv.org/abs/1611.06130}
  {arXiv:1611.06130 [astro-ph.CO]} \BibitemShut {NoStop}%
\bibitem [{\citenamefont {Orlofsky}\ \emph {et~al.}(2017)\citenamefont
  {Orlofsky}, \citenamefont {Pierce},\ and\ \citenamefont
  {Wells}}]{Orlofsky:2016vbd}%
  \BibitemOpen
  \bibfield  {author} {\bibinfo {author} {\bibfnamefont {N.}~\bibnamefont
  {Orlofsky}}, \bibinfo {author} {\bibfnamefont {A.}~\bibnamefont {Pierce}}, \
  and\ \bibinfo {author} {\bibfnamefont {J.~D.}\ \bibnamefont {Wells}},\ }\href
  {\doibase 10.1103/PhysRevD.95.063518} {\bibfield  {journal} {\bibinfo
  {journal} {Phys. Rev.}\ }\textbf {\bibinfo {volume} {D95}},\ \bibinfo {pages}
  {063518} (\bibinfo {year} {2017})},\ \Eprint
  {http://arxiv.org/abs/1612.05279} {arXiv:1612.05279 [astro-ph.CO]}
  \BibitemShut {NoStop}%
\bibitem [{\citenamefont {Di}\ and\ \citenamefont {Gong}(2018)}]{Di:2017ndc}%
  \BibitemOpen
  \bibfield  {author} {\bibinfo {author} {\bibfnamefont {H.}~\bibnamefont
  {Di}}\ and\ \bibinfo {author} {\bibfnamefont {Y.}~\bibnamefont {Gong}},\
  }\href {\doibase 10.1088/1475-7516/2018/07/007} {\bibfield  {journal}
  {\bibinfo  {journal} {JCAP}\ }\textbf {\bibinfo {volume} {07}},\ \bibinfo
  {pages} {007} (\bibinfo {year} {2018})},\ \Eprint
  {http://arxiv.org/abs/1707.09578} {arXiv:1707.09578 [astro-ph.CO]}
  \BibitemShut {NoStop}%
\bibitem [{\citenamefont {Ando}\ \emph
  {et~al.}(2018{\natexlab{a}})\citenamefont {Ando}, \citenamefont {Inomata},
  \citenamefont {Kawasaki}, \citenamefont {Mukaida},\ and\ \citenamefont
  {Yanagida}}]{Ando:2017veq}%
  \BibitemOpen
  \bibfield  {author} {\bibinfo {author} {\bibfnamefont {K.}~\bibnamefont
  {Ando}}, \bibinfo {author} {\bibfnamefont {K.}~\bibnamefont {Inomata}},
  \bibinfo {author} {\bibfnamefont {M.}~\bibnamefont {Kawasaki}}, \bibinfo
  {author} {\bibfnamefont {K.}~\bibnamefont {Mukaida}}, \ and\ \bibinfo
  {author} {\bibfnamefont {T.~T.}\ \bibnamefont {Yanagida}},\ }\href {\doibase
  10.1103/PhysRevD.97.123512} {\bibfield  {journal} {\bibinfo  {journal} {Phys.
  Rev.}\ }\textbf {\bibinfo {volume} {D97}},\ \bibinfo {pages} {123512}
  (\bibinfo {year} {2018}{\natexlab{a}})},\ \Eprint
  {http://arxiv.org/abs/1711.08956} {arXiv:1711.08956 [astro-ph.CO]}
  \BibitemShut {NoStop}%
\bibitem [{\citenamefont {Drees}\ and\ \citenamefont
  {Xu}(2021)}]{Drees:2019xpp}%
  \BibitemOpen
  \bibfield  {author} {\bibinfo {author} {\bibfnamefont {M.}~\bibnamefont
  {Drees}}\ and\ \bibinfo {author} {\bibfnamefont {Y.}~\bibnamefont {Xu}},\
  }\href {\doibase 10.1140/epjc/s10052-021-08976-2} {\bibfield  {journal}
  {\bibinfo  {journal} {Eur. Phys. J. C}\ }\textbf {\bibinfo {volume} {81}},\
  \bibinfo {pages} {182} (\bibinfo {year} {2021})},\ \Eprint
  {http://arxiv.org/abs/1905.13581} {arXiv:1905.13581 [hep-ph]} \BibitemShut
  {NoStop}%
\bibitem [{\citenamefont {Fu}\ and\ \citenamefont {Wang}(2023)}]{Fu:2022ypp}%
  \BibitemOpen
  \bibfield  {author} {\bibinfo {author} {\bibfnamefont {C.}~\bibnamefont
  {Fu}}\ and\ \bibinfo {author} {\bibfnamefont {S.-J.}\ \bibnamefont {Wang}},\
  }\href {\doibase 10.1088/1475-7516/2023/06/012} {\bibfield  {journal}
  {\bibinfo  {journal} {JCAP}\ }\textbf {\bibinfo {volume} {06}},\ \bibinfo
  {pages} {012} (\bibinfo {year} {2023})},\ \Eprint
  {http://arxiv.org/abs/2211.03523} {arXiv:2211.03523 [astro-ph.CO]}
  \BibitemShut {NoStop}%
\bibitem [{\citenamefont {Vaskonen}\ and\ \citenamefont
  {Veerm\"ae}(2021)}]{Vaskonen:2020lbd}%
  \BibitemOpen
  \bibfield  {author} {\bibinfo {author} {\bibfnamefont {V.}~\bibnamefont
  {Vaskonen}}\ and\ \bibinfo {author} {\bibfnamefont {H.}~\bibnamefont
  {Veerm\"ae}},\ }\href {\doibase 10.1103/PhysRevLett.126.051303} {\bibfield
  {journal} {\bibinfo  {journal} {Phys. Rev. Lett.}\ }\textbf {\bibinfo
  {volume} {126}},\ \bibinfo {pages} {051303} (\bibinfo {year} {2021})},\
  \Eprint {http://arxiv.org/abs/2009.07832} {arXiv:2009.07832 [astro-ph.CO]}
  \BibitemShut {NoStop}%
\bibitem [{\citenamefont {De~Luca}\ \emph {et~al.}(2021)\citenamefont
  {De~Luca}, \citenamefont {Franciolini},\ and\ \citenamefont
  {Riotto}}]{DeLuca:2020agl}%
  \BibitemOpen
  \bibfield  {author} {\bibinfo {author} {\bibfnamefont {V.}~\bibnamefont
  {De~Luca}}, \bibinfo {author} {\bibfnamefont {G.}~\bibnamefont
  {Franciolini}}, \ and\ \bibinfo {author} {\bibfnamefont {A.}~\bibnamefont
  {Riotto}},\ }\href {\doibase 10.1103/PhysRevLett.126.041303} {\bibfield
  {journal} {\bibinfo  {journal} {Phys. Rev. Lett.}\ }\textbf {\bibinfo
  {volume} {126}},\ \bibinfo {pages} {041303} (\bibinfo {year} {2021})},\
  \Eprint {http://arxiv.org/abs/2009.08268} {arXiv:2009.08268 [astro-ph.CO]}
  \BibitemShut {NoStop}%
\bibitem [{\citenamefont {Kohri}\ and\ \citenamefont
  {Terada}(2021)}]{Kohri:2020qqd}%
  \BibitemOpen
  \bibfield  {author} {\bibinfo {author} {\bibfnamefont {K.}~\bibnamefont
  {Kohri}}\ and\ \bibinfo {author} {\bibfnamefont {T.}~\bibnamefont {Terada}},\
  }\href {\doibase 10.1016/j.physletb.2020.136040} {\bibfield  {journal}
  {\bibinfo  {journal} {Phys. Lett. B}\ }\textbf {\bibinfo {volume} {813}},\
  \bibinfo {pages} {136040} (\bibinfo {year} {2021})},\ \Eprint
  {http://arxiv.org/abs/2009.11853} {arXiv:2009.11853 [astro-ph.CO]}
  \BibitemShut {NoStop}%
\bibitem [{\citenamefont {Dom\`enech}\ and\ \citenamefont
  {Pi}(2022)}]{Domenech:2020ers}%
  \BibitemOpen
  \bibfield  {author} {\bibinfo {author} {\bibfnamefont {G.}~\bibnamefont
  {Dom\`enech}}\ and\ \bibinfo {author} {\bibfnamefont {S.}~\bibnamefont
  {Pi}},\ }\href {\doibase 10.1007/s11433-021-1839-6} {\bibfield  {journal}
  {\bibinfo  {journal} {Sci. China Phys. Mech. Astron.}\ }\textbf {\bibinfo
  {volume} {65}},\ \bibinfo {pages} {230411} (\bibinfo {year} {2022})},\
  \Eprint {http://arxiv.org/abs/2010.03976} {arXiv:2010.03976 [astro-ph.CO]}
  \BibitemShut {NoStop}%
\bibitem [{\citenamefont {Inomata}\ \emph {et~al.}(2021)\citenamefont
  {Inomata}, \citenamefont {Kawasaki}, \citenamefont {Mukaida},\ and\
  \citenamefont {Yanagida}}]{Inomata:2020xad}%
  \BibitemOpen
  \bibfield  {author} {\bibinfo {author} {\bibfnamefont {K.}~\bibnamefont
  {Inomata}}, \bibinfo {author} {\bibfnamefont {M.}~\bibnamefont {Kawasaki}},
  \bibinfo {author} {\bibfnamefont {K.}~\bibnamefont {Mukaida}}, \ and\
  \bibinfo {author} {\bibfnamefont {T.~T.}\ \bibnamefont {Yanagida}},\ }\href
  {\doibase 10.1103/PhysRevLett.126.131301} {\bibfield  {journal} {\bibinfo
  {journal} {Phys. Rev. Lett.}\ }\textbf {\bibinfo {volume} {126}},\ \bibinfo
  {pages} {131301} (\bibinfo {year} {2021})},\ \Eprint
  {http://arxiv.org/abs/2011.01270} {arXiv:2011.01270 [astro-ph.CO]}
  \BibitemShut {NoStop}%
\bibitem [{\citenamefont {Kawasaki}\ and\ \citenamefont
  {Nakatsuka}(2021)}]{Kawasaki:2021ycf}%
  \BibitemOpen
  \bibfield  {author} {\bibinfo {author} {\bibfnamefont {M.}~\bibnamefont
  {Kawasaki}}\ and\ \bibinfo {author} {\bibfnamefont {H.}~\bibnamefont
  {Nakatsuka}},\ }\href {\doibase 10.1088/1475-7516/2021/05/023} {\bibfield
  {journal} {\bibinfo  {journal} {JCAP}\ }\textbf {\bibinfo {volume} {05}},\
  \bibinfo {pages} {023} (\bibinfo {year} {2021})},\ \Eprint
  {http://arxiv.org/abs/2101.11244} {arXiv:2101.11244 [astro-ph.CO]}
  \BibitemShut {NoStop}%
\bibitem [{\citenamefont {Dandoy}\ \emph {et~al.}(2023)\citenamefont {Dandoy},
  \citenamefont {Domcke},\ and\ \citenamefont {Rompineve}}]{Dandoy:2023jot}%
  \BibitemOpen
  \bibfield  {author} {\bibinfo {author} {\bibfnamefont {V.}~\bibnamefont
  {Dandoy}}, \bibinfo {author} {\bibfnamefont {V.}~\bibnamefont {Domcke}}, \
  and\ \bibinfo {author} {\bibfnamefont {F.}~\bibnamefont {Rompineve}},\
  }\href@noop {} {\  (\bibinfo {year} {2023})},\ \Eprint
  {http://arxiv.org/abs/2302.07901} {arXiv:2302.07901 [astro-ph.CO]}
  \BibitemShut {NoStop}%
\bibitem [{\citenamefont {Madge}\ \emph {et~al.}(2023)\citenamefont {Madge},
  \citenamefont {Morgante}, \citenamefont {Puchades-Ib\'a\~nez}, \citenamefont
  {Ramberg}, \citenamefont {Ratzinger}, \citenamefont {Schenk},\ and\
  \citenamefont {Schwaller}}]{Madge:2023cak}%
  \BibitemOpen
  \bibfield  {author} {\bibinfo {author} {\bibfnamefont {E.}~\bibnamefont
  {Madge}}, \bibinfo {author} {\bibfnamefont {E.}~\bibnamefont {Morgante}},
  \bibinfo {author} {\bibfnamefont {C.}~\bibnamefont {Puchades-Ib\'a\~nez}},
  \bibinfo {author} {\bibfnamefont {N.}~\bibnamefont {Ramberg}}, \bibinfo
  {author} {\bibfnamefont {W.}~\bibnamefont {Ratzinger}}, \bibinfo {author}
  {\bibfnamefont {S.}~\bibnamefont {Schenk}}, \ and\ \bibinfo {author}
  {\bibfnamefont {P.}~\bibnamefont {Schwaller}},\ }\href@noop {} {\  (\bibinfo
  {year} {2023})},\ \Eprint {http://arxiv.org/abs/2306.14856} {arXiv:2306.14856
  [hep-ph]} \BibitemShut {NoStop}%
\bibitem [{\citenamefont {Franciolini}\ \emph
  {et~al.}(2023{\natexlab{a}})\citenamefont {Franciolini}, \citenamefont
  {Iovino}, \citenamefont {Vaskonen},\ and\ \citenamefont
  {Veermae}}]{Franciolini:2023pbf}%
  \BibitemOpen
  \bibfield  {author} {\bibinfo {author} {\bibfnamefont {G.}~\bibnamefont
  {Franciolini}}, \bibinfo {author} {\bibfnamefont {A.}~\bibnamefont {Iovino},
  \bibfnamefont {Junior.}}, \bibinfo {author} {\bibfnamefont {V.}~\bibnamefont
  {Vaskonen}}, \ and\ \bibinfo {author} {\bibfnamefont {H.}~\bibnamefont
  {Veermae}},\ }\href@noop {} {\  (\bibinfo {year} {2023}{\natexlab{a}})},\
  \Eprint {http://arxiv.org/abs/2306.17149} {arXiv:2306.17149 [astro-ph.CO]}
  \BibitemShut {NoStop}%
\bibitem [{\citenamefont {Cai}\ \emph {et~al.}(2023)\citenamefont {Cai},
  \citenamefont {He}, \citenamefont {Ma}, \citenamefont {Yan},\ and\
  \citenamefont {Yuan}}]{Cai:2023dls}%
  \BibitemOpen
  \bibfield  {author} {\bibinfo {author} {\bibfnamefont {Y.-F.}\ \bibnamefont
  {Cai}}, \bibinfo {author} {\bibfnamefont {X.-C.}\ \bibnamefont {He}},
  \bibinfo {author} {\bibfnamefont {X.}~\bibnamefont {Ma}}, \bibinfo {author}
  {\bibfnamefont {S.-F.}\ \bibnamefont {Yan}}, \ and\ \bibinfo {author}
  {\bibfnamefont {G.-W.}\ \bibnamefont {Yuan}},\ }\href@noop {} {\  (\bibinfo
  {year} {2023})},\ \Eprint {http://arxiv.org/abs/2306.17822} {arXiv:2306.17822
  [gr-qc]} \BibitemShut {NoStop}%
\bibitem [{\citenamefont {Wang}\ \emph {et~al.}(2023)\citenamefont {Wang},
  \citenamefont {Zhao}, \citenamefont {Li},\ and\ \citenamefont
  {Zhu}}]{Wang:2023ost}%
  \BibitemOpen
  \bibfield  {author} {\bibinfo {author} {\bibfnamefont {S.}~\bibnamefont
  {Wang}}, \bibinfo {author} {\bibfnamefont {Z.-C.}\ \bibnamefont {Zhao}},
  \bibinfo {author} {\bibfnamefont {J.-P.}\ \bibnamefont {Li}}, \ and\ \bibinfo
  {author} {\bibfnamefont {Q.-H.}\ \bibnamefont {Zhu}},\ }\href@noop {} {\
  (\bibinfo {year} {2023})},\ \Eprint {http://arxiv.org/abs/2307.00572}
  {arXiv:2307.00572 [astro-ph.CO]} \BibitemShut {NoStop}%
\bibitem [{\citenamefont {Liu}\ \emph {et~al.}(2023{\natexlab{a}})\citenamefont
  {Liu}, \citenamefont {Chen},\ and\ \citenamefont {Huang}}]{Liu:2023ymk}%
  \BibitemOpen
  \bibfield  {author} {\bibinfo {author} {\bibfnamefont {L.}~\bibnamefont
  {Liu}}, \bibinfo {author} {\bibfnamefont {Z.-C.}\ \bibnamefont {Chen}}, \
  and\ \bibinfo {author} {\bibfnamefont {Q.-G.}\ \bibnamefont {Huang}},\
  }\href@noop {} {\  (\bibinfo {year} {2023}{\natexlab{a}})},\ \Eprint
  {http://arxiv.org/abs/2307.01102} {arXiv:2307.01102 [astro-ph.CO]}
  \BibitemShut {NoStop}%
\bibitem [{\citenamefont {Abe}\ and\ \citenamefont {Tada}(2023)}]{Abe:2023yrw}%
  \BibitemOpen
  \bibfield  {author} {\bibinfo {author} {\bibfnamefont {K.~T.}\ \bibnamefont
  {Abe}}\ and\ \bibinfo {author} {\bibfnamefont {Y.}~\bibnamefont {Tada}},\
  }\href@noop {} {\  (\bibinfo {year} {2023})},\ \Eprint
  {http://arxiv.org/abs/2307.01653} {arXiv:2307.01653 [astro-ph.CO]}
  \BibitemShut {NoStop}%
\bibitem [{\citenamefont {Zhu}\ \emph {et~al.}(2023)\citenamefont {Zhu},
  \citenamefont {Zhao},\ and\ \citenamefont {Wang}}]{Zhu:2023faa}%
  \BibitemOpen
  \bibfield  {author} {\bibinfo {author} {\bibfnamefont {Q.-H.}\ \bibnamefont
  {Zhu}}, \bibinfo {author} {\bibfnamefont {Z.-C.}\ \bibnamefont {Zhao}}, \
  and\ \bibinfo {author} {\bibfnamefont {S.}~\bibnamefont {Wang}},\ }\href@noop
  {} {\  (\bibinfo {year} {2023})},\ \Eprint {http://arxiv.org/abs/2307.03095}
  {arXiv:2307.03095 [astro-ph.CO]} \BibitemShut {NoStop}%
\bibitem [{\citenamefont {Firouzjahi}\ and\ \citenamefont
  {Talebian}(2023)}]{Firouzjahi:2023lzg}%
  \BibitemOpen
  \bibfield  {author} {\bibinfo {author} {\bibfnamefont {H.}~\bibnamefont
  {Firouzjahi}}\ and\ \bibinfo {author} {\bibfnamefont {A.}~\bibnamefont
  {Talebian}},\ }\href@noop {} {\  (\bibinfo {year} {2023})},\ \Eprint
  {http://arxiv.org/abs/2307.03164} {arXiv:2307.03164 [gr-qc]} \BibitemShut
  {NoStop}%
\bibitem [{\citenamefont {Bari}\ \emph {et~al.}(2023)\citenamefont {Bari},
  \citenamefont {Bartolo}, \citenamefont {Dom\`enech},\ and\ \citenamefont
  {Matarrese}}]{Bari:2023rcw}%
  \BibitemOpen
  \bibfield  {author} {\bibinfo {author} {\bibfnamefont {P.}~\bibnamefont
  {Bari}}, \bibinfo {author} {\bibfnamefont {N.}~\bibnamefont {Bartolo}},
  \bibinfo {author} {\bibfnamefont {G.}~\bibnamefont {Dom\`enech}}, \ and\
  \bibinfo {author} {\bibfnamefont {S.}~\bibnamefont {Matarrese}},\ }\href@noop
  {} {\  (\bibinfo {year} {2023})},\ \Eprint {http://arxiv.org/abs/2307.05404}
  {arXiv:2307.05404 [astro-ph.CO]} \BibitemShut {NoStop}%
\bibitem [{\citenamefont {Hosseini~Mansoori}\ \emph {et~al.}(2023)\citenamefont
  {Hosseini~Mansoori}, \citenamefont {Felegray}, \citenamefont {Talebian},\
  and\ \citenamefont {Sami}}]{HosseiniMansoori:2023mqh}%
  \BibitemOpen
  \bibfield  {author} {\bibinfo {author} {\bibfnamefont {S.~A.}\ \bibnamefont
  {Hosseini~Mansoori}}, \bibinfo {author} {\bibfnamefont {F.}~\bibnamefont
  {Felegray}}, \bibinfo {author} {\bibfnamefont {A.}~\bibnamefont {Talebian}},
  \ and\ \bibinfo {author} {\bibfnamefont {M.}~\bibnamefont {Sami}},\
  }\href@noop {} {\  (\bibinfo {year} {2023})},\ \Eprint
  {http://arxiv.org/abs/2307.06757} {arXiv:2307.06757 [astro-ph.CO]}
  \BibitemShut {NoStop}%
\bibitem [{\citenamefont {Balaji}\ \emph {et~al.}(2023)\citenamefont {Balaji},
  \citenamefont {Dom\`enech},\ and\ \citenamefont
  {Franciolini}}]{Balaji:2023ehk}%
  \BibitemOpen
  \bibfield  {author} {\bibinfo {author} {\bibfnamefont {S.}~\bibnamefont
  {Balaji}}, \bibinfo {author} {\bibfnamefont {G.}~\bibnamefont {Dom\`enech}},
  \ and\ \bibinfo {author} {\bibfnamefont {G.}~\bibnamefont {Franciolini}},\
  }\href@noop {} {\  (\bibinfo {year} {2023})},\ \Eprint
  {http://arxiv.org/abs/2307.08552} {arXiv:2307.08552 [gr-qc]} \BibitemShut
  {NoStop}%
\bibitem [{\citenamefont {Zhao}\ \emph {et~al.}(2023)\citenamefont {Zhao},
  \citenamefont {Zhu}, \citenamefont {Wang},\ and\ \citenamefont
  {Zhang}}]{Zhao:2023joc}%
  \BibitemOpen
  \bibfield  {author} {\bibinfo {author} {\bibfnamefont {Z.-C.}\ \bibnamefont
  {Zhao}}, \bibinfo {author} {\bibfnamefont {Q.-H.}\ \bibnamefont {Zhu}},
  \bibinfo {author} {\bibfnamefont {S.}~\bibnamefont {Wang}}, \ and\ \bibinfo
  {author} {\bibfnamefont {X.}~\bibnamefont {Zhang}},\ }\href@noop {} {\
  (\bibinfo {year} {2023})},\ \Eprint {http://arxiv.org/abs/2307.13574}
  {arXiv:2307.13574 [astro-ph.CO]} \BibitemShut {NoStop}%
\bibitem [{\citenamefont {Liu}\ \emph {et~al.}(2023{\natexlab{b}})\citenamefont
  {Liu}, \citenamefont {Chen},\ and\ \citenamefont {Huang}}]{Liu:2023pau}%
  \BibitemOpen
  \bibfield  {author} {\bibinfo {author} {\bibfnamefont {L.}~\bibnamefont
  {Liu}}, \bibinfo {author} {\bibfnamefont {Z.-C.}\ \bibnamefont {Chen}}, \
  and\ \bibinfo {author} {\bibfnamefont {Q.-G.}\ \bibnamefont {Huang}},\
  }\href@noop {} {\  (\bibinfo {year} {2023}{\natexlab{b}})},\ \Eprint
  {http://arxiv.org/abs/2307.14911} {arXiv:2307.14911 [astro-ph.CO]}
  \BibitemShut {NoStop}%
\bibitem [{\citenamefont {Yi}\ \emph {et~al.}(2023{\natexlab{a}})\citenamefont
  {Yi}, \citenamefont {You},\ and\ \citenamefont {Wu}}]{Yi:2023tdk}%
  \BibitemOpen
  \bibfield  {author} {\bibinfo {author} {\bibfnamefont {Z.}~\bibnamefont
  {Yi}}, \bibinfo {author} {\bibfnamefont {Z.-Q.}\ \bibnamefont {You}}, \ and\
  \bibinfo {author} {\bibfnamefont {Y.}~\bibnamefont {Wu}},\ }\href@noop {} {\
  (\bibinfo {year} {2023}{\natexlab{a}})},\ \Eprint
  {http://arxiv.org/abs/2308.05632} {arXiv:2308.05632 [astro-ph.CO]}
  \BibitemShut {NoStop}%
\bibitem [{\citenamefont {Frosina}\ and\ \citenamefont
  {Urbano}(2023)}]{Frosina:2023nxu}%
  \BibitemOpen
  \bibfield  {author} {\bibinfo {author} {\bibfnamefont {L.}~\bibnamefont
  {Frosina}}\ and\ \bibinfo {author} {\bibfnamefont {A.}~\bibnamefont
  {Urbano}},\ }\href@noop {} {\  (\bibinfo {year} {2023})},\ \Eprint
  {http://arxiv.org/abs/2308.06915} {arXiv:2308.06915 [astro-ph.CO]}
  \BibitemShut {NoStop}%
\bibitem [{\citenamefont {Choudhury}\ \emph {et~al.}(2023)\citenamefont
  {Choudhury}, \citenamefont {Karde}, \citenamefont {Panda},\ and\
  \citenamefont {Sami}}]{Choudhury:2023wrm}%
  \BibitemOpen
  \bibfield  {author} {\bibinfo {author} {\bibfnamefont {S.}~\bibnamefont
  {Choudhury}}, \bibinfo {author} {\bibfnamefont {A.}~\bibnamefont {Karde}},
  \bibinfo {author} {\bibfnamefont {S.}~\bibnamefont {Panda}}, \ and\ \bibinfo
  {author} {\bibfnamefont {M.}~\bibnamefont {Sami}},\ }\href@noop {} {\
  (\bibinfo {year} {2023})},\ \Eprint {http://arxiv.org/abs/2308.09273}
  {arXiv:2308.09273 [astro-ph.CO]} \BibitemShut {NoStop}%
\bibitem [{\citenamefont {Kawasaki}\ and\ \citenamefont
  {Murai}(2023)}]{Kawasaki:2023rfx}%
  \BibitemOpen
  \bibfield  {author} {\bibinfo {author} {\bibfnamefont {M.}~\bibnamefont
  {Kawasaki}}\ and\ \bibinfo {author} {\bibfnamefont {K.}~\bibnamefont
  {Murai}},\ }\href@noop {} {\  (\bibinfo {year} {2023})},\ \Eprint
  {http://arxiv.org/abs/2308.13134} {arXiv:2308.13134 [astro-ph.CO]}
  \BibitemShut {NoStop}%
\bibitem [{\citenamefont {Yi}\ \emph {et~al.}(2023{\natexlab{b}})\citenamefont
  {Yi}, \citenamefont {You}, \citenamefont {Wu}, \citenamefont {Chen},\ and\
  \citenamefont {Liu}}]{Yi:2023npi}%
  \BibitemOpen
  \bibfield  {author} {\bibinfo {author} {\bibfnamefont {Z.}~\bibnamefont
  {Yi}}, \bibinfo {author} {\bibfnamefont {Z.-Q.}\ \bibnamefont {You}},
  \bibinfo {author} {\bibfnamefont {Y.}~\bibnamefont {Wu}}, \bibinfo {author}
  {\bibfnamefont {Z.-C.}\ \bibnamefont {Chen}}, \ and\ \bibinfo {author}
  {\bibfnamefont {L.}~\bibnamefont {Liu}},\ }\href@noop {} {\  (\bibinfo {year}
  {2023}{\natexlab{b}})},\ \Eprint {http://arxiv.org/abs/2308.14688}
  {arXiv:2308.14688 [astro-ph.CO]} \BibitemShut {NoStop}%
\bibitem [{\citenamefont {Saito}\ and\ \citenamefont
  {Yokoyama}(2009)}]{Saito:2008jc}%
  \BibitemOpen
  \bibfield  {author} {\bibinfo {author} {\bibfnamefont {R.}~\bibnamefont
  {Saito}}\ and\ \bibinfo {author} {\bibfnamefont {J.}~\bibnamefont
  {Yokoyama}},\ }\href {\doibase 10.1103/PhysRevLett.102.161101,
  10.1103/PhysRevLett.107.069901} {\bibfield  {journal} {\bibinfo  {journal}
  {Phys. Rev. Lett.}\ }\textbf {\bibinfo {volume} {102}},\ \bibinfo {pages}
  {161101} (\bibinfo {year} {2009})},\ \bibinfo {note} {[Erratum: Phys. Rev.
  Lett.107,069901(2011)]},\ \Eprint {http://arxiv.org/abs/0812.4339}
  {arXiv:0812.4339 [astro-ph]} \BibitemShut {NoStop}%
\bibitem [{\citenamefont {Saito}\ and\ \citenamefont
  {Yokoyama}(2010)}]{Saito:2009jt}%
  \BibitemOpen
  \bibfield  {author} {\bibinfo {author} {\bibfnamefont {R.}~\bibnamefont
  {Saito}}\ and\ \bibinfo {author} {\bibfnamefont {J.}~\bibnamefont
  {Yokoyama}},\ }\href {\doibase 10.1143/PTP.126.351, 10.1143/PTP.123.867}
  {\bibfield  {journal} {\bibinfo  {journal} {Prog. Theor. Phys.}\ }\textbf
  {\bibinfo {volume} {123}},\ \bibinfo {pages} {867} (\bibinfo {year}
  {2010})},\ \bibinfo {note} {[Erratum: Prog. Theor. Phys.126,351(2011)]},\
  \Eprint {http://arxiv.org/abs/0912.5317} {arXiv:0912.5317 [astro-ph.CO]}
  \BibitemShut {NoStop}%
\bibitem [{\citenamefont {Bugaev}\ and\ \citenamefont
  {Klimai}(2010)}]{Bugaev:2009zh}%
  \BibitemOpen
  \bibfield  {author} {\bibinfo {author} {\bibfnamefont {E.}~\bibnamefont
  {Bugaev}}\ and\ \bibinfo {author} {\bibfnamefont {P.}~\bibnamefont
  {Klimai}},\ }\href {\doibase 10.1103/PhysRevD.81.023517} {\bibfield
  {journal} {\bibinfo  {journal} {Phys. Rev.}\ }\textbf {\bibinfo {volume}
  {D81}},\ \bibinfo {pages} {023517} (\bibinfo {year} {2010})},\ \Eprint
  {http://arxiv.org/abs/0908.0664} {arXiv:0908.0664 [astro-ph.CO]} \BibitemShut
  {NoStop}%
\bibitem [{\citenamefont {Garcia-Bellido}\ \emph {et~al.}(2017)\citenamefont
  {Garcia-Bellido}, \citenamefont {Peloso},\ and\ \citenamefont
  {Unal}}]{Garcia-Bellido:2017aan}%
  \BibitemOpen
  \bibfield  {author} {\bibinfo {author} {\bibfnamefont {J.}~\bibnamefont
  {Garcia-Bellido}}, \bibinfo {author} {\bibfnamefont {M.}~\bibnamefont
  {Peloso}}, \ and\ \bibinfo {author} {\bibfnamefont {C.}~\bibnamefont
  {Unal}},\ }\href {\doibase 10.1088/1475-7516/2017/09/013} {\bibfield
  {journal} {\bibinfo  {journal} {JCAP}\ }\textbf {\bibinfo {volume} {1709}},\
  \bibinfo {pages} {013} (\bibinfo {year} {2017})},\ \Eprint
  {http://arxiv.org/abs/1707.02441} {arXiv:1707.02441 [astro-ph.CO]}
  \BibitemShut {NoStop}%
\bibitem [{\citenamefont {Inomata}\ \emph {et~al.}(2018)\citenamefont
  {Inomata}, \citenamefont {Kawasaki}, \citenamefont {Mukaida},\ and\
  \citenamefont {Yanagida}}]{Inomata:2017vxo}%
  \BibitemOpen
  \bibfield  {author} {\bibinfo {author} {\bibfnamefont {K.}~\bibnamefont
  {Inomata}}, \bibinfo {author} {\bibfnamefont {M.}~\bibnamefont {Kawasaki}},
  \bibinfo {author} {\bibfnamefont {K.}~\bibnamefont {Mukaida}}, \ and\
  \bibinfo {author} {\bibfnamefont {T.~T.}\ \bibnamefont {Yanagida}},\ }\href
  {\doibase 10.1103/PhysRevD.97.043514} {\bibfield  {journal} {\bibinfo
  {journal} {Phys. Rev.}\ }\textbf {\bibinfo {volume} {D97}},\ \bibinfo {pages}
  {043514} (\bibinfo {year} {2018})},\ \Eprint
  {http://arxiv.org/abs/1711.06129} {arXiv:1711.06129 [astro-ph.CO]}
  \BibitemShut {NoStop}%
\bibitem [{\citenamefont {Bartolo}\ \emph {et~al.}(2019)\citenamefont
  {Bartolo}, \citenamefont {De~Luca}, \citenamefont {Franciolini},
  \citenamefont {Peloso}, \citenamefont {Racco},\ and\ \citenamefont
  {Riotto}}]{Bartolo:2018rku}%
  \BibitemOpen
  \bibfield  {author} {\bibinfo {author} {\bibfnamefont {N.}~\bibnamefont
  {Bartolo}}, \bibinfo {author} {\bibfnamefont {V.}~\bibnamefont {De~Luca}},
  \bibinfo {author} {\bibfnamefont {G.}~\bibnamefont {Franciolini}}, \bibinfo
  {author} {\bibfnamefont {M.}~\bibnamefont {Peloso}}, \bibinfo {author}
  {\bibfnamefont {D.}~\bibnamefont {Racco}}, \ and\ \bibinfo {author}
  {\bibfnamefont {A.}~\bibnamefont {Riotto}},\ }\href {\doibase
  10.1103/PhysRevD.99.103521} {\bibfield  {journal} {\bibinfo  {journal} {Phys.
  Rev. D}\ }\textbf {\bibinfo {volume} {99}},\ \bibinfo {pages} {103521}
  (\bibinfo {year} {2019})},\ \Eprint {http://arxiv.org/abs/1810.12224}
  {arXiv:1810.12224 [astro-ph.CO]} \BibitemShut {NoStop}%
\bibitem [{\citenamefont {Kawasaki}\ and\ \citenamefont
  {Tada}(2016)}]{Kawasaki:2015ppx}%
  \BibitemOpen
  \bibfield  {author} {\bibinfo {author} {\bibfnamefont {M.}~\bibnamefont
  {Kawasaki}}\ and\ \bibinfo {author} {\bibfnamefont {Y.}~\bibnamefont
  {Tada}},\ }\href {\doibase 10.1088/1475-7516/2016/08/041} {\bibfield
  {journal} {\bibinfo  {journal} {JCAP}\ }\textbf {\bibinfo {volume} {08}},\
  \bibinfo {pages} {041} (\bibinfo {year} {2016})},\ \Eprint
  {http://arxiv.org/abs/1512.03515} {arXiv:1512.03515 [astro-ph.CO]}
  \BibitemShut {NoStop}%
\bibitem [{\citenamefont {Pattison}\ \emph {et~al.}(2017)\citenamefont
  {Pattison}, \citenamefont {Vennin}, \citenamefont {Assadullahi},\ and\
  \citenamefont {Wands}}]{Pattison:2017mbe}%
  \BibitemOpen
  \bibfield  {author} {\bibinfo {author} {\bibfnamefont {C.}~\bibnamefont
  {Pattison}}, \bibinfo {author} {\bibfnamefont {V.}~\bibnamefont {Vennin}},
  \bibinfo {author} {\bibfnamefont {H.}~\bibnamefont {Assadullahi}}, \ and\
  \bibinfo {author} {\bibfnamefont {D.}~\bibnamefont {Wands}},\ }\href
  {\doibase 10.1088/1475-7516/2017/10/046} {\bibfield  {journal} {\bibinfo
  {journal} {JCAP}\ }\textbf {\bibinfo {volume} {10}},\ \bibinfo {pages} {046}
  (\bibinfo {year} {2017})},\ \Eprint {http://arxiv.org/abs/1707.00537}
  {arXiv:1707.00537 [hep-th]} \BibitemShut {NoStop}%
\bibitem [{\citenamefont {Ezquiaga}\ \emph {et~al.}(2020)\citenamefont
  {Ezquiaga}, \citenamefont {Garc\'\i{}a-Bellido},\ and\ \citenamefont
  {Vennin}}]{Ezquiaga:2019ftu}%
  \BibitemOpen
  \bibfield  {author} {\bibinfo {author} {\bibfnamefont {J.~M.}\ \bibnamefont
  {Ezquiaga}}, \bibinfo {author} {\bibfnamefont {J.}~\bibnamefont
  {Garc\'\i{}a-Bellido}}, \ and\ \bibinfo {author} {\bibfnamefont
  {V.}~\bibnamefont {Vennin}},\ }\href {\doibase 10.1088/1475-7516/2020/03/029}
  {\bibfield  {journal} {\bibinfo  {journal} {JCAP}\ }\textbf {\bibinfo
  {volume} {03}},\ \bibinfo {pages} {029} (\bibinfo {year} {2020})},\ \Eprint
  {http://arxiv.org/abs/1912.05399} {arXiv:1912.05399 [astro-ph.CO]}
  \BibitemShut {NoStop}%
\bibitem [{\citenamefont {Tada}\ and\ \citenamefont
  {Yamada}(2023)}]{Tada:2023pue}%
  \BibitemOpen
  \bibfield  {author} {\bibinfo {author} {\bibfnamefont {Y.}~\bibnamefont
  {Tada}}\ and\ \bibinfo {author} {\bibfnamefont {M.}~\bibnamefont {Yamada}},\
  }\href {\doibase 10.1103/PhysRevD.107.123539} {\bibfield  {journal} {\bibinfo
   {journal} {Phys. Rev. D}\ }\textbf {\bibinfo {volume} {107}},\ \bibinfo
  {pages} {123539} (\bibinfo {year} {2023})},\ \Eprint
  {http://arxiv.org/abs/2304.01249} {arXiv:2304.01249 [astro-ph.CO]}
  \BibitemShut {NoStop}%
\bibitem [{\citenamefont {Ando}\ \emph
  {et~al.}(2018{\natexlab{b}})\citenamefont {Ando}, \citenamefont {Inomata},\
  and\ \citenamefont {Kawasaki}}]{Ando:2018qdb}%
  \BibitemOpen
  \bibfield  {author} {\bibinfo {author} {\bibfnamefont {K.}~\bibnamefont
  {Ando}}, \bibinfo {author} {\bibfnamefont {K.}~\bibnamefont {Inomata}}, \
  and\ \bibinfo {author} {\bibfnamefont {M.}~\bibnamefont {Kawasaki}},\ }\href
  {\doibase 10.1103/PhysRevD.97.103528} {\bibfield  {journal} {\bibinfo
  {journal} {Phys. Rev.}\ }\textbf {\bibinfo {volume} {D97}},\ \bibinfo {pages}
  {103528} (\bibinfo {year} {2018}{\natexlab{b}})},\ \Eprint
  {http://arxiv.org/abs/1802.06393} {arXiv:1802.06393 [astro-ph.CO]}
  \BibitemShut {NoStop}%
\bibitem [{\citenamefont {Young}(2019)}]{Young:2019osy}%
  \BibitemOpen
  \bibfield  {author} {\bibinfo {author} {\bibfnamefont {S.}~\bibnamefont
  {Young}},\ }\href {\doibase 10.1142/S0218271820300025} {\bibfield  {journal}
  {\bibinfo  {journal} {Int. J. Mod. Phys. D}\ }\textbf {\bibinfo {volume}
  {29}},\ \bibinfo {pages} {2030002} (\bibinfo {year} {2019})},\ \Eprint
  {http://arxiv.org/abs/1905.01230} {arXiv:1905.01230 [astro-ph.CO]}
  \BibitemShut {NoStop}%
\bibitem [{\citenamefont {Yoo}\ \emph {et~al.}(2021)\citenamefont {Yoo},
  \citenamefont {Harada}, \citenamefont {Hirano},\ and\ \citenamefont
  {Kohri}}]{Yoo:2020dkz}%
  \BibitemOpen
  \bibfield  {author} {\bibinfo {author} {\bibfnamefont {C.-M.}\ \bibnamefont
  {Yoo}}, \bibinfo {author} {\bibfnamefont {T.}~\bibnamefont {Harada}},
  \bibinfo {author} {\bibfnamefont {S.}~\bibnamefont {Hirano}}, \ and\ \bibinfo
  {author} {\bibfnamefont {K.}~\bibnamefont {Kohri}},\ }\href {\doibase
  10.1093/ptep/ptaa155} {\bibfield  {journal} {\bibinfo  {journal} {PTEP}\
  }\textbf {\bibinfo {volume} {2021}},\ \bibinfo {pages} {013E02} (\bibinfo
  {year} {2021})},\ \Eprint {http://arxiv.org/abs/2008.02425} {arXiv:2008.02425
  [astro-ph.CO]} \BibitemShut {NoStop}%
\bibitem [{\citenamefont {De~Luca}\ \emph {et~al.}(2023)\citenamefont
  {De~Luca}, \citenamefont {Kehagias},\ and\ \citenamefont
  {Riotto}}]{DeLuca:2023tun}%
  \BibitemOpen
  \bibfield  {author} {\bibinfo {author} {\bibfnamefont {V.}~\bibnamefont
  {De~Luca}}, \bibinfo {author} {\bibfnamefont {A.}~\bibnamefont {Kehagias}}, \
  and\ \bibinfo {author} {\bibfnamefont {A.}~\bibnamefont {Riotto}},\
  }\href@noop {} {\  (\bibinfo {year} {2023})},\ \Eprint
  {http://arxiv.org/abs/2307.13633} {arXiv:2307.13633 [astro-ph.CO]}
  \BibitemShut {NoStop}%
\bibitem [{\citenamefont {Gorji}\ \emph {et~al.}(2023)\citenamefont {Gorji},
  \citenamefont {Sasaki},\ and\ \citenamefont {Suyama}}]{Gorji:2023sil}%
  \BibitemOpen
  \bibfield  {author} {\bibinfo {author} {\bibfnamefont {M.~A.}\ \bibnamefont
  {Gorji}}, \bibinfo {author} {\bibfnamefont {M.}~\bibnamefont {Sasaki}}, \
  and\ \bibinfo {author} {\bibfnamefont {T.}~\bibnamefont {Suyama}},\
  }\href@noop {} {\  (\bibinfo {year} {2023})},\ \Eprint
  {http://arxiv.org/abs/2307.13109} {arXiv:2307.13109 [astro-ph.CO]}
  \BibitemShut {NoStop}%
\bibitem [{\citenamefont {Inomata}\ \emph {et~al.}(2023)\citenamefont
  {Inomata}, \citenamefont {Kohri},\ and\ \citenamefont
  {Terada}}]{Inomata:2023zup}%
  \BibitemOpen
  \bibfield  {author} {\bibinfo {author} {\bibfnamefont {K.}~\bibnamefont
  {Inomata}}, \bibinfo {author} {\bibfnamefont {K.}~\bibnamefont {Kohri}}, \
  and\ \bibinfo {author} {\bibfnamefont {T.}~\bibnamefont {Terada}},\
  }\href@noop {} {\  (\bibinfo {year} {2023})},\ \Eprint
  {http://arxiv.org/abs/2306.17834} {arXiv:2306.17834 [astro-ph.CO]}
  \BibitemShut {NoStop}%
\bibitem [{\citenamefont {Babichev}\ \emph {et~al.}(2023)\citenamefont
  {Babichev}, \citenamefont {Gorbunov}, \citenamefont {Ramazanov},
  \citenamefont {Samanta},\ and\ \citenamefont {Vikman}}]{Babichev:2023pbf}%
  \BibitemOpen
  \bibfield  {author} {\bibinfo {author} {\bibfnamefont {E.}~\bibnamefont
  {Babichev}}, \bibinfo {author} {\bibfnamefont {D.}~\bibnamefont {Gorbunov}},
  \bibinfo {author} {\bibfnamefont {S.}~\bibnamefont {Ramazanov}}, \bibinfo
  {author} {\bibfnamefont {R.}~\bibnamefont {Samanta}}, \ and\ \bibinfo
  {author} {\bibfnamefont {A.}~\bibnamefont {Vikman}},\ }\href@noop {} {\
  (\bibinfo {year} {2023})},\ \Eprint {http://arxiv.org/abs/2307.04582}
  {arXiv:2307.04582 [hep-ph]} \BibitemShut {NoStop}%
\bibitem [{\citenamefont {Cai}\ \emph {et~al.}(2020)\citenamefont {Cai},
  \citenamefont {Pi},\ and\ \citenamefont {Sasaki}}]{Cai:2019cdl}%
  \BibitemOpen
  \bibfield  {author} {\bibinfo {author} {\bibfnamefont {R.-G.}\ \bibnamefont
  {Cai}}, \bibinfo {author} {\bibfnamefont {S.}~\bibnamefont {Pi}}, \ and\
  \bibinfo {author} {\bibfnamefont {M.}~\bibnamefont {Sasaki}},\ }\href
  {\doibase 10.1103/PhysRevD.102.083528} {\bibfield  {journal} {\bibinfo
  {journal} {Phys. Rev. D}\ }\textbf {\bibinfo {volume} {102}},\ \bibinfo
  {pages} {083528} (\bibinfo {year} {2020})},\ \Eprint
  {http://arxiv.org/abs/1909.13728} {arXiv:1909.13728 [astro-ph.CO]}
  \BibitemShut {NoStop}%
\bibitem [{\citenamefont {Yuan}\ \emph {et~al.}(2020)\citenamefont {Yuan},
  \citenamefont {Chen},\ and\ \citenamefont {Huang}}]{Yuan:2019wwo}%
  \BibitemOpen
  \bibfield  {author} {\bibinfo {author} {\bibfnamefont {C.}~\bibnamefont
  {Yuan}}, \bibinfo {author} {\bibfnamefont {Z.-C.}\ \bibnamefont {Chen}}, \
  and\ \bibinfo {author} {\bibfnamefont {Q.-G.}\ \bibnamefont {Huang}},\ }\href
  {\doibase 10.1103/PhysRevD.101.043019} {\bibfield  {journal} {\bibinfo
  {journal} {Phys. Rev. D}\ }\textbf {\bibinfo {volume} {101}},\ \bibinfo
  {pages} {043019} (\bibinfo {year} {2020})},\ \Eprint
  {http://arxiv.org/abs/1910.09099} {arXiv:1910.09099 [astro-ph.CO]}
  \BibitemShut {NoStop}%
\bibitem [{\citenamefont {Dom\`enech}\ \emph {et~al.}(2020)\citenamefont
  {Dom\`enech}, \citenamefont {Pi},\ and\ \citenamefont
  {Sasaki}}]{Domenech:2020kqm}%
  \BibitemOpen
  \bibfield  {author} {\bibinfo {author} {\bibfnamefont {G.}~\bibnamefont
  {Dom\`enech}}, \bibinfo {author} {\bibfnamefont {S.}~\bibnamefont {Pi}}, \
  and\ \bibinfo {author} {\bibfnamefont {M.}~\bibnamefont {Sasaki}},\ }\href
  {\doibase 10.1088/1475-7516/2020/08/017} {\bibfield  {journal} {\bibinfo
  {journal} {JCAP}\ }\textbf {\bibinfo {volume} {08}},\ \bibinfo {pages} {017}
  (\bibinfo {year} {2020})},\ \Eprint {http://arxiv.org/abs/2005.12314}
  {arXiv:2005.12314 [gr-qc]} \BibitemShut {NoStop}%
\bibitem [{\citenamefont {Spokoiny}(1993)}]{Spokoiny:1993kt}%
  \BibitemOpen
  \bibfield  {author} {\bibinfo {author} {\bibfnamefont {B.}~\bibnamefont
  {Spokoiny}},\ }\href {\doibase 10.1016/0370-2693(93)90155-B} {\bibfield
  {journal} {\bibinfo  {journal} {Phys. Lett. B}\ }\textbf {\bibinfo {volume}
  {315}},\ \bibinfo {pages} {40} (\bibinfo {year} {1993})},\ \Eprint
  {http://arxiv.org/abs/gr-qc/9306008} {arXiv:gr-qc/9306008} \BibitemShut
  {NoStop}%
\bibitem [{\citenamefont {Joyce}(1997)}]{Joyce:1996cp}%
  \BibitemOpen
  \bibfield  {author} {\bibinfo {author} {\bibfnamefont {M.}~\bibnamefont
  {Joyce}},\ }\href {\doibase 10.1103/PhysRevD.55.1875} {\bibfield  {journal}
  {\bibinfo  {journal} {Phys. Rev. D}\ }\textbf {\bibinfo {volume} {55}},\
  \bibinfo {pages} {1875} (\bibinfo {year} {1997})},\ \Eprint
  {http://arxiv.org/abs/hep-ph/9606223} {arXiv:hep-ph/9606223} \BibitemShut
  {NoStop}%
\bibitem [{\citenamefont {Ferreira}\ and\ \citenamefont
  {Joyce}(1998)}]{Ferreira:1997hj}%
  \BibitemOpen
  \bibfield  {author} {\bibinfo {author} {\bibfnamefont {P.~G.}\ \bibnamefont
  {Ferreira}}\ and\ \bibinfo {author} {\bibfnamefont {M.}~\bibnamefont
  {Joyce}},\ }\href {\doibase 10.1103/PhysRevD.58.023503} {\bibfield  {journal}
  {\bibinfo  {journal} {Phys. Rev. D}\ }\textbf {\bibinfo {volume} {58}},\
  \bibinfo {pages} {023503} (\bibinfo {year} {1998})},\ \Eprint
  {http://arxiv.org/abs/astro-ph/9711102} {arXiv:astro-ph/9711102} \BibitemShut
  {NoStop}%
\bibitem [{\citenamefont {Co}\ \emph {et~al.}(2022)\citenamefont {Co},
  \citenamefont {Dunsky}, \citenamefont {Fernandez}, \citenamefont {Ghalsasi},
  \citenamefont {Hall}, \citenamefont {Harigaya},\ and\ \citenamefont
  {Shelton}}]{Co:2021lkc}%
  \BibitemOpen
  \bibfield  {author} {\bibinfo {author} {\bibfnamefont {R.~T.}\ \bibnamefont
  {Co}}, \bibinfo {author} {\bibfnamefont {D.}~\bibnamefont {Dunsky}}, \bibinfo
  {author} {\bibfnamefont {N.}~\bibnamefont {Fernandez}}, \bibinfo {author}
  {\bibfnamefont {A.}~\bibnamefont {Ghalsasi}}, \bibinfo {author}
  {\bibfnamefont {L.~J.}\ \bibnamefont {Hall}}, \bibinfo {author}
  {\bibfnamefont {K.}~\bibnamefont {Harigaya}}, \ and\ \bibinfo {author}
  {\bibfnamefont {J.}~\bibnamefont {Shelton}},\ }\href {\doibase
  10.1007/JHEP09(2022)116} {\bibfield  {journal} {\bibinfo  {journal} {JHEP}\
  }\textbf {\bibinfo {volume} {09}},\ \bibinfo {pages} {116} (\bibinfo {year}
  {2022})},\ \Eprint {http://arxiv.org/abs/2108.09299} {arXiv:2108.09299
  [hep-ph]} \BibitemShut {NoStop}%
\bibitem [{\citenamefont {Gouttenoire}\ \emph {et~al.}(2021)\citenamefont
  {Gouttenoire}, \citenamefont {Servant},\ and\ \citenamefont
  {Simakachorn}}]{Gouttenoire:2021jhk}%
  \BibitemOpen
  \bibfield  {author} {\bibinfo {author} {\bibfnamefont {Y.}~\bibnamefont
  {Gouttenoire}}, \bibinfo {author} {\bibfnamefont {G.}~\bibnamefont
  {Servant}}, \ and\ \bibinfo {author} {\bibfnamefont {P.}~\bibnamefont
  {Simakachorn}},\ }\href@noop {} {\  (\bibinfo {year} {2021})},\ \Eprint
  {http://arxiv.org/abs/2111.01150} {arXiv:2111.01150 [hep-ph]} \BibitemShut
  {NoStop}%
\bibitem [{\citenamefont {Haque}\ \emph {et~al.}(2021)\citenamefont {Haque},
  \citenamefont {Maity}, \citenamefont {Paul},\ and\ \citenamefont
  {Sriramkumar}}]{Haque:2021dha}%
  \BibitemOpen
  \bibfield  {author} {\bibinfo {author} {\bibfnamefont {M.~R.}\ \bibnamefont
  {Haque}}, \bibinfo {author} {\bibfnamefont {D.}~\bibnamefont {Maity}},
  \bibinfo {author} {\bibfnamefont {T.}~\bibnamefont {Paul}}, \ and\ \bibinfo
  {author} {\bibfnamefont {L.}~\bibnamefont {Sriramkumar}},\ }\href {\doibase
  10.1103/PhysRevD.104.063513} {\bibfield  {journal} {\bibinfo  {journal}
  {Phys. Rev. D}\ }\textbf {\bibinfo {volume} {104}},\ \bibinfo {pages}
  {063513} (\bibinfo {year} {2021})},\ \Eprint
  {http://arxiv.org/abs/2105.09242} {arXiv:2105.09242 [astro-ph.CO]}
  \BibitemShut {NoStop}%
\bibitem [{\citenamefont {Chowdhury}\ \emph {et~al.}(2023)\citenamefont
  {Chowdhury}, \citenamefont {Tasinato},\ and\ \citenamefont
  {Zavala}}]{Chowdhury:2023opo}%
  \BibitemOpen
  \bibfield  {author} {\bibinfo {author} {\bibfnamefont {D.}~\bibnamefont
  {Chowdhury}}, \bibinfo {author} {\bibfnamefont {G.}~\bibnamefont {Tasinato}},
  \ and\ \bibinfo {author} {\bibfnamefont {I.}~\bibnamefont {Zavala}},\
  }\href@noop {} {\  (\bibinfo {year} {2023})},\ \Eprint
  {http://arxiv.org/abs/2307.01188} {arXiv:2307.01188 [hep-th]} \BibitemShut
  {NoStop}%
\bibitem [{\citenamefont {Ben-Dayan}\ \emph {et~al.}(2023)\citenamefont
  {Ben-Dayan}, \citenamefont {Kumar}, \citenamefont {Thattarampilly},\ and\
  \citenamefont {Verma}}]{Ben-Dayan:2023lwd}%
  \BibitemOpen
  \bibfield  {author} {\bibinfo {author} {\bibfnamefont {I.}~\bibnamefont
  {Ben-Dayan}}, \bibinfo {author} {\bibfnamefont {U.}~\bibnamefont {Kumar}},
  \bibinfo {author} {\bibfnamefont {U.}~\bibnamefont {Thattarampilly}}, \ and\
  \bibinfo {author} {\bibfnamefont {A.}~\bibnamefont {Verma}},\ }\href@noop {}
  {\  (\bibinfo {year} {2023})},\ \Eprint {http://arxiv.org/abs/2307.15123}
  {arXiv:2307.15123 [astro-ph.CO]} \BibitemShut {NoStop}%
\bibitem [{\citenamefont {Co}\ and\ \citenamefont
  {Harigaya}(2020)}]{Co:2019wyp}%
  \BibitemOpen
  \bibfield  {author} {\bibinfo {author} {\bibfnamefont {R.~T.}\ \bibnamefont
  {Co}}\ and\ \bibinfo {author} {\bibfnamefont {K.}~\bibnamefont {Harigaya}},\
  }\href {\doibase 10.1103/PhysRevLett.124.111602} {\bibfield  {journal}
  {\bibinfo  {journal} {Phys. Rev. Lett.}\ }\textbf {\bibinfo {volume} {124}},\
  \bibinfo {pages} {111602} (\bibinfo {year} {2020})},\ \Eprint
  {http://arxiv.org/abs/1910.02080} {arXiv:1910.02080 [hep-ph]} \BibitemShut
  {NoStop}%
\bibitem [{\citenamefont {Carr}(1975)}]{Carr:1975qj}%
  \BibitemOpen
  \bibfield  {author} {\bibinfo {author} {\bibfnamefont {B.~J.}\ \bibnamefont
  {Carr}},\ }\href {\doibase 10.1086/153853} {\bibfield  {journal} {\bibinfo
  {journal} {Astrophys. J.}\ }\textbf {\bibinfo {volume} {201}},\ \bibinfo
  {pages} {1} (\bibinfo {year} {1975})}\BibitemShut {NoStop}%
\bibitem [{\citenamefont {Musco}\ and\ \citenamefont
  {Miller}(2013)}]{Musco:2012au}%
  \BibitemOpen
  \bibfield  {author} {\bibinfo {author} {\bibfnamefont {I.}~\bibnamefont
  {Musco}}\ and\ \bibinfo {author} {\bibfnamefont {J.~C.}\ \bibnamefont
  {Miller}},\ }\href {\doibase 10.1088/0264-9381/30/14/145009} {\bibfield
  {journal} {\bibinfo  {journal} {Class. Quant. Grav.}\ }\textbf {\bibinfo
  {volume} {30}},\ \bibinfo {pages} {145009} (\bibinfo {year} {2013})},\
  \Eprint {http://arxiv.org/abs/1201.2379} {arXiv:1201.2379 [gr-qc]}
  \BibitemShut {NoStop}%
\bibitem [{\citenamefont {Harada}\ \emph {et~al.}(2013)\citenamefont {Harada},
  \citenamefont {Yoo},\ and\ \citenamefont {Kohri}}]{Harada:2013epa}%
  \BibitemOpen
  \bibfield  {author} {\bibinfo {author} {\bibfnamefont {T.}~\bibnamefont
  {Harada}}, \bibinfo {author} {\bibfnamefont {C.-M.}\ \bibnamefont {Yoo}}, \
  and\ \bibinfo {author} {\bibfnamefont {K.}~\bibnamefont {Kohri}},\ }\href
  {\doibase 10.1103/PhysRevD.88.084051, 10.1103/PhysRevD.89.029903} {\bibfield
  {journal} {\bibinfo  {journal} {Phys. Rev.}\ }\textbf {\bibinfo {volume}
  {D88}},\ \bibinfo {pages} {084051} (\bibinfo {year} {2013})},\ \bibinfo
  {note} {[Erratum: Phys. Rev.D89,no.2,029903(2014)]},\ \Eprint
  {http://arxiv.org/abs/1309.4201} {arXiv:1309.4201 [astro-ph.CO]} \BibitemShut
  {NoStop}%
\bibitem [{\citenamefont {Escriv\`a}\ \emph {et~al.}(2021)\citenamefont
  {Escriv\`a}, \citenamefont {Germani},\ and\ \citenamefont
  {Sheth}}]{Escriva:2020tak}%
  \BibitemOpen
  \bibfield  {author} {\bibinfo {author} {\bibfnamefont {A.}~\bibnamefont
  {Escriv\`a}}, \bibinfo {author} {\bibfnamefont {C.}~\bibnamefont {Germani}},
  \ and\ \bibinfo {author} {\bibfnamefont {R.~K.}\ \bibnamefont {Sheth}},\
  }\href {\doibase 10.1088/1475-7516/2021/01/030} {\bibfield  {journal}
  {\bibinfo  {journal} {JCAP}\ }\textbf {\bibinfo {volume} {01}},\ \bibinfo
  {pages} {030} (\bibinfo {year} {2021})},\ \Eprint
  {http://arxiv.org/abs/2007.05564} {arXiv:2007.05564 [gr-qc]} \BibitemShut
  {NoStop}%
\bibitem [{\citenamefont {Dom\`enech}(2020)}]{Domenech:2019quo}%
  \BibitemOpen
  \bibfield  {author} {\bibinfo {author} {\bibfnamefont {G.}~\bibnamefont
  {Dom\`enech}},\ }\href {\doibase 10.1142/S0218271820500285} {\bibfield
  {journal} {\bibinfo  {journal} {Int. J. Mod. Phys. D}\ }\textbf {\bibinfo
  {volume} {29}},\ \bibinfo {pages} {2050028} (\bibinfo {year} {2020})},\
  \Eprint {http://arxiv.org/abs/1912.05583} {arXiv:1912.05583 [gr-qc]}
  \BibitemShut {NoStop}%
\bibitem [{\citenamefont {Witkowski}\ \emph {et~al.}(2022)\citenamefont
  {Witkowski}, \citenamefont {Dom\`enech}, \citenamefont {Fumagalli},\ and\
  \citenamefont {Renaux-Petel}}]{Witkowski:2021raz}%
  \BibitemOpen
  \bibfield  {author} {\bibinfo {author} {\bibfnamefont {L.~T.}\ \bibnamefont
  {Witkowski}}, \bibinfo {author} {\bibfnamefont {G.}~\bibnamefont
  {Dom\`enech}}, \bibinfo {author} {\bibfnamefont {J.}~\bibnamefont
  {Fumagalli}}, \ and\ \bibinfo {author} {\bibfnamefont {S.}~\bibnamefont
  {Renaux-Petel}},\ }\href {\doibase 10.1088/1475-7516/2022/05/028} {\bibfield
  {journal} {\bibinfo  {journal} {JCAP}\ }\textbf {\bibinfo {volume} {05}},\
  \bibinfo {pages} {028} (\bibinfo {year} {2022})},\ \Eprint
  {http://arxiv.org/abs/2110.09480} {arXiv:2110.09480 [astro-ph.CO]}
  \BibitemShut {NoStop}%
\bibitem [{\citenamefont {Espinosa}\ \emph {et~al.}(2018)\citenamefont
  {Espinosa}, \citenamefont {Racco},\ and\ \citenamefont
  {Riotto}}]{Espinosa:2018eve}%
  \BibitemOpen
  \bibfield  {author} {\bibinfo {author} {\bibfnamefont {J.~R.}\ \bibnamefont
  {Espinosa}}, \bibinfo {author} {\bibfnamefont {D.}~\bibnamefont {Racco}}, \
  and\ \bibinfo {author} {\bibfnamefont {A.}~\bibnamefont {Riotto}},\ }\href
  {\doibase 10.1088/1475-7516/2018/09/012} {\bibfield  {journal} {\bibinfo
  {journal} {JCAP}\ }\textbf {\bibinfo {volume} {1809}},\ \bibinfo {pages}
  {012} (\bibinfo {year} {2018})},\ \Eprint {http://arxiv.org/abs/1804.07732}
  {arXiv:1804.07732 [hep-ph]} \BibitemShut {NoStop}%
\bibitem [{\citenamefont {Pi}\ and\ \citenamefont {Sasaki}(2020)}]{Pi:2020otn}%
  \BibitemOpen
  \bibfield  {author} {\bibinfo {author} {\bibfnamefont {S.}~\bibnamefont
  {Pi}}\ and\ \bibinfo {author} {\bibfnamefont {M.}~\bibnamefont {Sasaki}},\
  }\href {\doibase 10.1088/1475-7516/2020/09/037} {\bibfield  {journal}
  {\bibinfo  {journal} {JCAP}\ }\textbf {\bibinfo {volume} {09}},\ \bibinfo
  {pages} {037} (\bibinfo {year} {2020})},\ \Eprint
  {http://arxiv.org/abs/2005.12306} {arXiv:2005.12306 [gr-qc]} \BibitemShut
  {NoStop}%
\bibitem [{\citenamefont {Harigaya}\ \emph {et~al.}(2023)\citenamefont
  {Harigaya}, \citenamefont {Inomata},\ and\ \citenamefont
  {Terada}}]{Harigaya:2023ecg}%
  \BibitemOpen
  \bibfield  {author} {\bibinfo {author} {\bibfnamefont {K.}~\bibnamefont
  {Harigaya}}, \bibinfo {author} {\bibfnamefont {K.}~\bibnamefont {Inomata}}, \
  and\ \bibinfo {author} {\bibfnamefont {T.}~\bibnamefont {Terada}},\
  }\href@noop {} {\  (\bibinfo {year} {2023})},\ \Eprint
  {http://arxiv.org/abs/2305.14242} {arXiv:2305.14242 [hep-ph]} \BibitemShut
  {NoStop}%
\bibitem [{\citenamefont {Dom\`enech}(2021)}]{Domenech:2021ztg}%
  \BibitemOpen
  \bibfield  {author} {\bibinfo {author} {\bibfnamefont {G.}~\bibnamefont
  {Dom\`enech}},\ }\href {\doibase 10.3390/universe7110398} {\bibfield
  {journal} {\bibinfo  {journal} {Universe}\ }\textbf {\bibinfo {volume} {7}},\
  \bibinfo {pages} {398} (\bibinfo {year} {2021})},\ \Eprint
  {http://arxiv.org/abs/2109.01398} {arXiv:2109.01398 [gr-qc]} \BibitemShut
  {NoStop}%
\bibitem [{\citenamefont {Inomata}\ \emph {et~al.}(2020)\citenamefont
  {Inomata}, \citenamefont {Kawasaki}, \citenamefont {Mukaida}, \citenamefont
  {Terada},\ and\ \citenamefont {Yanagida}}]{Inomata:2020lmk}%
  \BibitemOpen
  \bibfield  {author} {\bibinfo {author} {\bibfnamefont {K.}~\bibnamefont
  {Inomata}}, \bibinfo {author} {\bibfnamefont {M.}~\bibnamefont {Kawasaki}},
  \bibinfo {author} {\bibfnamefont {K.}~\bibnamefont {Mukaida}}, \bibinfo
  {author} {\bibfnamefont {T.}~\bibnamefont {Terada}}, \ and\ \bibinfo {author}
  {\bibfnamefont {T.~T.}\ \bibnamefont {Yanagida}},\ }\href {\doibase
  10.1103/PhysRevD.101.123533} {\bibfield  {journal} {\bibinfo  {journal}
  {Phys. Rev. D}\ }\textbf {\bibinfo {volume} {101}},\ \bibinfo {pages}
  {123533} (\bibinfo {year} {2020})},\ \Eprint
  {http://arxiv.org/abs/2003.10455} {arXiv:2003.10455 [astro-ph.CO]}
  \BibitemShut {NoStop}%
\bibitem [{\citenamefont {Saikawa}\ and\ \citenamefont
  {Shirai}(2018)}]{Saikawa:2018rcs}%
  \BibitemOpen
  \bibfield  {author} {\bibinfo {author} {\bibfnamefont {K.}~\bibnamefont
  {Saikawa}}\ and\ \bibinfo {author} {\bibfnamefont {S.}~\bibnamefont
  {Shirai}},\ }\href {\doibase 10.1088/1475-7516/2018/05/035} {\bibfield
  {journal} {\bibinfo  {journal} {JCAP}\ }\textbf {\bibinfo {volume} {1805}},\
  \bibinfo {pages} {035} (\bibinfo {year} {2018})},\ \Eprint
  {http://arxiv.org/abs/1803.01038} {arXiv:1803.01038 [hep-ph]} \BibitemShut
  {NoStop}%
\bibitem [{\citenamefont {Aristizabal~Sierra}\ \emph
  {et~al.}(2023)\citenamefont {Aristizabal~Sierra}, \citenamefont {Gariazzo},\
  and\ \citenamefont {Villanueva}}]{AristizabalSierra:2023bah}%
  \BibitemOpen
  \bibfield  {author} {\bibinfo {author} {\bibfnamefont {D.}~\bibnamefont
  {Aristizabal~Sierra}}, \bibinfo {author} {\bibfnamefont {S.}~\bibnamefont
  {Gariazzo}}, \ and\ \bibinfo {author} {\bibfnamefont {A.}~\bibnamefont
  {Villanueva}},\ }\href@noop {} {\  (\bibinfo {year} {2023})},\ \Eprint
  {http://arxiv.org/abs/2308.15531} {arXiv:2308.15531 [astro-ph.CO]}
  \BibitemShut {NoStop}%
\bibitem [{\citenamefont {Kolb}\ and\ \citenamefont
  {Turner}(1990)}]{Kolb:206230}%
  \BibitemOpen
  \bibfield  {author} {\bibinfo {author} {\bibfnamefont {E.~W.}\ \bibnamefont
  {Kolb}}\ and\ \bibinfo {author} {\bibfnamefont {M.~S.}\ \bibnamefont
  {Turner}},\ }\href {https://cds.cern.ch/record/206230} {\emph {\bibinfo
  {title} {{The early universe}}}},\ Frontiers in Physics\ (\bibinfo
  {publisher} {Westview Press},\ \bibinfo {address} {Boulder, CO},\ \bibinfo
  {year} {1990})\BibitemShut {NoStop}%
\bibitem [{\citenamefont {Aghanim}\ \emph {et~al.}(2020)\citenamefont {Aghanim}
  \emph {et~al.}}]{Planck:2018vyg}%
  \BibitemOpen
  \bibfield  {author} {\bibinfo {author} {\bibfnamefont {N.}~\bibnamefont
  {Aghanim}} \emph {et~al.} (\bibinfo {collaboration} {Planck}),\ }\href
  {\doibase 10.1051/0004-6361/201833910} {\bibfield  {journal} {\bibinfo
  {journal} {Astron. Astrophys.}\ }\textbf {\bibinfo {volume} {641}},\ \bibinfo
  {pages} {A6} (\bibinfo {year} {2020})},\ \bibinfo {note} {[Erratum:
  Astron.Astrophys. 652, C4 (2021)]},\ \Eprint
  {http://arxiv.org/abs/1807.06209} {arXiv:1807.06209 [astro-ph.CO]}
  \BibitemShut {NoStop}%
\bibitem [{\citenamefont {Kohri}\ and\ \citenamefont
  {Terada}(2018)}]{Kohri:2018awv}%
  \BibitemOpen
  \bibfield  {author} {\bibinfo {author} {\bibfnamefont {K.}~\bibnamefont
  {Kohri}}\ and\ \bibinfo {author} {\bibfnamefont {T.}~\bibnamefont {Terada}},\
  }\href {\doibase 10.1103/PhysRevD.97.123532} {\bibfield  {journal} {\bibinfo
  {journal} {Phys. Rev.}\ }\textbf {\bibinfo {volume} {D97}},\ \bibinfo {pages}
  {123532} (\bibinfo {year} {2018})},\ \Eprint
  {http://arxiv.org/abs/1804.08577} {arXiv:1804.08577 [gr-qc]} \BibitemShut
  {NoStop}%
\bibitem [{\citenamefont {Clarke}\ \emph {et~al.}(2020)\citenamefont {Clarke},
  \citenamefont {Copeland},\ and\ \citenamefont {Moss}}]{Clarke:2020bil}%
  \BibitemOpen
  \bibfield  {author} {\bibinfo {author} {\bibfnamefont {T.~J.}\ \bibnamefont
  {Clarke}}, \bibinfo {author} {\bibfnamefont {E.~J.}\ \bibnamefont
  {Copeland}}, \ and\ \bibinfo {author} {\bibfnamefont {A.}~\bibnamefont
  {Moss}},\ }\href {\doibase 10.1088/1475-7516/2020/10/002} {\bibfield
  {journal} {\bibinfo  {journal} {JCAP}\ }\textbf {\bibinfo {volume} {10}},\
  \bibinfo {pages} {002} (\bibinfo {year} {2020})},\ \Eprint
  {http://arxiv.org/abs/2004.11396} {arXiv:2004.11396 [astro-ph.CO]}
  \BibitemShut {NoStop}%
\bibitem [{\citenamefont {Mitridate}(2023)}]{andrea_mitridate_2023}%
  \BibitemOpen
  \bibfield  {author} {\bibinfo {author} {\bibfnamefont {A.}~\bibnamefont
  {Mitridate}},\ }\href {\doibase 10.5281/zenodo.7876430} {\  (\bibinfo {year}
  {2023}),\ 10.5281/zenodo.7876430}\BibitemShut {NoStop}%
\bibitem [{\citenamefont {Mitridate}\ \emph {et~al.}(2023)\citenamefont
  {Mitridate}, \citenamefont {Wright}, \citenamefont {von Eckardstein},
  \citenamefont {Schr\"oder}, \citenamefont {Nay}, \citenamefont {Olum},
  \citenamefont {Schmitz},\ and\ \citenamefont {Trickle}}]{Mitridate:2023oar}%
  \BibitemOpen
  \bibfield  {author} {\bibinfo {author} {\bibfnamefont {A.}~\bibnamefont
  {Mitridate}}, \bibinfo {author} {\bibfnamefont {D.}~\bibnamefont {Wright}},
  \bibinfo {author} {\bibfnamefont {R.}~\bibnamefont {von Eckardstein}},
  \bibinfo {author} {\bibfnamefont {T.}~\bibnamefont {Schr\"oder}}, \bibinfo
  {author} {\bibfnamefont {J.}~\bibnamefont {Nay}}, \bibinfo {author}
  {\bibfnamefont {K.}~\bibnamefont {Olum}}, \bibinfo {author} {\bibfnamefont
  {K.}~\bibnamefont {Schmitz}}, \ and\ \bibinfo {author} {\bibfnamefont
  {T.}~\bibnamefont {Trickle}},\ }\href@noop {} {\  (\bibinfo {year} {2023})},\
  \Eprint {http://arxiv.org/abs/2306.16377} {arXiv:2306.16377 [hep-ph]}
  \BibitemShut {NoStop}%
\bibitem [{\citenamefont {Lamb}\ \emph {et~al.}(2023)\citenamefont {Lamb},
  \citenamefont {Taylor},\ and\ \citenamefont {van Haasteren}}]{lamb2023need}%
  \BibitemOpen
  \bibfield  {author} {\bibinfo {author} {\bibfnamefont {W.~G.}\ \bibnamefont
  {Lamb}}, \bibinfo {author} {\bibfnamefont {S.~R.}\ \bibnamefont {Taylor}}, \
  and\ \bibinfo {author} {\bibfnamefont {R.}~\bibnamefont {van Haasteren}},\
  }\href@noop {} {\enquote {\bibinfo {title} {The need for speed: Rapid
  refitting techniques for bayesian spectral characterization of the
  gravitational wave background using ptas},}\ } (\bibinfo {year} {2023}),\
  \Eprint {http://arxiv.org/abs/2303.15442} {arXiv:2303.15442 [astro-ph.HE]}
  \BibitemShut {NoStop}%
\bibitem [{\citenamefont {Young}\ \emph {et~al.}(2014)\citenamefont {Young},
  \citenamefont {Byrnes},\ and\ \citenamefont {Sasaki}}]{Young:2014ana}%
  \BibitemOpen
  \bibfield  {author} {\bibinfo {author} {\bibfnamefont {S.}~\bibnamefont
  {Young}}, \bibinfo {author} {\bibfnamefont {C.~T.}\ \bibnamefont {Byrnes}}, \
  and\ \bibinfo {author} {\bibfnamefont {M.}~\bibnamefont {Sasaki}},\ }\href
  {\doibase 10.1088/1475-7516/2014/07/045} {\bibfield  {journal} {\bibinfo
  {journal} {JCAP}\ }\textbf {\bibinfo {volume} {1407}},\ \bibinfo {pages}
  {045} (\bibinfo {year} {2014})},\ \Eprint {http://arxiv.org/abs/1405.7023}
  {arXiv:1405.7023 [gr-qc]} \BibitemShut {NoStop}%
\bibitem [{\citenamefont {Papanikolaou}\ \emph {et~al.}(2021)\citenamefont
  {Papanikolaou}, \citenamefont {Vennin},\ and\ \citenamefont
  {Langlois}}]{Papanikolaou:2020qtd}%
  \BibitemOpen
  \bibfield  {author} {\bibinfo {author} {\bibfnamefont {T.}~\bibnamefont
  {Papanikolaou}}, \bibinfo {author} {\bibfnamefont {V.}~\bibnamefont
  {Vennin}}, \ and\ \bibinfo {author} {\bibfnamefont {D.}~\bibnamefont
  {Langlois}},\ }\href {\doibase 10.1088/1475-7516/2021/03/053} {\bibfield
  {journal} {\bibinfo  {journal} {JCAP}\ }\textbf {\bibinfo {volume} {03}},\
  \bibinfo {pages} {053} (\bibinfo {year} {2021})},\ \Eprint
  {http://arxiv.org/abs/2010.11573} {arXiv:2010.11573 [astro-ph.CO]}
  \BibitemShut {NoStop}%
\bibitem [{\citenamefont {Bhaumik}\ \emph {et~al.}(2023)\citenamefont
  {Bhaumik}, \citenamefont {Jain},\ and\ \citenamefont
  {Lewicki}}]{Bhaumik:2023wmw}%
  \BibitemOpen
  \bibfield  {author} {\bibinfo {author} {\bibfnamefont {N.}~\bibnamefont
  {Bhaumik}}, \bibinfo {author} {\bibfnamefont {R.~K.}\ \bibnamefont {Jain}}, \
  and\ \bibinfo {author} {\bibfnamefont {M.}~\bibnamefont {Lewicki}},\
  }\href@noop {} {\  (\bibinfo {year} {2023})},\ \Eprint
  {http://arxiv.org/abs/2308.07912} {arXiv:2308.07912 [astro-ph.CO]}
  \BibitemShut {NoStop}%
\bibitem [{\citenamefont {Inomata}\ \emph
  {et~al.}(2019{\natexlab{a}})\citenamefont {Inomata}, \citenamefont {Kohri},
  \citenamefont {Nakama},\ and\ \citenamefont {Terada}}]{Inomata:2019ivs}%
  \BibitemOpen
  \bibfield  {author} {\bibinfo {author} {\bibfnamefont {K.}~\bibnamefont
  {Inomata}}, \bibinfo {author} {\bibfnamefont {K.}~\bibnamefont {Kohri}},
  \bibinfo {author} {\bibfnamefont {T.}~\bibnamefont {Nakama}}, \ and\ \bibinfo
  {author} {\bibfnamefont {T.}~\bibnamefont {Terada}},\ }\href {\doibase
  10.1103/PhysRevD.100.043532} {\bibfield  {journal} {\bibinfo  {journal}
  {Phys. Rev.}\ }\textbf {\bibinfo {volume} {D100}},\ \bibinfo {pages} {043532}
  (\bibinfo {year} {2019}{\natexlab{a}})},\ \bibinfo {note} {[Erratum:
  Phys.~Rev.~D 108, 049901 (2023)]},\ \Eprint {http://arxiv.org/abs/1904.12879}
  {arXiv:1904.12879 [astro-ph.CO]} \BibitemShut {NoStop}%
\bibitem [{\citenamefont {Inomata}\ \emph
  {et~al.}(2019{\natexlab{b}})\citenamefont {Inomata}, \citenamefont {Kohri},
  \citenamefont {Nakama},\ and\ \citenamefont {Terada}}]{Inomata:2019zqy}%
  \BibitemOpen
  \bibfield  {author} {\bibinfo {author} {\bibfnamefont {K.}~\bibnamefont
  {Inomata}}, \bibinfo {author} {\bibfnamefont {K.}~\bibnamefont {Kohri}},
  \bibinfo {author} {\bibfnamefont {T.}~\bibnamefont {Nakama}}, \ and\ \bibinfo
  {author} {\bibfnamefont {T.}~\bibnamefont {Terada}},\ }\href {\doibase
  10.1088/1475-7516/2019/10/071} {\bibfield  {journal} {\bibinfo  {journal}
  {JCAP}\ }\textbf {\bibinfo {volume} {10}},\ \bibinfo {pages} {071} (\bibinfo
  {year} {2019}{\natexlab{b}})},\ \bibinfo {note} {[Erratum: JCAP 08, E01
  (2023)]},\ \Eprint {http://arxiv.org/abs/1904.12878} {arXiv:1904.12878
  [astro-ph.CO]} \BibitemShut {NoStop}%
\bibitem [{\citenamefont {Kawasaki}\ and\ \citenamefont
  {Nakatsuka}(2019)}]{Kawasaki:2019mbl}%
  \BibitemOpen
  \bibfield  {author} {\bibinfo {author} {\bibfnamefont {M.}~\bibnamefont
  {Kawasaki}}\ and\ \bibinfo {author} {\bibfnamefont {H.}~\bibnamefont
  {Nakatsuka}},\ }\href {\doibase 10.1103/PhysRevD.99.123501} {\bibfield
  {journal} {\bibinfo  {journal} {Phys. Rev. D}\ }\textbf {\bibinfo {volume}
  {99}},\ \bibinfo {pages} {123501} (\bibinfo {year} {2019})},\ \Eprint
  {http://arxiv.org/abs/1903.02994} {arXiv:1903.02994 [astro-ph.CO]}
  \BibitemShut {NoStop}%
\bibitem [{\citenamefont {Young}\ \emph {et~al.}(2019)\citenamefont {Young},
  \citenamefont {Musco},\ and\ \citenamefont {Byrnes}}]{Young:2019yug}%
  \BibitemOpen
  \bibfield  {author} {\bibinfo {author} {\bibfnamefont {S.}~\bibnamefont
  {Young}}, \bibinfo {author} {\bibfnamefont {I.}~\bibnamefont {Musco}}, \ and\
  \bibinfo {author} {\bibfnamefont {C.~T.}\ \bibnamefont {Byrnes}},\ }\href
  {\doibase 10.1088/1475-7516/2019/11/012} {\bibfield  {journal} {\bibinfo
  {journal} {JCAP}\ }\textbf {\bibinfo {volume} {11}},\ \bibinfo {pages} {012}
  (\bibinfo {year} {2019})},\ \Eprint {http://arxiv.org/abs/1904.00984}
  {arXiv:1904.00984 [astro-ph.CO]} \BibitemShut {NoStop}%
\bibitem [{\citenamefont {De~Luca}\ \emph {et~al.}(2019)\citenamefont
  {De~Luca}, \citenamefont {Franciolini}, \citenamefont {Kehagias},
  \citenamefont {Peloso}, \citenamefont {Riotto},\ and\ \citenamefont
  {\"Unal}}]{DeLuca:2019qsy}%
  \BibitemOpen
  \bibfield  {author} {\bibinfo {author} {\bibfnamefont {V.}~\bibnamefont
  {De~Luca}}, \bibinfo {author} {\bibfnamefont {G.}~\bibnamefont
  {Franciolini}}, \bibinfo {author} {\bibfnamefont {A.}~\bibnamefont
  {Kehagias}}, \bibinfo {author} {\bibfnamefont {M.}~\bibnamefont {Peloso}},
  \bibinfo {author} {\bibfnamefont {A.}~\bibnamefont {Riotto}}, \ and\ \bibinfo
  {author} {\bibfnamefont {C.}~\bibnamefont {\"Unal}},\ }\href {\doibase
  10.1088/1475-7516/2019/07/048} {\bibfield  {journal} {\bibinfo  {journal}
  {JCAP}\ }\textbf {\bibinfo {volume} {07}},\ \bibinfo {pages} {048} (\bibinfo
  {year} {2019})},\ \Eprint {http://arxiv.org/abs/1904.00970} {arXiv:1904.00970
  [astro-ph.CO]} \BibitemShut {NoStop}%
\bibitem [{\citenamefont {Nakama}\ \emph {et~al.}(2017)\citenamefont {Nakama},
  \citenamefont {Silk},\ and\ \citenamefont {Kamionkowski}}]{Nakama:2016gzw}%
  \BibitemOpen
  \bibfield  {author} {\bibinfo {author} {\bibfnamefont {T.}~\bibnamefont
  {Nakama}}, \bibinfo {author} {\bibfnamefont {J.}~\bibnamefont {Silk}}, \ and\
  \bibinfo {author} {\bibfnamefont {M.}~\bibnamefont {Kamionkowski}},\ }\href
  {\doibase 10.1103/PhysRevD.95.043511} {\bibfield  {journal} {\bibinfo
  {journal} {Phys. Rev.}\ }\textbf {\bibinfo {volume} {D95}},\ \bibinfo {pages}
  {043511} (\bibinfo {year} {2017})},\ \Eprint
  {http://arxiv.org/abs/1612.06264} {arXiv:1612.06264 [astro-ph.CO]}
  \BibitemShut {NoStop}%
\bibitem [{\citenamefont {Cai}\ \emph {et~al.}(2019)\citenamefont {Cai},
  \citenamefont {Pi},\ and\ \citenamefont {Sasaki}}]{Cai:2018dig}%
  \BibitemOpen
  \bibfield  {author} {\bibinfo {author} {\bibfnamefont {R.-g.}\ \bibnamefont
  {Cai}}, \bibinfo {author} {\bibfnamefont {S.}~\bibnamefont {Pi}}, \ and\
  \bibinfo {author} {\bibfnamefont {M.}~\bibnamefont {Sasaki}},\ }\href
  {\doibase 10.1103/PhysRevLett.122.201101} {\bibfield  {journal} {\bibinfo
  {journal} {Phys. Rev. Lett.}\ }\textbf {\bibinfo {volume} {122}},\ \bibinfo
  {pages} {201101} (\bibinfo {year} {2019})},\ \Eprint
  {http://arxiv.org/abs/1810.11000} {arXiv:1810.11000 [astro-ph.CO]}
  \BibitemShut {NoStop}%
\bibitem [{\citenamefont {Unal}(2019)}]{Unal:2018yaa}%
  \BibitemOpen
  \bibfield  {author} {\bibinfo {author} {\bibfnamefont {C.}~\bibnamefont
  {Unal}},\ }\href {\doibase 10.1103/PhysRevD.99.041301} {\bibfield  {journal}
  {\bibinfo  {journal} {Phys. Rev. D}\ }\textbf {\bibinfo {volume} {99}},\
  \bibinfo {pages} {041301} (\bibinfo {year} {2019})},\ \Eprint
  {http://arxiv.org/abs/1811.09151} {arXiv:1811.09151 [astro-ph.CO]}
  \BibitemShut {NoStop}%
\bibitem [{\citenamefont {Yuan}\ and\ \citenamefont
  {Huang}(2021)}]{Yuan:2020iwf}%
  \BibitemOpen
  \bibfield  {author} {\bibinfo {author} {\bibfnamefont {C.}~\bibnamefont
  {Yuan}}\ and\ \bibinfo {author} {\bibfnamefont {Q.-G.}\ \bibnamefont
  {Huang}},\ }\href {\doibase 10.1016/j.physletb.2021.136606} {\bibfield
  {journal} {\bibinfo  {journal} {Phys. Lett. B}\ }\textbf {\bibinfo {volume}
  {821}},\ \bibinfo {pages} {136606} (\bibinfo {year} {2021})},\ \Eprint
  {http://arxiv.org/abs/2007.10686} {arXiv:2007.10686 [astro-ph.CO]}
  \BibitemShut {NoStop}%
\bibitem [{\citenamefont {Atal}\ and\ \citenamefont
  {Dom\`enech}(2021)}]{Atal:2021jyo}%
  \BibitemOpen
  \bibfield  {author} {\bibinfo {author} {\bibfnamefont {V.}~\bibnamefont
  {Atal}}\ and\ \bibinfo {author} {\bibfnamefont {G.}~\bibnamefont
  {Dom\`enech}},\ }\href {\doibase 10.1088/1475-7516/2021/06/001} {\bibfield
  {journal} {\bibinfo  {journal} {JCAP}\ }\textbf {\bibinfo {volume} {06}},\
  \bibinfo {pages} {001} (\bibinfo {year} {2021})},\ \Eprint
  {http://arxiv.org/abs/2103.01056} {arXiv:2103.01056 [astro-ph.CO]}
  \BibitemShut {NoStop}%
\bibitem [{\citenamefont {Adshead}\ \emph {et~al.}(2021)\citenamefont
  {Adshead}, \citenamefont {Lozanov},\ and\ \citenamefont
  {Weiner}}]{Adshead:2021hnm}%
  \BibitemOpen
  \bibfield  {author} {\bibinfo {author} {\bibfnamefont {P.}~\bibnamefont
  {Adshead}}, \bibinfo {author} {\bibfnamefont {K.~D.}\ \bibnamefont
  {Lozanov}}, \ and\ \bibinfo {author} {\bibfnamefont {Z.~J.}\ \bibnamefont
  {Weiner}},\ }\href {\doibase 10.1088/1475-7516/2021/10/080} {\bibfield
  {journal} {\bibinfo  {journal} {JCAP}\ }\textbf {\bibinfo {volume} {10}},\
  \bibinfo {pages} {080} (\bibinfo {year} {2021})},\ \Eprint
  {http://arxiv.org/abs/2105.01659} {arXiv:2105.01659 [astro-ph.CO]}
  \BibitemShut {NoStop}%
\bibitem [{\citenamefont {Garcia-Saenz}\ \emph {et~al.}(2023)\citenamefont
  {Garcia-Saenz}, \citenamefont {Pinol}, \citenamefont {Renaux-Petel},\ and\
  \citenamefont {Werth}}]{Garcia-Saenz:2022tzu}%
  \BibitemOpen
  \bibfield  {author} {\bibinfo {author} {\bibfnamefont {S.}~\bibnamefont
  {Garcia-Saenz}}, \bibinfo {author} {\bibfnamefont {L.}~\bibnamefont {Pinol}},
  \bibinfo {author} {\bibfnamefont {S.}~\bibnamefont {Renaux-Petel}}, \ and\
  \bibinfo {author} {\bibfnamefont {D.}~\bibnamefont {Werth}},\ }\href
  {\doibase 10.1088/1475-7516/2023/03/057} {\bibfield  {journal} {\bibinfo
  {journal} {JCAP}\ }\textbf {\bibinfo {volume} {03}},\ \bibinfo {pages} {057}
  (\bibinfo {year} {2023})},\ \Eprint {http://arxiv.org/abs/2207.14267}
  {arXiv:2207.14267 [astro-ph.CO]} \BibitemShut {NoStop}%
\bibitem [{\citenamefont {Li}\ \emph {et~al.}(2023)\citenamefont {Li},
  \citenamefont {Wang}, \citenamefont {Zhao},\ and\ \citenamefont
  {Kohri}}]{Li:2023qua}%
  \BibitemOpen
  \bibfield  {author} {\bibinfo {author} {\bibfnamefont {J.-P.}\ \bibnamefont
  {Li}}, \bibinfo {author} {\bibfnamefont {S.}~\bibnamefont {Wang}}, \bibinfo
  {author} {\bibfnamefont {Z.-C.}\ \bibnamefont {Zhao}}, \ and\ \bibinfo
  {author} {\bibfnamefont {K.}~\bibnamefont {Kohri}},\ }\href@noop {} {\
  (\bibinfo {year} {2023})},\ \Eprint {http://arxiv.org/abs/2305.19950}
  {arXiv:2305.19950 [astro-ph.CO]} \BibitemShut {NoStop}%
\bibitem [{\citenamefont {Yuan}\ \emph {et~al.}(2023)\citenamefont {Yuan},
  \citenamefont {Meng},\ and\ \citenamefont {Huang}}]{Yuan:2023ofl}%
  \BibitemOpen
  \bibfield  {author} {\bibinfo {author} {\bibfnamefont {C.}~\bibnamefont
  {Yuan}}, \bibinfo {author} {\bibfnamefont {D.-S.}\ \bibnamefont {Meng}}, \
  and\ \bibinfo {author} {\bibfnamefont {Q.-G.}\ \bibnamefont {Huang}},\
  }\href@noop {} {\  (\bibinfo {year} {2023})},\ \Eprint
  {http://arxiv.org/abs/2308.07155} {arXiv:2308.07155 [astro-ph.CO]}
  \BibitemShut {NoStop}%
\bibitem [{\citenamefont {Abe}\ \emph {et~al.}(2021)\citenamefont {Abe},
  \citenamefont {Tada},\ and\ \citenamefont {Ueda}}]{Abe:2020sqb}%
  \BibitemOpen
  \bibfield  {author} {\bibinfo {author} {\bibfnamefont {K.~T.}\ \bibnamefont
  {Abe}}, \bibinfo {author} {\bibfnamefont {Y.}~\bibnamefont {Tada}}, \ and\
  \bibinfo {author} {\bibfnamefont {I.}~\bibnamefont {Ueda}},\ }\href {\doibase
  10.1088/1475-7516/2021/06/048} {\bibfield  {journal} {\bibinfo  {journal}
  {JCAP}\ }\textbf {\bibinfo {volume} {06}},\ \bibinfo {pages} {048} (\bibinfo
  {year} {2021})},\ \Eprint {http://arxiv.org/abs/2010.06193} {arXiv:2010.06193
  [astro-ph.CO]} \BibitemShut {NoStop}%
\bibitem [{\citenamefont {Franciolini}\ \emph
  {et~al.}(2023{\natexlab{b}})\citenamefont {Franciolini}, \citenamefont
  {Racco},\ and\ \citenamefont {Rompineve}}]{Franciolini:2023wjm}%
  \BibitemOpen
  \bibfield  {author} {\bibinfo {author} {\bibfnamefont {G.}~\bibnamefont
  {Franciolini}}, \bibinfo {author} {\bibfnamefont {D.}~\bibnamefont {Racco}},
  \ and\ \bibinfo {author} {\bibfnamefont {F.}~\bibnamefont {Rompineve}},\
  }\href@noop {} {\  (\bibinfo {year} {2023}{\natexlab{b}})},\ \Eprint
  {http://arxiv.org/abs/2306.17136} {arXiv:2306.17136 [astro-ph.CO]}
  \BibitemShut {NoStop}%
\bibitem [{\citenamefont {Yuan}\ \emph {et~al.}(2019)\citenamefont {Yuan},
  \citenamefont {Chen},\ and\ \citenamefont {Huang}}]{Yuan:2019udt}%
  \BibitemOpen
  \bibfield  {author} {\bibinfo {author} {\bibfnamefont {C.}~\bibnamefont
  {Yuan}}, \bibinfo {author} {\bibfnamefont {Z.-C.}\ \bibnamefont {Chen}}, \
  and\ \bibinfo {author} {\bibfnamefont {Q.-G.}\ \bibnamefont {Huang}},\ }\href
  {\doibase 10.1103/PhysRevD.100.081301} {\bibfield  {journal} {\bibinfo
  {journal} {Phys. Rev. D}\ }\textbf {\bibinfo {volume} {100}},\ \bibinfo
  {pages} {081301} (\bibinfo {year} {2019})},\ \Eprint
  {http://arxiv.org/abs/1906.11549} {arXiv:1906.11549 [astro-ph.CO]}
  \BibitemShut {NoStop}%
\bibitem [{\citenamefont {Zhou}\ \emph {et~al.}(2022)\citenamefont {Zhou},
  \citenamefont {Zhang}, \citenamefont {Zhu},\ and\ \citenamefont
  {Chang}}]{Zhou:2021vcw}%
  \BibitemOpen
  \bibfield  {author} {\bibinfo {author} {\bibfnamefont {J.-Z.}\ \bibnamefont
  {Zhou}}, \bibinfo {author} {\bibfnamefont {X.}~\bibnamefont {Zhang}},
  \bibinfo {author} {\bibfnamefont {Q.-H.}\ \bibnamefont {Zhu}}, \ and\
  \bibinfo {author} {\bibfnamefont {Z.}~\bibnamefont {Chang}},\ }\href
  {\doibase 10.1088/1475-7516/2022/05/013} {\bibfield  {journal} {\bibinfo
  {journal} {JCAP}\ }\textbf {\bibinfo {volume} {05}},\ \bibinfo {pages} {013}
  (\bibinfo {year} {2022})},\ \Eprint {http://arxiv.org/abs/2106.01641}
  {arXiv:2106.01641 [astro-ph.CO]} \BibitemShut {NoStop}%
\bibitem [{\citenamefont {Chang}\ \emph {et~al.}(2023)\citenamefont {Chang},
  \citenamefont {Kuang}, \citenamefont {Zhang},\ and\ \citenamefont
  {Zhou}}]{Chang:2022nzu}%
  \BibitemOpen
  \bibfield  {author} {\bibinfo {author} {\bibfnamefont {Z.}~\bibnamefont
  {Chang}}, \bibinfo {author} {\bibfnamefont {Y.-T.}\ \bibnamefont {Kuang}},
  \bibinfo {author} {\bibfnamefont {X.}~\bibnamefont {Zhang}}, \ and\ \bibinfo
  {author} {\bibfnamefont {J.-Z.}\ \bibnamefont {Zhou}},\ }\href {\doibase
  10.1088/1674-1137/acc649} {\bibfield  {journal} {\bibinfo  {journal} {Chin.
  Phys. C}\ }\textbf {\bibinfo {volume} {47}},\ \bibinfo {pages} {055104}
  (\bibinfo {year} {2023})},\ \Eprint {http://arxiv.org/abs/2209.12404}
  {arXiv:2209.12404 [astro-ph.CO]} \BibitemShut {NoStop}%
\bibitem [{\citenamefont {{Mr{\'o}z}}\ \emph {et~al.}(2017)\citenamefont
  {{Mr{\'o}z}}, \citenamefont {{Udalski}}, \citenamefont {{Skowron}},
  \citenamefont {{Poleski}}, \citenamefont {{Koz{\l}owski}}, \citenamefont
  {{Szyma{\'n}ski}}, \citenamefont {{Soszy{\'n}ski}}, \citenamefont
  {{Wyrzykowski}}, \citenamefont {{Pietrukowicz}}, \citenamefont {{Ulaczyk}},
  \citenamefont {{Skowron}},\ and\ \citenamefont
  {{Pawlak}}}]{2017Natur.548..183M}%
  \BibitemOpen
  \bibfield  {author} {\bibinfo {author} {\bibfnamefont {P.}~\bibnamefont
  {{Mr{\'o}z}}}, \bibinfo {author} {\bibfnamefont {A.}~\bibnamefont
  {{Udalski}}}, \bibinfo {author} {\bibfnamefont {J.}~\bibnamefont
  {{Skowron}}}, \bibinfo {author} {\bibfnamefont {R.}~\bibnamefont
  {{Poleski}}}, \bibinfo {author} {\bibfnamefont {S.}~\bibnamefont
  {{Koz{\l}owski}}}, \bibinfo {author} {\bibfnamefont {M.~K.}\ \bibnamefont
  {{Szyma{\'n}ski}}}, \bibinfo {author} {\bibfnamefont {I.}~\bibnamefont
  {{Soszy{\'n}ski}}}, \bibinfo {author} {\bibfnamefont {{\L}.}~\bibnamefont
  {{Wyrzykowski}}}, \bibinfo {author} {\bibfnamefont {P.}~\bibnamefont
  {{Pietrukowicz}}}, \bibinfo {author} {\bibfnamefont {K.}~\bibnamefont
  {{Ulaczyk}}}, \bibinfo {author} {\bibfnamefont {D.}~\bibnamefont
  {{Skowron}}}, \ and\ \bibinfo {author} {\bibfnamefont {M.}~\bibnamefont
  {{Pawlak}}},\ }\href {\doibase 10.1038/nature23276} {\bibfield  {journal}
  {\bibinfo  {journal} {\nat}\ }\textbf {\bibinfo {volume} {548}},\ \bibinfo
  {pages} {183} (\bibinfo {year} {2017})},\ \Eprint
  {http://arxiv.org/abs/1707.07634} {arXiv:1707.07634 [astro-ph.EP]}
  \BibitemShut {NoStop}%
\bibitem [{\citenamefont {Niikura}\ \emph
  {et~al.}(2019{\natexlab{a}})\citenamefont {Niikura}, \citenamefont {Takada},
  \citenamefont {Yasuda}, \citenamefont {Lupton}, \citenamefont {Sumi},
  \citenamefont {More}, \citenamefont {More}, \citenamefont {Oguri},\ and\
  \citenamefont {Chiba}}]{Niikura:2017zjd}%
  \BibitemOpen
  \bibfield  {author} {\bibinfo {author} {\bibfnamefont {H.}~\bibnamefont
  {Niikura}}, \bibinfo {author} {\bibfnamefont {M.}~\bibnamefont {Takada}},
  \bibinfo {author} {\bibfnamefont {N.}~\bibnamefont {Yasuda}}, \bibinfo
  {author} {\bibfnamefont {R.~H.}\ \bibnamefont {Lupton}}, \bibinfo {author}
  {\bibfnamefont {T.}~\bibnamefont {Sumi}}, \bibinfo {author} {\bibfnamefont
  {S.}~\bibnamefont {More}}, \bibinfo {author} {\bibfnamefont {A.}~\bibnamefont
  {More}}, \bibinfo {author} {\bibfnamefont {M.}~\bibnamefont {Oguri}}, \ and\
  \bibinfo {author} {\bibfnamefont {M.}~\bibnamefont {Chiba}},\ }\href
  {\doibase 10.1038/s41550-019-0723-1} {\bibfield  {journal} {\bibinfo
  {journal} {Nature Astron.}\ }\textbf {\bibinfo {volume} {3}},\ \bibinfo
  {pages} {524} (\bibinfo {year} {2019}{\natexlab{a}})},\ \Eprint
  {http://arxiv.org/abs/1701.02151} {arXiv:1701.02151 [astro-ph.CO]}
  \BibitemShut {NoStop}%
\bibitem [{\citenamefont {Niikura}\ \emph
  {et~al.}(2019{\natexlab{b}})\citenamefont {Niikura}, \citenamefont {Takada},
  \citenamefont {Yokoyama}, \citenamefont {Sumi},\ and\ \citenamefont
  {Masaki}}]{Niikura:2019kqi}%
  \BibitemOpen
  \bibfield  {author} {\bibinfo {author} {\bibfnamefont {H.}~\bibnamefont
  {Niikura}}, \bibinfo {author} {\bibfnamefont {M.}~\bibnamefont {Takada}},
  \bibinfo {author} {\bibfnamefont {S.}~\bibnamefont {Yokoyama}}, \bibinfo
  {author} {\bibfnamefont {T.}~\bibnamefont {Sumi}}, \ and\ \bibinfo {author}
  {\bibfnamefont {S.}~\bibnamefont {Masaki}},\ }\href {\doibase
  10.1103/PhysRevD.99.083503} {\bibfield  {journal} {\bibinfo  {journal} {Phys.
  Rev. D}\ }\textbf {\bibinfo {volume} {99}},\ \bibinfo {pages} {083503}
  (\bibinfo {year} {2019}{\natexlab{b}})},\ \Eprint
  {http://arxiv.org/abs/1901.07120} {arXiv:1901.07120 [astro-ph.CO]}
  \BibitemShut {NoStop}%
\bibitem [{\citenamefont {Wang}\ \emph {et~al.}(2019)\citenamefont {Wang},
  \citenamefont {Terada},\ and\ \citenamefont {Kohri}}]{Wang:2019kaf}%
  \BibitemOpen
  \bibfield  {author} {\bibinfo {author} {\bibfnamefont {S.}~\bibnamefont
  {Wang}}, \bibinfo {author} {\bibfnamefont {T.}~\bibnamefont {Terada}}, \ and\
  \bibinfo {author} {\bibfnamefont {K.}~\bibnamefont {Kohri}},\ }\href
  {\doibase 10.1103/PhysRevD.99.103531} {\bibfield  {journal} {\bibinfo
  {journal} {Phys. Rev. D}\ }\textbf {\bibinfo {volume} {99}},\ \bibinfo
  {pages} {103531} (\bibinfo {year} {2019})},\ \bibinfo {note} {[Erratum:
  Phys.Rev.D 101, 069901 (2020)]},\ \Eprint {http://arxiv.org/abs/1903.05924}
  {arXiv:1903.05924 [astro-ph.CO]} \BibitemShut {NoStop}%
\bibitem [{\citenamefont {Co}\ \emph {et~al.}(2020)\citenamefont {Co},
  \citenamefont {Hall},\ and\ \citenamefont {Harigaya}}]{Co:2019jts}%
  \BibitemOpen
  \bibfield  {author} {\bibinfo {author} {\bibfnamefont {R.~T.}\ \bibnamefont
  {Co}}, \bibinfo {author} {\bibfnamefont {L.~J.}\ \bibnamefont {Hall}}, \ and\
  \bibinfo {author} {\bibfnamefont {K.}~\bibnamefont {Harigaya}},\ }\href
  {\doibase 10.1103/PhysRevLett.124.251802} {\bibfield  {journal} {\bibinfo
  {journal} {Phys. Rev. Lett.}\ }\textbf {\bibinfo {volume} {124}},\ \bibinfo
  {pages} {251802} (\bibinfo {year} {2020})},\ \Eprint
  {http://arxiv.org/abs/1910.14152} {arXiv:1910.14152 [hep-ph]} \BibitemShut
  {NoStop}%
\bibitem [{\citenamefont {Er\"oncel}\ \emph {et~al.}(2022)\citenamefont
  {Er\"oncel}, \citenamefont {Sato}, \citenamefont {Servant},\ and\
  \citenamefont {S\o{}rensen}}]{Eroncel:2022vjg}%
  \BibitemOpen
  \bibfield  {author} {\bibinfo {author} {\bibfnamefont {C.}~\bibnamefont
  {Er\"oncel}}, \bibinfo {author} {\bibfnamefont {R.}~\bibnamefont {Sato}},
  \bibinfo {author} {\bibfnamefont {G.}~\bibnamefont {Servant}}, \ and\
  \bibinfo {author} {\bibfnamefont {P.}~\bibnamefont {S\o{}rensen}},\ }\href
  {\doibase 10.1088/1475-7516/2022/10/053} {\bibfield  {journal} {\bibinfo
  {journal} {JCAP}\ }\textbf {\bibinfo {volume} {10}},\ \bibinfo {pages} {053}
  (\bibinfo {year} {2022})},\ \Eprint {http://arxiv.org/abs/2206.14259}
  {arXiv:2206.14259 [hep-ph]} \BibitemShut {NoStop}%
\end{thebibliography}%

\end{document}